\documentclass [11pt, twoside] {uwthesis}
 
\usepackage{alltt}  %
\usepackage{amsmath,amssymb}
\usepackage{graphicx}
\usepackage{cite}

\def\N{{\mathcal{N}}}
\newcommand{\beq}{\begin{equation}}
\newcommand{\eeq}{\end{equation}}
\newcommand{\bea}{\begin{eqnarray}}
\newcommand{\eea}{\end{eqnarray}}
\newcommand{\bdm}{\begin{displaymath}}
\newcommand{\edm}{\end{displaymath}}

\newcommand{\ra}{\rightarrow}
\renewcommand{\theequation}{\thesection.\arabic{equation}}
\def\<{\langle}
\def\>{\rangle}
\def\O{\mathcal{O}}
\def\Om{\O_m}
\def\Omv{\langle \Om \rangle}
\def\<{\langle}
\def\>{\rangle}

\def\a{\alpha}
\def\b{\beta}
\def\d{\delta}
\def\e{\epsilon}           
  
\def\g{\gamma}

\def\lam{\lambda}
\def\m{\mu}
\def\n{\nu}
  
\def\p{\pi}                
\def\th{\theta}                   
\def\s{\sigma}                                   

\def\G{\Gamma}

\begin{document}
 
%

\prelimpages
 
%
%
\Title{Holographic Thermodynamics and Transport of Flavor Fields}
\Author{Andrew Hill O'Bannon}
\Year{2008}
\Program{Physics}

{\Degreetext{A dissertation
  submitted in partial fulfillment of\\
  the requirements for the degree of}
 \def\thefootnote{\fnsymbol{footnote}}
 \let\footnoterule\relax
 \titlepage
 }
\setcounter{footnote}{0}
 
%
%
%
\Chair{Andreas Karch}{Assistant Professor}{Physics}

\Signature{Andreas Karch}
\Signature{Dam Son}
\Signature{Laurence Yaffe}
\signaturepage

%
%

\setcounter{page}{-1}
\abstract{We use gauge-gravity duality to study a strongly-coupled non-Abelian gauge theory with flavor fields, \textit{i.e.} fields transforming in the fundamental representation of the gauge group. We first study the thermodynamics of the flavor fields. In the grand canonical ensemble at zero temperature, we find a second-order transition when the mass of the flavor fields equals the chemical potential. We then study the transport properties of the flavor fields at finite temperature and density. We introduce external electric and magnetic fields and compute the resulting current of flavor charge. From this current we extract the conductivity, using Ohm's law. In addition, we compute the drag force on the flavor fields at large mass, in the presence of a finite baryon density and external electric and magnetic fields.
}
 
%
%
\tableofcontents
\listoffigures

%
%

\textpages

\chapter {Introduction}\label{intro}

\section{Motivation: QCD}

All known matter interacts via only four forces: gravity, electromagnetism, the weak nuclear force, and the strong nuclear force. Gravity is perhaps the most mysterious. Why is the cosmological constant so small? What is dark energy? How do we quantize gravity? The Standard Model of particle physics describes the remaining three forces, at least at energies below about $10^{12}$ electron-Volts. Electromagnetism is extremely well-understood, thanks to the spectacular success of Quantum Electrodynamics (QED). The weak force is also well-understood, though big questions remain, for example about the exact nature of neutrinos and whether the Higgs exists.

The strong nuclear force also remains mysterious. The accepted theory of the strong nuclear force is Quantum Chromodynamics (QCD). QCD is a non-Abelian gauge theory with gauge group $SU(3)$. The theory thus has gauge bosons, massless spin-1 fields called ``gluons,'' transforming in the adjoint representation of the gauge group. The theory also has matter fields, massive spin-1/2 fields called ``quarks,'' transforming in the fundamental representation of the gauge group.

QCD has ``asymptotic freedom'' \cite{Gross:1973id,Politzer:1973fx}. An asymptotically free theory is a theory whose coupling strength \textit{decreases} as the renormalization scale \textit{increases}, and vice-versa. Physically, in an asymptotically free theory, particles scattering at higher and higher energy will interact more and more weakly, while particles scattering at lower and lower energy will interact more and more strongly. In QCD, the renormalization scale at which the coupling becomes order one is called $\Lambda_{QCD}$. This is one of the fundamental energy scales of QCD, in addition to the quark masses. When discussing QCD, we will refer to ``low-energy'' and ``high-energy'' regimes, where ``low'' and ``high'' are with respect to $\Lambda_{QCD}$, which is roughly $200$ million electron-Volts.

Standard ``textbook'' methods for computing observables in quantum field theories rely on a valid perturbative expansion: assume the theory is almost free and calculate observables in a term-by-term expansion in a small parameter, namely the coupling. These methods were invented for QED, whose coupling remains perturbatively small over a very wide range of energies (far above the Planck scale, if we na\"{i}vely extrapolate up in energy).

For QCD, on the other hand, perturbation theory is only reliable in high-energy regimes, due to asymptotic freedom. Theorists thus go looking for dials to turn to reach high energy. For example, we can introduce a temperature in QCD. At very high temperatures, far above $\Lambda_{QCD}$, we expect excitations to scatter with energies on the order of the temperature, and hence to be weakly interacting. We can then ask questions and compute the answers reliably using perturbation theory. For instance, we can ask the questions of thermodynamics: what is the pressure of a thermal bath of quarks and gluons? The pressure of hot QCD has in fact been computed perturbatively to very high order in the coupling \cite{Kajantie:2002wa,Vuorinen:2003fs,DiRenzo:2006nh}. Similarly, we can introduce a finite chemical potential in QCD. For a chemical potential far above $\Lambda_{QCD}$ (and far above the temperature), we expect to find a Fermi surface of quarks, and we expect excitations about that Fermi surface to scatter with energies on the order of the chemical potential. Perturbation theory should thus be reliable again, and indeed, has provided truly amazing insights into the nature of extremely dense quark matter \cite{Rajagopal:2000wf}. These systems are not purely theoretical: the early universe was very hot and neutron stars are very dense.

Of course, an enormous amount of physics occurs in the low-energy regime where QCD is strongly coupled and perturbation theory is no longer reliable. We know from experiments (for example, deep inelastic scattering) that at high energy the appropriate degrees of freedom of QCD are quarks and gluons, but at low energy the particles observed in accelerators are not quarks and gluons, but hadrons: baryons and mesons. We thus say that at low energy quarks are ``confined'' in hadrons, that is, a single, isolated quark has never been observed in vacuum, in dramatic contrast to, say, electrons or muons.

The fundamental problem of QCD is easy to pose: given the high-energy theory, namely the path integral of QCD, derive the low-energy spectrum, \textit{i.e.} the observed spectrum of hadrons, as well as all of their interactions. To date, all attempts to solve this problem have failed. We still do not understand the dynamical origin of confinement in QCD. In one form or another, the obstacle to solving the problem is always the fact that the theory is strongly interacting at low energy.

Perturbation theory is not the only tool theorists have, of course. A number of non-perturbative techniques have been developed over the past few decades. One of the most versatile, and successful, is lattice QCD. In lattice QCD, spacetime is represented by a finite number of lattice sites, and quantum fields as degrees of freedom on that lattice\footnote{Some standard textbooks on lattice QCD are refs. \cite{Creutz:1984mg,Montvay:1994cy,Smit:2002ug,Rothe:2005nw}.}. The QCD path integral is then a finite number of finite-dimensional integrals. We can thus compute observables by ``brute force,'' using a computer. Indeed, in this fashion we can compute any observable we want, at least in principle.

The many non-perturbative methods developed for QCD can answer an enormous number of questions about hadronic matter, but, unfortunately, they have yet to solve the fundamental problem of QCD. With lattice QCD, for example, we can compute the low-energy spectrum of QCD, but while a numerical simulation may give us an answer, it does not \textit{explain} the dynamical mechanism of confinement.

Moreover, even for less ambitious questions, many non-perturbative methods have technical problems that prohibit their use. For example, the lattice is very good at answering questions about thermal physics of \textit{equilibrium} states, but not very good at answering questions about thermal physics out of equilibrium, that is, about time-dependent dynamics. Furthermore, lattice QCD also faces technical obstacles even for certain equilibrium states: currently, no one knows how to simulate a finite density of quarks in lattice QCD, for instance. The problem in both cases is essentially the same. Lattice simulations are done in Euclidean signature via Monte-Carlo sampling of field configurations, with weighting factor $e^{-S}$, with $S$ the Euclidean QCD action. Real-time dynamics requires Lorentzian signature, hence the weighting factor becomes $e^{iS}$ with $S$ the Lorentzian-signature action, and similarly a chemical potential introduces a factor of $i$ into the action, so that in both cases the weighting factor oscillates, making Monte-Carlo impossible in practice (though, strictly speaking, not in principle)\footnote{In the actual practice of lattice QCD, the fermions are integrated out and hence represented as a determinant in the path integral. This just shifts the problems around, though. Introducing a chemical potential, for example, makes the fermion determinant complex, producing the so-called ``sign problem,'' which is equivalent to what we have described.}.

The situation is especially dire given the recent experimental results from the Relativistic Heavy Ion Collider\footnote{A good pedagogical review of heavy-ion physics is ref. \cite{Heinz:2004qz}.} (RHIC). The basic result of RHIC was that, for temperatures around $250$ million electron-Volts, the best description of QCD is ideal hydrodynamics \cite{Kolb:2000fha,Teaney:2001av,Kolb:2003dz,Shuryak:2003xe,Hirano:2004er,Shuryak:2004cy}. In other words, the simulations that best fit (certain appropriate) observables were not performed in lattice QCD, but rather in hydrodynamic models, and in fact in hydrodynamic models where the shear viscosity was set to \textit{zero} (hence ``ideal'').

That hydrodynamics was a good description is not difficult to understand. Many people were surprised that \textit{ideal} hydrodynamics worked so well, though. To explain why, we need to review some of the physics of hydrodynamics.

From a modern perspective, hydrodynamics is just an effective field theory. In the ocean, the mean free path of a single water molecule is on the order of nanometers, while ocean waves have wavelengths on the order of meters or longer. If we only want to ask questions about the physics of the low-energy degrees of freedom, the water waves, then a description using the microscopic degrees of freedom, the water molecules, will probably not be easy to use, given that the interaction between water molecules may be very complicated.

We have a large separation of scales, however, so we can follow the rules of effective field theory and write equations of motion\footnote{We do not start with an effective Lagrangian because of dissipation.} including \textit{every possible} term consistent with the symmetries. For liquid water the symmetries are rotation and translation invariance. The effective theory will be valid up to some UV cutoff, that is, down to some length scale somewhere between a meter and a nanometer. Higher-derivative terms will be suppressed by inverse powers of the cutoff, so we can keep leading terms up to some order in the cutoff, giving us a theory with a manageable (read finite) number of terms. All these terms will \textit{a priori} have undetermined coefficients, however. In hydrodynamics these coefficients are called transport coefficients, an example being the shear viscosity, $\eta$. We cannot compute anything with our effective theory until we determine these coefficients. In principle, we can compute the transport coefficients from a microscopic theory. For most real materials, though, such as water, the microscopic theory is very complicated, so no one would even try to compute the transport coefficients. Fortunately, in the real world, we can usually just \textit{measure} transport coefficients, and then use the effective theory to make predictions.

The fact that hydrodynamic modeling of RHIC collisions works at all indicates that the mean free path inside the RHIC plasma must be much smaller than the size of the plasma \cite{Shuryak:2003xe,Shuryak:2004cy}, which is about\footnote{At RHIC, gold nuclei are used in collisions. A gold nucleus is about $15\times10^{-15}$ meters in size, so that their collision produces a plasma of the size quoted above. The exact size in a particular collision depends on how much the gold nuclei overlap when they collide (\textit{i.e.} on the impact parameter, or equivalently the ``centrality'' of the collision).} $10^{-14}$ meters. A short mean free path in turn suggests that the system is strongly-coupled \cite{Shuryak:2003xe,Shuryak:2004cy}, so that an excitation will not travel very far before experiencing a collision. That the RHIC plasma is strongly-interacting is not surprising, given that RHIC produces a plasma at temperatures on the order of $\Lambda_{QCD}$, where we expect the coupling to be order-one anyway.

The fact that \textit{ideal} hydrodynamics worked so well was totally unexpected, however. Before the experiments at RHIC, the only estimates for the viscosity came from perturbation theory \cite{Arnold:2000dr,Arnold:2003zc}. The actual visocsity appears to be much smaller than these estimates. The big question that arose from RHIC was thus: why does the viscosity of QCD appear to be so small at RHIC temperatures?

Theorists would love to answer this question by \textit{computing} the viscosity from the underlying microscopic theory, QCD. The problem, once again, is strong coupling. Perturbation theory will not be reliable. RHIC collisions involve real-time, out-of-equilibrium physics that lattice QCD cannot reliably address. In short, as this point in history, \textit{no one knows} how to compute the viscosity of the RHIC plasma from QCD!

That could have been the end of the story. At this point, however, like a bolt out of the blue, \textit{quantum gravity} enters the picture, via the anti-de Sitter / Conformal Field Theory (AdS/CFT) correspondence \cite{Maldacena:1997re}. Let us pause to introduce AdS/CFT\footnote{Two standard reviews of the AdS/CFT correspondence are refs. \cite{magoo,DHoker:2002aw}.}, and then discuss what it has to do with the viscosity of QCD.

The original AdS/CFT conjecture was that a conformal field theory, $\N = 4$ supersymmetric $SU(N_c)$ Yang-Mills theory (SYM), is in fact \textit{equivalent} to a theory of quantum gravity, type IIB string theory, formulated on background geometries that asymptotically approach $AdS_5 \times S^5$, where $AdS_5$ is (4+1)-dimensional anti-de Sitter space and $S^5$ is a five-sphere \cite{Maldacena:1997re}. Anti- de Sitter spaces are defined as spaces of constant negative curvature. They are naturally equipped with a radial coordinate, and also with a boundary at some asymptotic value of that radial coordinate. The SYM theory in some sense ``lives'' on this (3+1)-dimensional boundary. We will review $\N=4$ SYM theory and AdS spaces below, in sections \ref{n4symreview} and \ref{adsspace}.

What do we mean that the two theories are ``equivalent?'' AdS/CFT is the statement that the two theories, $\N=4$ SYM theory and type IIB string theory on $AdS_5 \times S^5$, are \textit{identical}: they are merely two different \textit{descriptions} of the \textit{same} physics. If this is true, then all of the physics of one description must map onto all of the physics of the other description somehow. We will review the ``dictionary'' of how to translate between the two theories in section \ref{statement}.

What makes AdS/CFT \textit{useful} is that it is a \textit{duality}: when one theory is weakly-coupled, the other is strongly-coupled, and vice-versa. AdS/CFT is thus especially useful for answering questions about the $\N=4$ SYM theory \textit{at strong coupling}. To see this explicitly, we start with two key entries in the AdS/CFT dictionary \cite{Maldacena:1997re},
\beq
\frac{L^4}{\alpha'^2} = 4 \pi g_s N_c, \qquad 2 \pi \, g_s = g_{YM}^2 \nonumber
\eeq
Here $L$ is the radius of curvature of both the $AdS_5$ and $S^5$ factors of the background, $g_s$ is the string coupling, $N_c$ is the number of colors of the $\N=4$ SYM theory, and $g_{YM}$ is the coupling of the $\N=4$ SYM theory\footnote{We follow the conventions of D-brane physics (which we review in section \ref{dbranes}), for which the definition of $g_{YM}^2$ is actually half the conventional value. The discrepancy comes from the normalization of the generators of the gauge group (in our case, $SU(N_c)$): $Tr(T_a T_b) = d \, \delta_{ab}$. The usual choice in D-brane physics is $d=1$, whereas the conventional choice in field theory is $d=1/2$.}. $\alpha'$ is related to the tension, $T_s$, or energy per unit length, of strings in string theory by $T_s = 1/2\pi\alpha'$. The string tension is a fundamental dimensionful scale in string theory, which, via some dimensional analysis, gives us a length scale, the string length $\ell_s \equiv \sqrt{\alpha'}$. Notice that not all string theory strings have the length $\ell_s$, rather, it is a characteristic scale at which the finite length of strings becomes important.

Currently, only perturbative string theory is under good theoretical control, so we want to take $g_s \ll 1$. At leading order, we then have classical string theory, in the sense of ``tree-level in the string genus expansion.'' If we keep $N_c$ fixed as we take $g_s \ll 1$, then we see that $L^4 \ll \alpha'^2$, meaning that the background is very highly curved: the radius of curvature is smaller than the string scale!

Currently, no one knows the full string spectrum for strings propagating in $AdS_5 \times S^5$, because string theory is difficult to quantize in this background, but dimensional analysis tells us that the mass-squareds must be proportional to $\alpha'^{-1}$. When $L^4 \ll \alpha'^2$, the massive string modes will be very light relative to the natural mass scale of the background, $L^{-1}$. Since we do not know the spectrum of these modes, we would like to avoid this limit.

We are thus motivated to take $N_c \ra \infty$ as we take $g_s \ra 0$ in such a fashion that $g_s N_c$ remains fixed, and then take an additional limit in which $g_s N_c \gg 1$, so that $L^4 \gg \alpha'^2$. The massive string states will then be very heavy, and we can justifiably write an effective theory of massless modes alone. In this classical, low-energy limit, the effective theory of type IIB string theory is known: it is type IIB supergravity.

Notice what this limit translates into in the gauge theory: the 't Hooft limit \cite{tHooft:1973jz}, with large 't Hooft coupling. In other words, we define the 't Hooft coupling $\lambda \equiv 2 \, g_{YM}^2 N_c$, which is the physical coupling of the gauge theory. We then take two limits. First, we take $N_c \ra \infty$ with $g_{YM}^2 \ra 0$ such that $\lambda$ is fixed. Second, we take a strong-coupling limit, $\lambda \ra \infty$. AdS/CFT is telling us that this \textit{strongly-coupled} gauge theory is in fact equivalent to a \textit{classical} theory of \textit{gravity}. We thus have a \textit{new method} for computing observables in strongly-coupled gauge theories: classical gravity.

The first objection, of course, is that $\N=4$ SYM is not QCD! Indeed, the two theories are vastly different. $\N=4$ SYM has much more symmetry than QCD, namely supersymmetry and conformal symmetry. Supersymmetry is why $\N=4$ SYM has many more fields than QCD, namely fermions and scalars in the adjoint representation of the gauge group (in a supermultiplet with the gluons). Conformal symmetry is why, where the QCD coupling shrinks with increasing energy scale, the coupling of $\N=4$ SYM does not change at all as the energy scale changes. Furthermore, in $\N=4$ SYM, as in any conformal theory, we cannot even \textit{define} an S-matrix, or even excitations that we can call particles, since we cannot define asymptotic states.

The original AdS/CFT conjecture, however, motivates the more general conjecture of ``gauge-gravity duality,'' which is the statement that \textit{any} theory of quantum gravity (it need not be a string theory), if formulated on a spacetime that approaches an AdS space asymptotically, is equivalent to a gauge theory formulated on the boundary of that space \cite{Witten:1998qj}.  In other words, gauge-gravity duality proposes that many theories of quantum gravity are identical to gauge theories in one lower dimension, again in the sense that the two theories will merely be two different descriptions of the same physics.

Gauge-gravity duality systems are concrete examples of ``holography'' \cite{tHooft:1993gx,Susskind:1994vu,Bousso:2002ju}, which, roughly speaking, is the statement that the physics of any theory of quantum gravity in some region of spacetime must be described by degrees of freedom formulated on the boundary of that spacetime. In other words, any theory of quantum gravity may be described by some degrees of freedom ``living'' in one lower dimension, hence the name ``holography.'' A big hint for holography comes from black hole physics, where the entropy of a black hole is proportional to the \textit{area} of its event horizon, rather than its volume. Notice that holography is not constructive, however, asserting only the \textit{existence} of some lower-dimensional degrees of freedom, but saying nothing about their dynamics. Gauge-gravity duality tells us that, in some cases, the lower-dimensional degrees of freedom arrange themselves into a gauge theory.

Gauge-gravity duality is an extremely profound statement, raising deep questions about both gauge theories and quantum gravity. How do certain gauge theories, which include no dynamical gravity whatsoever, ``know'' about quantum gravity, and, on top of that, emergent extra dimensions? Where in the Hilbert space of a gauge theory are the microstates of a black hole? If gauge theories are unitary, does that mean black hole evaporation must be unitary? (By now even Stephen Hawking concedes that the answer is yes, because of AdS/CFT \cite{Hawking:2005kf}.) Indeed, many people believe that, ``Gauge-gravity duality is the biggest conceptual change in \textit{quantum mechanics} since the 1930's'' \cite{larryquote}.

Gauge-gravity duality is wonderful and good, but in the interest of full disclosure we present a caveat: it has not been \textit{proven} to be true. In other words, nothing like a rigorous mathematical proof of an equality between theories exists\footnote{A good question is what would really constitute such a proof. Before attempting to prove that two theories are equivalent, however, a good idea is to \textit{define} the two theories involved. We can in principle provide a lattice definition for a quantum field theory, but defining a theory of quantum gravity is much harder. Given this state of affairs, some people have actually proposed that gauge-gravity duality \textit{itself} could serve as the definition of the theory on the gravity side of the correspondence \cite{Maldacena:1997re}.}, and hence gauge-gravity duality remains a ``conjecture.'' Nevertheless, an enormous amount of evidence has accumulated suggesting that gauge-gravity duality is true. For a partial list of such evidence, see ref. \cite{magoo}. The growing mountain of evidence has laid to rest any doubts about the validity of the conjecture.

$\N=4$ SYM does not describe the real world at all, of course, so we would like some way to determine what theories have holographic duals. In particular, we would like to know whether QCD has a holographic dual. If not, then gauge-gravity duality is just a theoretical exercise. Most examples of gauge-gravity duality come from D-brane physics in string theory, which is one method to generate specific dualities (gauge theory $X$ is dual to string theory $Y$). If someone hands you an arbitrary gauge theory, however, and asks whether it has a dual description as a theory of quantum gravity, how do you answer the question? What tests could you do? These are open problems; a complete classification of which theories have dual descriptions does not yet exist.

A key clue, though, is the large-$N_c$ limit. For theories with a well-defined large-$N_c$ limit, 't Hooft noticed that the Feynman diagrams of the theory, when drawn in a double-line notation (basically meaning one line per gauge index), naturally arrange themselves in powers of $N_c$ \cite{tHooft:1973jz}. The power of $N_c$ is, in fact, the genus of the Riemann surface on which the diagram can be drawn without any lines crossing: the leading diagrams are order $N_c^2$ and can be drawn on a sphere, the sub-leading diagrams are order $N_c^0$ and can be drawn on a torus, and so on. In other words, the large-$N_c$ expansion is shockingly similar to the genus expansion in perturbative string theory, if we identify $1/N_c$ with $g_s$. The original AdS/CFT conjecture provides an explicit example where the large-$N_c$ genus expansion is, in fact, a perturbative string expansion: using the dictionary, and with $L^4/\alpha'^2$ fixed, we find $g_s \propto 1/N_c$. 't Hooft discovered the large-$N_c$ genus expansion over thirty years ago, but it remains the biggest piece of evidence suggesting that gauge theories with a well-defined large-$N_c$ limit \textit{may} have dual descriptions as theories of quantum gravity. In particular, this is one of the main reasons why QCD is believed to have a dual description.

The holographic dual of QCD will be hard to find, if it exists, however. Perhaps the biggest obstacle is that QCD has no large separation of scales. In the original AdS/CFT conjecture, the gauge theory was conformal, meaning in particular that the coupling $\lambda$ was a free parameter, so we could, quite artificially, dial $\lambda$ to be as big as we like. On the string theory side, this made massive string states arbitrarily heavy and the background geometry arbitrarily weakly curved, so that classical gravity was then a good approximation to the string theory. In QCD, however, the coupling is not a free parameter: it is determined dynamically. We cannot artifically produce a large separation of scales. We will be forced to deal with a fully quantum theory of gravity, with high spacetime curvatures, many excited states (the massive string states), etc.

What should we do, then? One option is to forge ahead and attempt to construct the holographic dual of QCD. We can take a ``top-down'' approach, deriving from a full quantum gravity theory, such as a string theory, a geometry whose dual is QCD, or perhaps whose dual approaches QCD in some limit. An example of such a top-down approach appears in ref. \cite{Sakai:2004cn}. Another approach is ``bottom-up,'' that is, \textit{demanding} that the gauge theory be QCD and \textit{inferring} the holographic dual \cite{Erlich:2005qh,Karch:2006pv}.

Yet another approach is to \textit{abandon} QCD and work with a theory whose dual we know precisely, namely $\N=4$ SYM. Given how different $\N=4$ SYM and QCD are, though, what could this approach possibly teach us about QCD?

Our philosophy is this: \textit{try to make ``universal'' statements that may be applicable to QCD and its holographic dual}. These ``universal'' lessons come in two flavors: qualitative and quantitative.

What are some examples of qualitative universal statements? A state in a gauge theory with nonzero entropy will be dual to a geometry with a black hole horizon: the black hole's Hawking temperature and entropy are identified with the gauge theory temperature and entropy \cite{Gubser:1996de,Witten:1998zw}. \textit{Topology changes} in the geometry, for example a Hawking-Page transition in which a black hole condenses, generically map to \textit{first-order} phase transitions in the gauge theory, for example a deconfinement transition: this makes sense since on both sides a discontinuous change occurs \cite{Witten:1998zw}. As we can see, qualitative universal statements usually involve translating gauge theory questions into gravity questions. What does a finite quark density look like holographically \cite{Kobayashi:2006sb}? What does a quark moving through a gauge-theory plasma look like holographically \cite{Herzog:2006gh,Gubser:2006bz}? What does gluon scattering look like holographically \cite{Alday:2007hr,Alday:2007he}?

What are examples of quantitative universal statements? Here is where we return to the RHIC story. The canonical example of a quantitative universal statement concerns the ratio of shear viscosity, $\eta$, to entropy density, $s$. Hydrodynamics is the correct effective description of low-energy exictations of finite-temperature, strongly-coupled $\N=4$ SYM for essentially the same reasons as in QCD. In particular, at strong coupling the mean free path of excitations will be short. We may not know how to compute $\eta$ for QCD at RHIC temperatures, but we do know how to compute $\eta$ for $\N=4$ SYM theory using AdS/CFT \cite{Policastro:2001yc,Policastro:2002se,Kovtun:2003wp,Kovtun:2004de}. The result is usually expressed as the ratio $\eta/s = 1/4\pi$. This value is far smaller than that of any known substance. The big surprise here, though, was that the value $\eta/s = 1/4\pi$ is actually the result for \textit{any theory with a gravitational dual} \cite{Buchel:2003tz,Buchel:2004qq,Kovtun:2004de}. More precisely, $1/4\pi$ will be the value of $\eta/s$ in the limit of any gauge theory for which the holographic dual is classical gravity, so that the dual geometry includes a classical black hole (finite-coupling corrections, for example, which translate into curvature corrections in gravity, may alter the value of $\eta/s$). The smallness and universality of $\eta/s$ are suggestive, given that QCD has a very small viscosity and, most likely, a holographic dual. Though neither a prediction nor an explanation of the RHIC results, the precise number $\eta/s = 1/4\pi$ serves as an order-of-magnitude starting point for thinking about QCD. Universal quantities such as $\eta/s$ are hard to find, however.

In the big picture, then, the goal is not to compute observables in $\N=4$ SYM theory and then make a prediction about QCD. Comparing quantitative results between $\N=4$ SYM theory and QCD is a very difficult task, and probably not fruitful. Instead, we look for qualitative universal statements that should be true for the holographic dual of QCD, so that if today someone handed us the holographic dual of QCD and asked us to compute the phase diagram or the viscosity, we would at least know where to start: a state of nonzero entropy is dual to a black hole, deconfinement is dual to black hole condensation, etc. In other words, we develop \textit{methods} for computing gauge theory observables using gravity. Along the way, we try to extract quantitative universal statements, which may teach us how much of QCD physics is in fact insensitive to the details of its dynamics, that is, how much of QCD physics is in fact \textit{the same} as the physics of \textit{any} gauge theory with a gravitational/holographic dual.

We summarize our whole philosophy by saying that gauge-gravity duality is a \textit{tool} for studying strongly-coupled gauge theories. We do not ask questions about why the duality works. We do not ask questions about quantum gravity and compute the answer using the gauge theory. Instead, we begin with a question in a strongly-coupled gauge theory, translate to gravity, perform the calculation, and then translate back to gauge theory. In this spirit, AdS/CFT is a ``black box,'' and $\N=4$ SYM serves as a ``toy model'' of QCD. A good name for this approach, one that is only partly facetious, is ``Applied String Theory.''

In this dissertation we focus on questions in strongly-coupled $\N=4$ SYM theory that would be difficult to answer for QCD, using existing methods. First, we will study the theory at zero temperature and finite quark density. Second, we will compute a transport coefficient, a conductivity. We will not attempt to compare our results to QCD. As we have tried to argue, the method is more important than the result!

Hopefully we have convinced the reader that this line of research is worth pursuing, and in the process justified the existence of this dissertation. The other main lesson we hope to impart in the sequel is that AdS/CFT is not only useful, it's also fun!

\section{Outline}

In chapter \ref{adscft} we review the original AdS/CFT correspondence. We begin with a review of $\N=4$ SYM theory, and then of AdS spaces. We then repeat, in summary, Maldacena's original D-brane construction of AdS/CFT \cite{Maldacena:1997re}, and how to compute correlators in the gauge theory from the gravity theory \cite{Witten:1998qj,Gubser:1998bc}. We also review the finite-temperature version of the correspondence \cite{Gubser:1996de,Witten:1998zw}.

In chapter \ref{flavor} we review how to introduce fundamental-representation ``flavor'' fields, into the AdS/CFT correspondence, using probe D7-branes \cite{Karch:2000gx,Karch:2002sh}. The on-shell action of these D7-branes is infinite: it must be regulated and renormalized. We review the procedure of ``holographic renormalization'' of probe D7-branes in $AdS_5 \times S^5$, which was first described in ref. \cite{Karch:2005ms}.

In chapter \ref{thermo} we turn to the thermodynamics of the flavor fields. We review the known phase transitions associated with the flavor fields at finite temperature \cite{Babington:2003vm,Kirsch:2004km,Ghoroku:2005tf,Mateos:2006nu,Albash:2006ew,Karch:2006bv,Mateos:2007vn}, finite density \cite{Kobayashi:2006sb,Mateos:2007vc,Ghoroku:2007re}, and with finite electric and magnetic fields \cite{Filev:2007gb,Filev:2007qu,Albash:2007bk,Erdmenger:2007bn,Albash:2007bq}. In all of these cases, the calculations on the supergravity side were numerical. The point of this chapter is to present an \textit{exact} supergravity solution representing a state of the gauge theory with zero temperature but a finite density of flavor fields, as described in ref. \cite{Karch:2007br}. Using this description, we find a second order transition, in the grand canonical ensemble, when the mass of the flavor fields equals the chemical potential.

In chapter \ref{transport}, we turn to the transport properties of the flavor fields. We present supergravity solutions representing a state of the gauge theory with finite temperature, finite density and finite electric and magnetic fields. The external fields produce a current of flavor fields. We compute this current and from it extract a conductivity tensor. Along the way, we compute the drag force on the flavor fields. These results were first presented in refs. \cite{Karch:2007pd,OBannon:2007in}.

We conclude in chapter \ref{conclusion} with open questions and suggestions for future research.

\bigskip

\textbf{IMPORTANT NOTE ABOUT UNITS:} We only write the $AdS_5$ radius of curvature, $L$, explicitly in chapter \ref{adscft}. In the subsequent chapters, we measure lengths in units of $L$, or in other words, we set the radius of $AdS_5$ to be one: $L \equiv 1$. An important conversion from supergravity quantities to field theory quantities is then $\alpha'^{-2} = \lambda$.
%
%
%
%
%
%
%
%
%
%
%
%
%
%
%
\chapter {The AdS/CFT Correspondence}\label{adscft}

In this chapter we review the AdS/CFT correspondence. We first review the ingredients that enter into the correspondence: $\N=4$ SYM and AdS spaces, with emphasis on physics rather than equations, although we present a few AdS metrics explicitly. We then recall some very basic facts about strings and D-branes, so that we can review Maldacena's original ``derivation'' of the correspondence, using D3-branes. We review the precise statement of the correspondence, as an equivalence of partition functions, and then review the finite-temperature version of the correspondence. Most of the material in this chapter can be found in the AdS/CFT reviews refs. \cite{magoo,DHoker:2002aw} and the string theory textbooks refs. \cite{Green:1987sp,Polchinski:1998rq,Johnson:2003gi,Becker:2007zj}.

\section{$\N=4$ Super-Yang-Mills Theory}
\label{n4symreview}

In this section we will review some important properties of $\N=4$ SYM theory. We will only consider the theory in flat (3+1)-dimensional space, with mostly-plus Minkowski metric $\eta_{\mu \nu}$ (as opposed to, say, the theory formulated on some curved manifold).

Supersymmetry is an extension of the Poincar\'{e} algebra to include anti-commuting symmetry generators. These anti-commuting generators are spinors, called ``supercharges.'' Let $\N$ denote the number of left-handed Weyl spinor supercharges. Weyl spinors include two components, \textit{i.e.} two complex numbers. The total number of \textit{real} supercharges is thus $2 \times 2 \times \N$. The super-Poincar\'{e} algebra is invariant under rotations of the supercharges into one another. These rotations form a group $U(\N)_R$, called the R-symmetry group (hence the subscript $R$).

In a supersymmetric quantum theory, the supercharges act on the Hilbert space as raising and lowering operators for helicity. If $\N$ becomes bigger than $4$, then the theory will have states of helicity greater than $1$. Some theories have $\N > 4$, and include fields of higher helicity. An example is $\N=8$ supergravity, in which one field is the metric, with helicity $2$.

We are interested in theories with states of helicity $\leq 1$ only. More specifically, we are interested in gauge theories without dynamical gravity, hence $\N=4$ is the maximal number of supersymmetries we can allow. The maximally supersymmetric \textit{gauge} theory is $\N=4$ supersymmetric Yang-Mills theory ($\N=4$ SYM). The theory of course describes massless spin-1 fields, the gauge fields. Supersymmetry then demands that fields of lower spin also be included. The $\N=4$ gauge supermultiplet includes the gauge fields, as well as four left-handed Weyl fermions and six real scalars (or equivalently three complex scalars). By supersymmetry, all must transform in the same representation of the gauge group, namely the adjoint representation, and all must have the same mass. Gauge invariance forbids a mass for the gauge fields, hence the fermion and scalar fields are massless.

Remarkably, the Lagrangian for $\N=4$ SYM theory is unique! That is, if we demand that the theory be renormalizable, then gauge invariance and supersymmetry completely determine the form of the action. Contrast this with $\N=1$ supersymmetric theories, for example, where we have a lot of freedom in choosing the (super)potential.

We will not reproduce the $\N=4$ SYM Lagrangian. It is written explicitly in ref. \cite{DHoker:2002aw}. It includes the usual kinetic term and theta term for the gauge fields, the usual kinetic terms for the fermions and scalars (with gauge covariant derivatives), Yukawa couplings between the scalars and fermions, and quartic potential terms for the scalars. Supersymmetry of course restricts the interaction terms, and $\N=4$ supersymmetry is so restrictive that only one coupling, the gauge coupling $g_{YM}$, appears in the $\N=4$ SYM Lagrangian. In other words, $g_{YM}$ is not only the strength of the interaction of the fermions and scalars with the gauge fields, but also the strength of the Yukawa interactions and the scalars' quartic self-interactions.

Classically, $\N=4$ SYM is scale-invariant. The coupling $g_{YM}$ is dimensionless, so to specify the classical $\N=4$ SYM theory we need only specify the gauge group, $g_{YM}$ and the theta angle. (Starting now, we will ignore the theta angle.) We do not need to specify any dimensionful scale. Of course, the same is true for classical \textit{pure} Yang-Mills theory, as well as classical Yang-Mills plus \textit{massless} quarks.

Any textbook on \textit{quantum} field theory will tell you, however, that to define a quantum field theory we must specify the value of the coupling at some renormalization scale. The dynamics of the theory then determines how the coupling strength changes with energy scale, that is, how the coupling ``runs.'' In particular, at some characteristic scale the coupling will run to a non-perturbative, \textit{i.e.} order-one, value. Notice what this means for any theory that is classically scale-invariant: quantum effects \textit{dynamically} generate a scale. The classical symmetry, scale invariance, is broken in the quantum theory. Such theories thus have ``anomalous scale invariance,'' or simply a ``scale anomaly.'' Pure Yang-Mills and QCD have scale anomalies. In QCD, the dynamically generated scale is called $\Lambda_{QCD}$.

Amazingly, $\N=4$ SYM theory remains scale-invariant even quantum mechanically! That is, once we specify the value of $g_{YM}$ at one scale, the coupling does not change: $g_{YM}$ will have the same value at any other scale. In other words, $\N=4$ SYM has no scale anomaly, so nothing like $\Lambda_{QCD}$ is generated.

In fact, scale invariance is part of a larger symmetry: conformal symmetry. A conformal transformation acts on the metric as $g_{\mu \nu} \ra \Omega(t,\vec{x})^2 \, g_{\mu \nu}$ where $\Omega(t,\vec{x})^2$ is a smooth, positive function (which is why we write it as a square).

A conformal field theory (CFT) is a theory that is invariant under conformal transformations. In (1+1) dimensions, theorems exist showing that, if a theory is scale invariant, it must in fact be conformally invariant \cite{Zamolodchikov:1986gt,Polchinski:1987dy}. In other words, in (1+1) dimensions, scale invariance implies full conformal invariance. In (3+1) dimensions, no such theorem exists (yet), but in all known examples scale invariance is accompanied by the full conformal group. An example is $\N=4$ SYM.

What is the full symmetry group of $\N=4$ SYM theory, then? In brief terms, we can say that the Poincar\'{e} group is enhanced to the conformal group, which is $SO(4,2)$. More explicitly, the bosonic symmetries are, for spacetime coordinate $x^{\mu}$,
\begin{itemize}
\item{Translations: $x^{\mu} \ra x^{\mu} + c$ for real constant $c$}
\item{Lorentz transformations: $x^{\mu} \ra \Lambda^{\mu}_{~\nu} \, x^{\nu}$ for Lorentz transformation matrix $\Lambda^{\mu}_{~\nu}$}, obeying $\eta_{\alpha \beta} \Lambda^{\alpha}_{~\mu} \Lambda^{\beta}_{~\nu} = \eta_{\mu \nu}$
\item{Scale transformations: $x^{\mu} \ra \lambda \, x^{\mu}$ for real, constant, positive $\lambda$}
\item{Inversions: $x^{\mu} \ra \frac{x^{\mu}}{|x|^2}$ with $|x|^2 = g_{\mu \nu} x^{\mu} x^{\nu}$}
\end{itemize}
The scale transformation and inversions are the ``extra'' transformations that take us from the Poincar\'{e} group to the full conformal group. The theory of course also has the R-symmetry $U(4)_R$. Locally, the $SU(4)_R$ subgroup is isomorphic to $SO(6)$. The fermions of the theory transform as a $4$, and the scalars as an antisymmetric $6$, of the $SU(4)_R$.

$\N=4$ SYM theory also has fermionic symmetries, of course, namely the sixteen real supercharges of $\N=4$ supersymmetry. The Poincar\'{e} group is extended to include these supercharges as well as the generators of conformal symmetry, so the complete symmetry algebra is called the ``superconformal'' algebra. If that were all, however, the superconformal algebra would not close, so in fact sixteen \textit{additional} fermionic generators are required. These sixteen additional fermionic generators are called superconformal generators.

Conformal symmetry severely restricts observables. For example, in a CFT, the two-point functions of operators of definite scaling dimension are completely fixed, and the three-point functions of such operators are fixed up to some overall constants. An S-matrix cannot be defined in a CFT. In scattering, we begin with some infinitely far-separated wave packets (particles) that we ``shoot in'' to some region where the interaction occurs, and we then measure some infinitely far-separated wave packets that come out. With scale and inversion invariance, however, in which we can exchange the interaction point with the point at infinity, we cannot define infinitely far-separated wave packets, and hence we cannot define an S-matrix.

Conformal symmetry does have an upside, though: in any problem, the only scales will be those that we introduce ourselves, which can greatly simplify things. For example, suppose we study a CFT at finite temperature (and in flat space), and want to find any phase transitions. Without doing any work, we know the answer: the theory has no dimensionful scale that can set a transition temperature, hence \textit{no finite-temperature phase transitions can occur}!

\section{Anti-de Sitter Space}
\label{adsspace}

In this section we review properties of anti-de Sitter (AdS) spaces. We will not attempt to be mathematically rigorous, rather, we hope to provide simple, intuitive pictures sufficient for understanding physics.

\textit{de Sitter} spaces are defined as spaces of constant \textit{positive} curvature. \textit{Anti-de Sitter} spaces are defined as spaces of constant \textit{negative} curvature. In physics language, AdS spaces are solutions of Einstein's equations with negative cosmological constant.

We will be interested in (4+1)-dimensional anti-de Sitter space, $AdS_5$. We know we can get a space of constant negative curvature from a hyperboloid, so we start in one higher dimension and write $AdS_5$ as a hypersurface. That is, we begin with the algebraic equation for a hyperboloid in $\mathbb{R}^6$,
\beq
\label{hyperboloid}
X_0^2 + X_{5}^2 - \sum_{i=1}^{4} X_i^2 = L^2
\eeq
where $X_{\mu}\in\mathbb{R}^6$. We can already see from the signs on the left-hand side that $AdS_5$ has a symmetry $SO(4,2)$, which will be the isometry group of the $AdS_5$ metric. Here $L$ is the radius of curvature of the hyperboloid.

One way to parameterize a solution of eq. (\ref{hyperboloid}) is to let
\bea
X_0 & = & \frac{L^2}{2u} \left( 1 + \frac{u^2}{L^4} \left( \vec{x}^2 - t^2 + L^2 \right) \right), \qquad X_5 = \frac{ut}{L}, \nonumber \\ X_4 & = & \frac{L^2}{2u} \left( 1 + \frac{u^2}{L^4} \left(\vec{x}^2 - t^2 - L^2 \right) \right), \qquad X_i = \frac{u x_i}{L}
\eea
where $i=1,2,3$, $x_i\in\mathbb{R}^3$, and $u>0$. These coordinates only cover half of the hyperboloid and are thus called ``Poincar\'{e} patch'' coordinates. Coordinates that cover the entire hyperboloid do exist, the so-called ``global'' coordinates, but these will not be important in the sequel. In Poincar\'{e} patch coordinates, the induced metric of the hyperboloid is
\beq
\label{adsmetric}
ds_{AdS}^2 = L^2 \, \frac{du^2}{u^2} \, + \, \frac{u^2}{L^2} \, \left( -dt^2 + d\vec{x}^2 \right)
\eeq
An explicit calculation of the Ricci scalar gives $-\frac{20}{L^2}$, so the space has constant negative curvature with radius of curvature $L$, as advertised.

\begin{figure}
{\begin{tabular}{cc}
\includegraphics[width=0.5\textwidth]{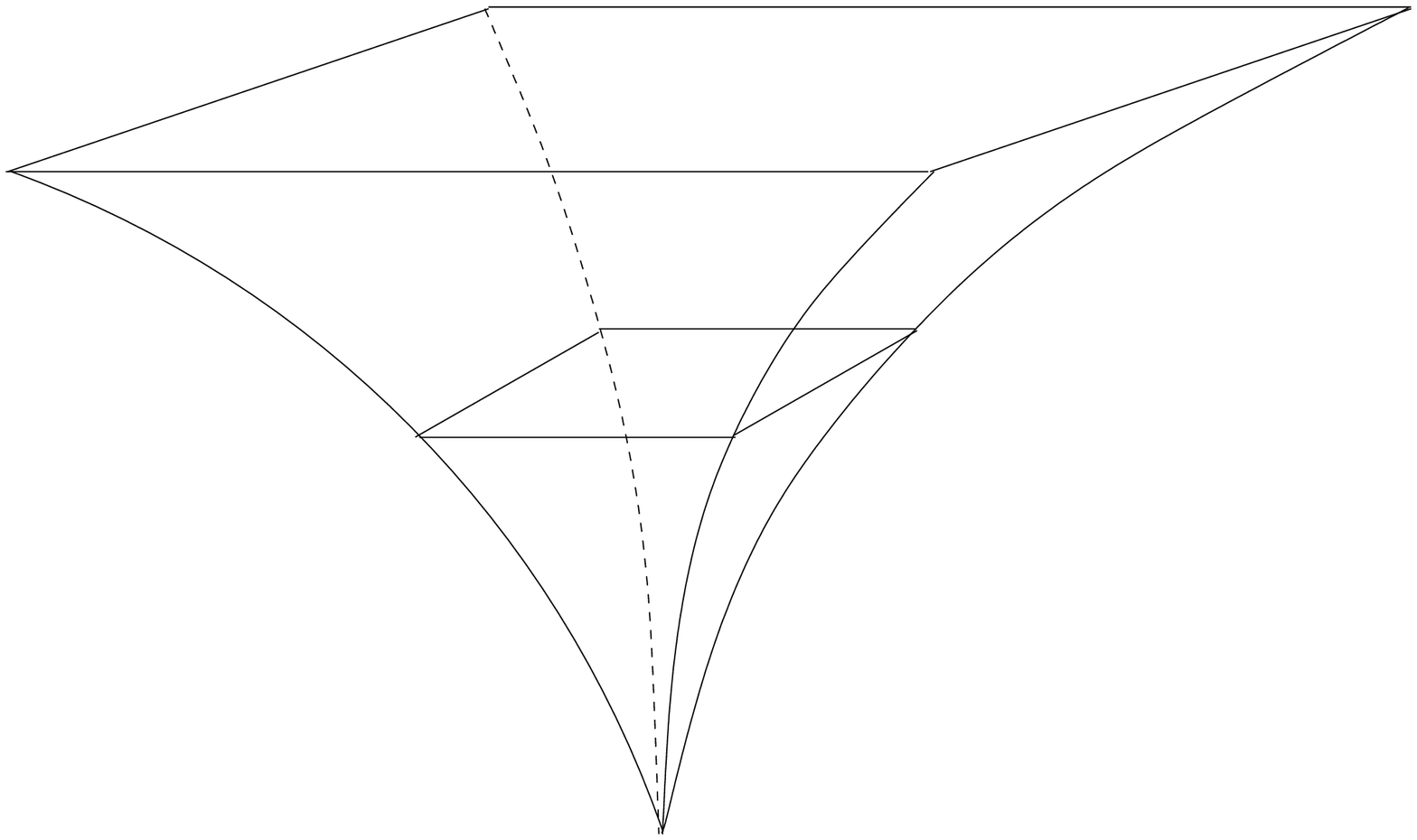} & \includegraphics[width=0.5\textwidth]{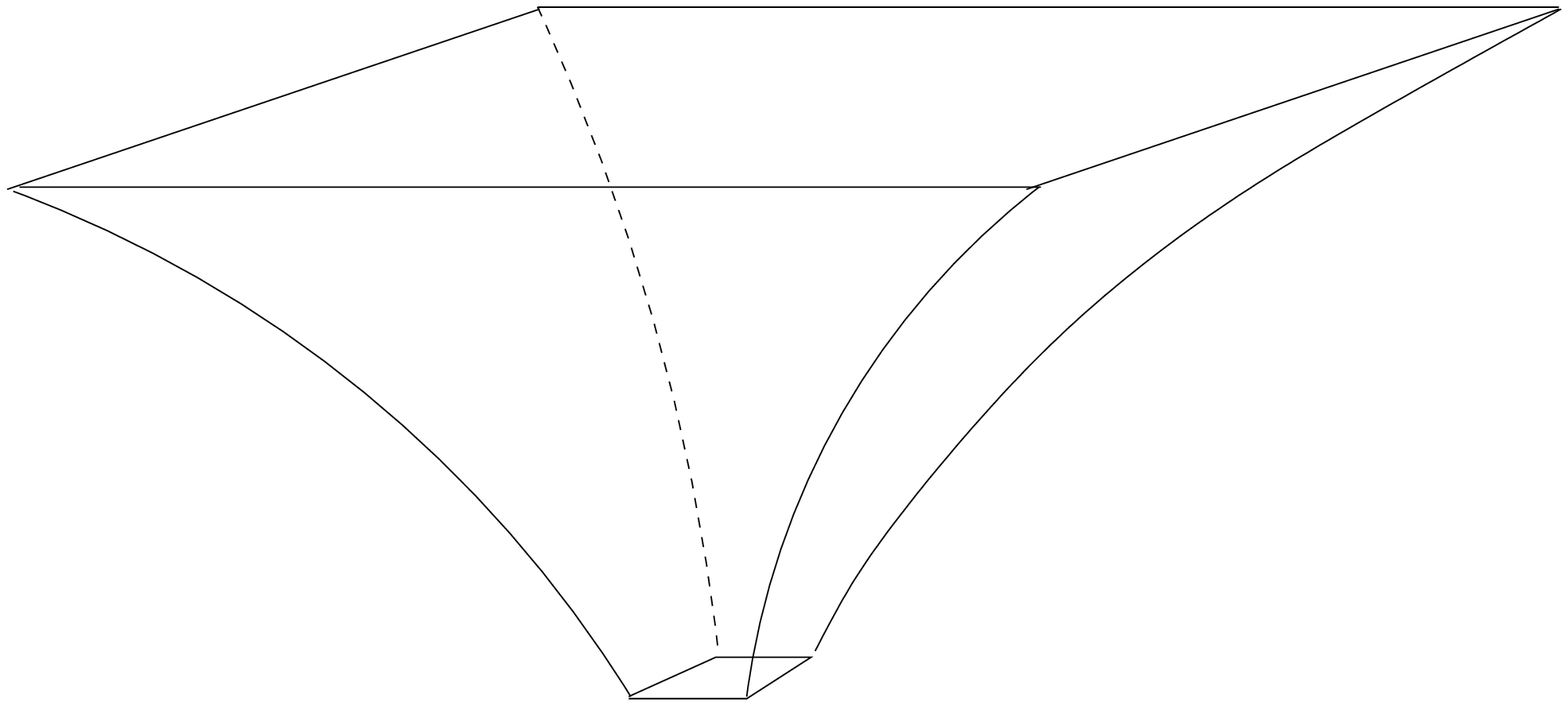} \\ (a.) & (b.)
\end{tabular}
\caption{\label{adsfig} (a.) Cartoon of AdS space. (b.) Cartoon of AdS-Schwarzschild.}
}
\end{figure}

We can see that the space is essentially (3+1)-dimensional flat space plus an extra ``warped'' direction, $u$. The coordinate $u$ is the radial coordinate of $AdS_5$. For any fixed value of $u$, we find a (3+1)-dimensional hypersurface whose metric is simply that of $\mathbb{R}^{3,1}$ times a conformal factor, $u^2/L^2$. A simple intuitive way to think of $AdS_5$ is thus as an onion, with each layer of the onion skin being flat $\mathbb{R}^{3,1}$. As $u$ shrinks, however, the layer of the onion skin shrinks as $u^2/L^2$. We depict AdS space (crudely) in figure \ref{adsfig} (a.).  In the figure, the vertical direction is $u$, with $u=0$ at the bottom and $u=\infty$ at the top. We have also drawn two spatial Minkowski directions and suppressed all other directions, and drawn a ``slice'' of the space at some finite value of $u$.

As $u \ra 0$ we find a horizon, since $g_{tt} \ra 0$. This horizon is called the Poincar\'{e} horizon. The Poincar\'{e} horizon has zero area because $g_{x_ix_i}\ra0$ as $u\ra0$. Nothing dramatic happens in crossing the Poincar\'{e} horizon: on the other side of the horizon is the other Poincar\'{e} patch, covering the other half of the hyperboloid. In figure \ref{adsfig} (a.), the Poincar\'{e} horizon is the point at the bottom.

The metric of any AdS space will always have a second-order pole for some value of the radial coordinate, \textit{i.e.} the metric will always diverge quadratically. The place where this occurs is called the \textit{boundary} of AdS. In our coordinates the boundary is at $u \ra \infty$. In figure \ref{adsfig} (a.) the boundary is at the top.

We can extend the full $AdS_5$ metric to the boundary, that is, we can extract a finite boundary metric from the $AdS_5$ metric, as follows. We must first choose a ``defining function,'' which is a function of $u$, $t$ and $\vec{x}$ that must have a second-order \textit{zero} at the boundary, and otherwise just needs to be smooth and positive. An example is $f(u,t,\vec{x}) = \Omega(t,\vec{x})^2 \frac{L^2}{u^2}$. If we multiply the Minkowski part of the metric by such an $f(u,t,\vec{x})$ and \textit{then} take the limit $u \ra \infty$, we recover a finite boundary metric, $ds^2_{\partial AdS} = \Omega^2(t,\vec{x}) \, \left(-dt^2 + d\vec{x}^2 \right)$. A particular choice of $f(u,t,\vec{x})$ gives a particular boundary metric, but any other choice is equally acceptable, so the boundary metric is only defined up to a conformal factor. In other words, the bulk metric only determines a \textit{conformal class} of boundary metrics.

The boundary ($u\ra\infty$) is in fact an infinite geodesic distance from any point in the bulk (any finite $u$). Physically, AdS spaces are gravitational potential wells. A massive object sitting at some value of the radial coordinate will naturally fall away from the boundary, which in the Poincar\'{e} patch means towards the Poincar\'{e} horizon.

Another metric that we will need is the AdS-Schwarzschild metric. As the name implies, AdS-Schwarzschild is a geometry that has a black hole horizon and that asymptotically approaches an AdS geometry. The AdS-Schwarzschild metric is
\beq
\label{originaladsbhmetric}
ds_{AdS+BH}^2 = L^2 \, \frac{1}{h(u)} \, \frac{du^2}{u^2} \, + \, \frac{u^2}{L^2} \, \left( -h(u) \, dt^2 + d\vec{x}^2 \right) 
\eeq
where $h(u) = 1 - u_h^4/u^4$ and $u_h = \pi T L^2$ is the location of the horizon, with $T$ being the Hawking temperature of the black hole. AdS-Schwarzschild is again a solution of Einstein's equation with negative cosmological constant, and with the same asymptotic boundary condition as pure AdS space. We depict AdS-Schwarzschild in figure \ref{adsfig} (b.). Notice that at $u=\infty$ the space looks the same as in figure \ref{adsfig} (a.). The space is now cut off at the horizon, at $u=u_h$, which now has finite area, as we have crudely depicted in the figure.

In what follows, another coordinate system will be very useful: so-called Fefferman-Graham coordinates \cite{feffermangraham}. These are coordinates in which we define a new radial coordinate, $z$, such that $g_{zz}$ is always just $1/z^2$, and the boundary is always at $z=0$. Explicitly, we change coordinates from $u$ to
\beq
z = \sqrt{2} \, L^2 \, \left( u^2 + \sqrt{u^4 - u_h^4} \right)^{-1/2}
\eeq
The horizon is then at $z_h = \sqrt{2} \, L^2 / u_h = \sqrt{2} / \pi T$, and the metric becomes
\beq
\label{adsbhmetric}
ds^2_{AdS+BH} = L^2 \left( \frac{dz^2}{z^2} \, - \, \frac{1}{z^2} \, \frac{\left( 1-z^4/z_h^4 \right)^2}{1+z^4/z_h^4} \, dt^2 \, + \, \frac{1}{z^2} \, \left( 1+z^4/z_h^4 \right) \, d\vec{x}^2 \right)
\eeq
The funny factors of $\sqrt{2}$ guarantee that when $T=0$ we have simply $z = L^2/u$. The Poincar\'{e} horizon is then at $z \ra \infty$.

Having learned a little about the symmetries of $\N=4$ SYM and about $AdS_5$, we can already see hints of the AdS/CFT correspondence. The $\N=4$ SYM theory has an $SO(4,2)$ symmetry, which is precisely the isometry group of $AdS_5$. The metric of $AdS_5$ only defines a conformal class of boundary metrics. Any physics that occurs on the boundary must be invariant under the choice of defining function, that is, must be conformally invariant. The AdS/CFT conjecture is, in fact, that a specific conformal field theory, $\N=4$ SYM, ``lives'' on the boundary. We can motivate the correspondence much better by studying strings and D-branes, though, so now we will quickly review some of their physics.

\section{Strings and D-branes}

Most of the material in this section can be found, in far greater detail, in standard string theory textbooks \cite{Green:1987sp,Polchinski:1998rq,Johnson:2003gi,Becker:2007zj}.

\subsection{String Basics}
\label{strings}

In its textbook formulation \cite{Green:1987sp,Polchinski:1998rq,Johnson:2003gi,Becker:2007zj}, string theory supposes that all fundamental particles are, in fact, not point particles but extended objects, namely one-dimensional objects called strings. Of course, we have yet to see evidence for such strings: to date, as far as we know, electrons, muons, neutrinos, etc., really are point particles. To be consistent with experiment, then, the length of strings must be very small, far smaller than any distance scale ever probed. Indeed, the string length, $\ell_s$ will be on the order of the Planck length\footnote{Actually, in perturbative string theory, the Planck length is much shorter than the string length: $\ell_{Planck} = g_s^{1/4} \ell_s$ for string coupling $g_s \ll 1$.}. For historical reasons, we usually work with its square, denoted $\alpha' \equiv \ell_s^2$. The tension of a string, $T_s$, is set by the fundamental scale, $\ell_s$, as $T_s = 1/(2\pi\alpha')$.

We hasten to note that not every string has a length $\ell_s$! Rather, the string length $\ell_s$ is a fundamental parameter of string theory. The point is that all string theory strings have an energy per unit length (tension) $T_s$, which encodes some characteristic length scale $\ell_s$, giving us an order-of-magnitude estimate of where we could detect that fundamental particles are actually extended objects.

Strings come in two species, closed and open. Closed strings are loops: the string has no boundary. Open strings are line segments: the string has a boundary, its two endpoints.

Strings may also be oriented or unoriented. We will discuss only oriented strings.

As time passes, a string will sweep out a two-dimensional surface, a Riemann surface, called the string worldsheet. As strings have a tension, physically we expect the worldsheet to be minimized. The action for a single string, the Nambu-Goto action, is just the area of the string worldsheet, which we extremize to find the string equations of motion.

Strings interact with coupling strength $g_s$. Physically, $g_s$ is the probability for a string to split into two strings, or for two strings to join if they cross one another.

In perturbative string theory, we represent string interactions with diagrams similar to Feynman diagrams, except now instead of point particles interacting at point vertices, we have strings sweeping out Riemann surfaces as time passes. Such diagrams are arranged according to topology. The topology of a Riemann surface is determined completely by its genus, so we arrive at the so-called ``genus expansion'' of string theory. If $g$ is the genus of the Riemann surface, $b$ the number of boundaries, and $\chi = 2-2g - b$ the Euler character, string diagrams are weighted by a factor $g_s^{-\chi}$. When $g_s \ll 1$, the genus-zero surface, topologically a sphere, dominates the expansion, having $\chi=2$. If we include open strings, the first sub-leading correction is the disk, with $g=0$, $b=1$ and hence $\chi = 1$. The next correction would be from closed strings, the torus, with $g=1$, $b=0$ and $\chi=0$.

\begin{figure}
{\begin{tabular}{cc}
\includegraphics[width=0.46\textwidth]{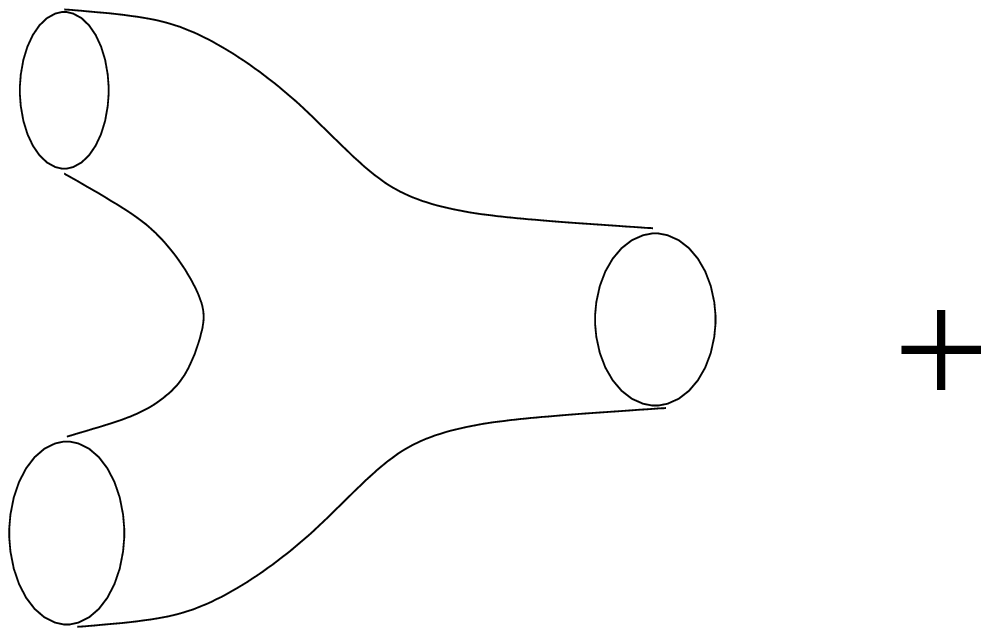} & \includegraphics[width=0.48\textwidth]{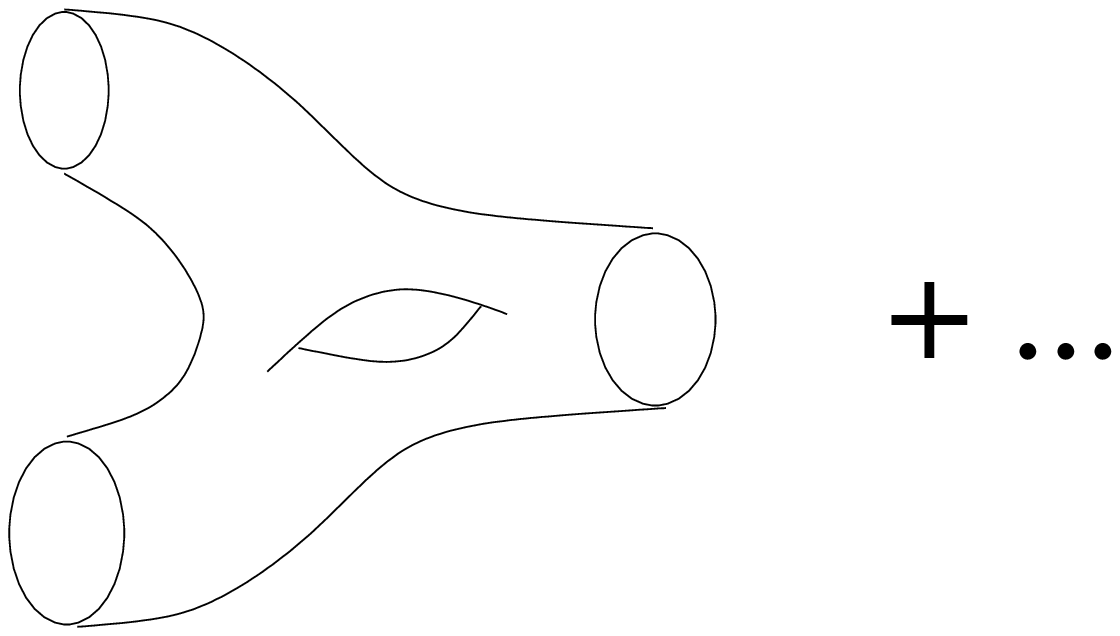}
\end{tabular}
\caption{\label{stringpert} Cartoon of leading and sub-leading contribution to perturbative string amplitude for the interaction of three closed strings.}
}
\end{figure}

We have depicted string perturbation theory in figure \ref{stringpert}. We have (crudely) drawn the interaction of three closed strings. The quantum mechanical amplitude for such a process will have a leading term from a Riemann surface with the topology of a sphere and a subleading term from a Riemann surface with the topology of a torus. The ``$+ \ldots$'' indicates the infinite number of higher-genus contributions.

At zero string coupling, for strings propagating in flat spacetime, we can quantize the excitations of the string and determine the string spectrum. We find massless modes and massive modes. In the closed string sector, one of the massless states has spin 2, so we identify this state as the graviton. In the open string sector, among the massless states we find a spin-1 excitation, which we identify as a gauge field. As for the massive modes, as the only dimensionful scale in string theory is $\ell_s$, we know without doing any work that the masses of the excited states must be proportional to $\ell_s^{-1}$. 

Suppose we consider string theory with only bosonic string excitations. Bosonic string theory has two fundamental problems. The first is that the spectrum includes a tachyon: a state whose mass-squared is \textit{negative}. The second is that the theory is only consistent quantum mechanically in twenty-six spacetime dimensions! Why is this so? String theory has a local invariance, \textit{i.e.} a gauge invariance, called Weyl invariance. Weyl transformations are essentially (but not exactly) conformal transformations on the string worldsheet. Weyl invariance is generically anomalous, which is very bad\footnote{An anomalous \textit{global symmetry} can have important physical consequences: recall the axial symmetry of QCD. An anomalous \textit{gauge invariance}, however, is a genuine sickness of a quantum theory.}. Only in special circumstances will the Weyl anomaly vanish, for example in twenty-six spacetime dimensions.

Supersymmetry, that is, adding fermionic excitations on the string, fixes some of the problems of the bosonic string. First, the supersymmetric string spectrum now contains no tachyon. Second, supersymmetric tring theory is consistently formulated in ten spacetime dimensions, a vast improvement over twenty-six (though still far from four). Five consistent superstring theories are known. Of importance to us will be the so-called type IIB theory. We will not need to know much about type IIB, except for some facts about D-branes which we will review in the next subsection.

Currently, string theory is only a ``first quantized'' theory: we only know how to quantize excitations on a \textit{single} string. We do not yet have a full understanding of string field theory, where individual strings become excitations of some ``string field,'' and where we could describe processes in which strings are created and destroyed.

String theory is a theory of quantum gravity, being quantum mechanical and including a graviton in its spectrum, and also describes gauge interactions. String theory is thus a candidate for a ``theory of everything.'' The problem string theorists would like to solve, then, is how string theory dynamically determines the geometry of spacetime, and all the low-energy physics in that spacetime. In other words, by what mechanism does string theory dynamically choose a vacuum state? This problem is incredibly difficult because an enormous number of possibilities exist.

For example, does string theory decide that spacetime should be ``compactified'' to (3+1) dimensions, that is, to $\mathbb{R}^{3,1} \times M$, where $M$ is some compact, very tiny six-dimensional manifold?  If we want $\N=1$ supersymmetry in $\mathbb{R}^{3,1}$, for example, we should choose $M$ to be a so-called Calabi-Yau manifold, of which about $10^{500}$ are known to exist! The problem is actually even worse, since string theory gives us more options. For example, we can choose ``fluxes'' on $M$, making the number of options even bigger than $10^{500}$.

The range of all possible vacua of string theory is called the string theory ``landscape.'' The landscape is the main reason that string theory is not yet a viable theory of everything. For us, the key point to take away is that, since we do not yet know how string theory dynamically determines a ten-dimensional geometry, the current formulation of string theory is not ``background-independent.'' In the current formulation of string theory, we must pick some ten-dimensional geometry and write an action for an individual string propagating in that geometry. In the back of our minds, we know the background is something like a ``coherent state of gravitons,'' although no one knows a precise meaning for those words.

Fortunately, somewhere in the landscape is $AdS_5 \times S^5$. The AdS/CFT correspondence then gives us a less ambitious approach to the real world: rather than a theory of everything, string theory can be a theory of strongly-coupled gauge theories.

Knowing that string theory is a ten-dimensional supersymmetric theory of gravity, the low-energy (meaning far below the string scale, $\ell_s^{-1}$), classical (meaning leading-order in $g_s$) effective description of string theory must be ten-dimensional supergravity. We understand supergravity fairly well, and in particular we actually know how to calculate things in supergravity, so it is usually the most useful tool for studying string theory, as either a theory of everything or in the AdS/CFT context.

More specifically, the low-energy effective theory of type IIB string theory is called type IIB supergravity (surprise, surprise). We will not need to know too much about type IIB supergravity, except that the theory includes the metric, whose kinetic term is the usual Einstein-Hilbert term, and some bosonic fields called Ramond-Ramond form fields, whose importance becomes clear in the context of D-brane physics.

\subsection{D-brane Basics}
\label{dbranes}

D-branes are fundamental objects in string theory. In perturbative string theory, D-branes appear as hypersurfaces on which open strings \textit{must} end: if we are studying open strings, some D-branes must be around; if we introduce D-branes, they are necessarily accompanied by open string degrees of freedom.

In the simplest contexts, a Dp-brane is just a (p+1)-dimensional generalization of a membrane. Like a soap bubble, or the head of a drum, a D-brane has a tension, and (part of) its action is simply its (p+1)-dimensional volume, or ``worldvolume,'' which it wants to minimize. Our objective in this section is to write this action explicitly.

In superstring theory, D-branes carry conserved charges, and hence are stable objects. The charges are called Ramond-Ramond (RR) charges. In particular, a Dp-brane will act as a source for a RR (p+1)-form field. In type IIA theory, only Dp-branes with \textit{even} p are stable, while in type IIB theory, only Dp-branes with \textit{odd} p are stable. In type IIB, then, D1-branes, D3-branes, D5-branes and D7-branes are stable\footnote{A D9-brane also exists in type IIB, but by itself is unstable, essentially by Gauss's law: the D9-brane will always fill all of space, but all of space cannot carry a net charge. We can cancel D9-brane charge in various ways, however, for example by introducing anti-D9-branes, or by introducing a (9+1)-dimensional orientifold, which can cancel sixteen units of D9-brane charge. In Euclidean signature, type IIB also includes D(-1)-branes, also known as D-instantons.}. A ``wrong-dimension'' Dp-brane can appear in either theory (odd in IIA or even in IIB), but will eventually decay into stable objects: lower-dimensional D-branes and strings.

Dp-branes with $p\leq3$ are electrically charged under the appropriate RR field, while  Dp-branes with $p>3$ are magnetically charged. Dp-branes thus come in electric-magnetic dual pairs. To explain this, we consider an example. D1-branes couple to the RR 2-form $A^{(2)}$ electrically, meaning with a term $\int A^{(2)}$ in their action, where the integral is over the D1-brane worldvolume. Let $F^{(3)} = dA^{(2)}$ be the 3-form field strength of $A^{(2)}$, and $\tilde{F}^{(7)} \equiv \star \, F^{(3)}$ the 7-form Hodge dual of $F^{(3)}$. We define a 6-form potential by $\tilde{F}^{(7)} \equiv d\tilde{A}^{(6)}$. The D5-brane then has a magnetic coupling, which is a coupling $\int \tilde{A}^{(6)}$, where the integral is over the D5-brane worldvolume. The magic number is six: Dp-branes and Dq-branes are electric-magnetic duals if $p+q = 6$.  D3-branes are special: they are self-dual. They couple to a 4-form $A^{(4)}$ whose 5-form field strength is self-dual, $F^{(5)} = dA^{(4)} = \star \, F^{(5)}$.

The Dp-brane action is known to leading order in $g_s$. Stable Dp-branes are supersymmetric, so the Dp-brane action includes fermionic and bosonic fields. We will not need the fermionic part in what follows, so we will not write it. The bosonic part of the action has terms of two kinds,
\beq
S_{Dp} = S_{DBI} + S_{WZ}
\eeq
Here $S_{DBI}$ is called the Dirac-Born-Infeld part of the action, and $S_{WZ}$ is called the Wess-Zumino part of the action. $S_{DBI}$ describes the Dp-brane's coupling to the closed-string fields of the background, namely the metric, dilaton and Neveu-Schwarz-Neveu-Schwarz (NS-NS) 2-form. The Wess-Zumino part of the action describes the Dp-brane's coupling to the RR forms, including for example the coupling $\int A^{(p+1)}$, as well as other terms. In later sections, we will only need $S_{DBI}$, and only with no dilaton or NS-NS 2-form, so $S_{DBI}$ is all we will write explicitly.

The full bosonic action is written in the textbooks, refs. \cite{Polchinski:1998rq,Johnson:2003gi,Becker:2007zj}. Explicit derivations of the DBI action from string theory appear in refs. \cite{Fradkin:1985qd,Leigh:1989jq,Tseytlin:1996it}. The fermionic terms in Dp-brane actions have been studied in refs. \cite{Marolf:2003ye,Marolf:2003vf,Martucci:2005rb}. 

To write the action, we first need a value for the tension, $T_{Dp}$. In terms of fundamental parameters \cite{Polchinski:1995mt},
\beq
T_{Dp} \, = \, \frac{1}{(2\pi)^{p}} \, \frac{1}{g_s} \, \frac{1}{\alpha'^{(p+1)/2}}.
\eeq
The Dp-brane action was originally computed from a calculation at leading order in the open string sector \cite{Fradkin:1985qd,Leigh:1989jq}, which is the disk, with $\chi = 1$, hence the factor $g_s^{-\chi} = g_s^{-1}$ in $T_{Dp}$. The factors of $\alpha'$ follow by dimensional analysis.

Next, we need to know what fields propagate on the Dp-brane. Dp-branes support two kinds of fields: scalars and gauge fields. We will explain the scalars first.

A Dp-brane will be extended in p+1 directions, but can move in directions orthogonal to its worldvolume. The analogy with a drumhead makes this intuitive: a drumhead extended in the $x$-$y$ plane can fluctuate in the $z$ direction. We would thus write a ``scalar field'' on the ``worldvolume'' of the drumhead, $z(t,x,y)$, to describe these fluctuations. We think of $z(t,x,y)$ as saying ``give me a point $(x,y)$ on the drumhead and I will tell you where it ends up in the transverse direction at time $t$.''

A Dp-brane is just the same as a drumhead, except it is relativistic and (possibly) higher-dimensional. Let $\xi^{\mu}$ denote the Dp-brane's worldvolume coordinates, with $\mu = 0, 1, \ldots, p$. We will always have some background geometry, with coordinates $X^M$ and metric $g_{M N}$, where $M,N = 0, 1, \ldots 9$. We embed the Dp-brane into the background by specifying the map $X^M(\xi^{\mu})$, where $p$ of these functions will describe directions along the Dp-brane and $9-p$ of them will describe the fluctuations orthogonal to the Dp-brane. The latter are just like $z(t,x,y)$ in the drumhead example. We actually have some ``gauge'' freedom in choosing the $X^M(\xi^{\mu})$. One physically sensible choice is to identify the time direction of the Dp-brane with the time direction of the background: $X^0(\xi^{\mu}) = \xi^0$. This is called ``static gauge.'' Additionally, in what follows, we will always identify those $X^M(\xi^{\mu})$ that have $M=1,\ldots,p$ with the worldvolume coordinates: $X^M(\xi^{\mu}) = \xi^{\mu}$ for $M=\mu=0,1,\ldots,p$. The remaining $9-p$ functions will then be scalar fields on the Dp-brane worldvolume.

For any gauge choice, the induced metric of the Dp-brane is
\beq
\label{dbranemetric}
g_{\mu \nu}^{Dp} = g_{M N} \frac{\partial X^M}{\partial \xi^{\mu}} \frac{\partial X^N}{\partial \xi^{\nu}}
\eeq
With just the scalar fluctuations, the Dp-brane action is 
\beq
S_{Dp} = - T_{Dp} \, \int \, d^{p+1} \xi \, \sqrt{-\det \, \left( g_{Dp} \right) },
\eeq
where the integral is just the worldvolume of the Dp-brane, the natural quantity for an object with tension to extremize (hopefully minimize). The action is thus the energy of the Dp-brane: the tension is the energy per unit worldvolume, and the integral is just the worldvolume. More specifically, if we think of the action as kinetic minus potential energy, then this Dp-brane action is just minus the potential energy. 

We also promised gauge fields. These arise very simply: suppose an open string has both ends on the same Dp-brane. The string has a tension, and (in this case) nothing forbids the string from minimizing its worldsheet area by collapsing to zero length. What is the low-energy effective description of such massless degrees of freedom? The answer is: a gauge field on the Dp-brane! In particular, with a single Dp-brane, we will have a $U(1)$ gauge field. Remarkably, however, if we have $N_c$ coincident Dp-branes, the gauge group is enhanced to a non-Abelian group, $U(N_c)$. The full non-Abelian action for such a stack of Dp-branes is not completely known. The action for the Abelian case, either a single Dp-brane or just the overall $U(1)$ subgroup of $U(N_c)$, is known, however. The action is of Born-Infeld type: for $U(1)$ field strength $F_{\mu \nu}$, the action is
\beq
\label{dbi}
S_{Dp} = - T_{Dp} \, \int \, d^{p+1} \xi \, \sqrt{-\det \, \left( g^{Dp}_{\mu \nu} + (2\pi\alpha') F_{\mu \nu} \right) }.
\eeq

If we have two D-branes that are separated in some direction, and a string stretching in that direction from one D-brane to the other, then on each D-brane the endpoint of the string will act as a point charge, \textit{i.e.} as a source for the worldvolume gauge field. With a stack of $N_c$ D-branes, the endpoint of a string will act as a source in the fundamental representation of $U(N_c)$ (or anti-fundamental, depending on the orientation of the string). The mass of the point charge will simply be the length of the string\footnote{We do not mean the ``string length'' $\ell_s$, which as we mentioned in section \ref{strings} is a fundamental scale of string theory. Indeed, strictly speaking, the string we are describing should have a length much greater than $\ell_s$, and hence be very heavy, such that we can treat it as a classical object, \textit{i.e.} ignore quantum fluctuations along the string.} times the string tension. The length of the string is just the distance between the D-branes, $l$, so the mass of the point charge will be $l/(2\pi\alpha')$.

Suppose we have a stack of $N_c$ D-branes and then separate some number of them, say $n$, in an orthognal direction, so that we end up with one stack of $N_c -n$, and another of stack of $n$, D-branes. We have thus introduced new open string degrees of freedom, namely the open strings stretched between the two stacks, which will be massive. We will have two gauge theories, one on each stack of D-branes, with gauge groups $U(N_c-n)$ and $U(n)$, as well as massive excitations in the fundamental (or anti-fundamental) representation in each theory. Such a procedure is the D-brane description of the Higgs mechanism. From the point of view of either stack's worldvolume gauge theory, we are studying a non-trivial point on the supersymmetric moduli space of the theory, where the gauge group is reduced and some degrees of freedom (the ``W bosons'') have become massive.

From a field theory perspective, the $U(1)$ part of the $U(N_c)$ gauge group always trivially decouples. From the D-brane perspective, this $U(1)$ corresponds to the overall center-of-mass motion of the entire stack of $N_c$ D-branes \cite{Dai:1989ua}. In cases where we are only interested in the motion (of strings, for example) relative to the stack, we will ignore the overall $U(1)$, and refer to the gauge group of the worldvolume D-brane theory as just $SU(N_c)$.

The Born-Infeld action was discovered in the 1930's \cite{Born:1934gh}. (Schr\"{o}dinger himself discovered that the Born-Infeld action was invariant under electric-magnetic duality \cite{schrodinger}.) At the time, people were worried that, according to Maxwell electrodynamics, the electric field near a point charge grows without bound. They did not know about QED or vacuum polarization, which solves this problem in the real world. Instead, they looked for a new theory of electrodynamics, one that preserved Lorentz and gauge invariance, that reduced to Maxwell theory for small electromagnetic fields, and that had some kind of maximum electric field, to prevent infinite fields at a point charge. They were thus led to the (3+1)-dimensional Born-Infeld action, which obeys all of these constraints. It describes a \textit{nonlinear} theory of \textit{interacting} photons, and so is often called nonlinear electrodynamics. The original Born-Infeld theory did not include fluctuations of the geometry, however (the worldvolume scalars of a Dp-brane): the metric was simply $\eta_{\mu \nu}$.

Notice a crucial assumption: where Maxwell theory is scale-invariant, Born-Infeld theory must have some scale, to set the value of the maximum electric field. The scale introduced in Born-Infeld theory always appears in front of the $F_{\mu \nu}$ factor. In string theory, this scale is the inverse string tension, $2\pi\alpha'$.

In string theory, the Born-Infeld action is an effective action. Suppose we want to write a low-energy action for a Dp-brane. We could imagine integrating out heavy modes: an open string ending on the Dp-brane may have a massive excitation on it, which ``tugs on'' the Dp-brane, exciting the worldvolume fields. At low energy, the appropriate worldvolume fields will be the massless excitations, namely the scalars and gauge fields. Following the rules of effective field theory, we would write every possible term allowed by the symmetries. The Dp-brane's local symmetries include coordinate transformations on the Dp-brane worldvolume and gauge invariance. We must therefore include all powers of $F_{\mu \nu}$, for example. We should think of the DBI action as the result of summing up all the powers of $F_{\mu \nu}$, with the nice result that the infinite number of terms sums up to form the determinant and the square root. The DBI action is thus valid for field strengths $F_{\mu \nu} < 1/(2\pi\alpha')$: the electric field pulls on the ends of an open string, so if the electric field grows larger than the string tension, the string will be ripped apart, and our effective description will break down. This is the sense in which Dp-branes have a ``maximum'' electric field.

In our effective action we must also include powers of \textit{derivatives} of $F_{\mu \nu}$, combined in gauge-invariant ways. The derivatives of the field strength must also remain below the string scale. If they approach the string scale, then the DBI action receives an infinite number of corrections from higher-dimension operators built from derivatives of $F_{\mu \nu}$.

At low energy, Born-Infeld theory reduces to Maxwell electrodynamics. Let us illustrate this with the Dp-brane action. If $F_{\mu \nu}$ and its derivatives are sufficiently below the string scale we can Taylor expand,
\beq
S_{Dp} = - T_{Dp} \, \int \, d^{p+1} \xi \, \sqrt{-\det \, g_{Dp}} \left [ 1 - (2\pi\alpha')^2 \, \frac{1}{4} \, F_{\mu \nu} F^{\mu \nu} + O\left( \alpha'^4 F^4 \right) \right],
\eeq
The first term is just the worldvolume again, and the leading term in $F_{\mu \nu}$ is just the Lagrangian of Maxwell electrodynamics. At low energy, then, the open string degrees of freedom are described just by Maxwell electrodynamics.

A similar statement applies in the non-Abelian case, with $U(1)$ becoming $U(N_c)$. In practical terms, all we do at order $F^2$ is add a trace over gauge indices. We then recover the action of $U(N_c)$ Yang-Mills theory\footnote{As mentioned in the Introduction, we are following the conventions of D-brane physics, for which the definition of $g_{YM}^2$ is actually half the conventional value. The discrepancy comes from the normalization of the generators of the gauge group: $Tr(T_a T_b) = d \, \delta_{ab}$. In D-brane physics, $d=1$, whereas the conventional choice is $d=1/2$.}, $\frac{1}{4 g_{YM}^2} \int \, tr \, F^2$, where we identify the Yang-Mills coupling in terms of fundamental string parameters,
\beq
\label{gym}
g_{YM}^2 \, = \, T_{Dp}^{-1} \, (2\pi \alpha')^{-2} \, = \, (2\pi)^{p-2} \, \alpha'^{(p-3)/2} \, g_s
\eeq
Such a simple prescription (just adding a trace) fails at higher orders in $F$, however, due to ambiguities in the ordering of gauge indices. Currently, the non-Abelian DBI action is only known with certainty up to order\footnote{Schematically, with $D$ representing the gauge-covariant derivative, the derivative corrections are known up to order $D^4 F^4$ and $D^8 F^2$.} $F^6$ \cite{Koerber:2002zb}.

\begin{figure}
{\centering
\includegraphics[width=0.50\textwidth]{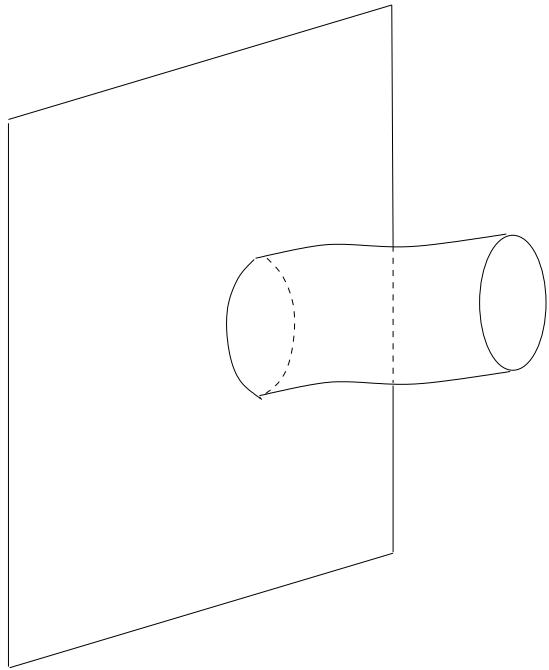}
\caption{\label{dbranefig} Cartoon of a D-brane emitting (or absorbing) a closed string.}
}
\end{figure}

Dp-branes are gravitating objects. We can see this easily: all Dp-branes have a tension, that is, energy per unit volume, and anything with energy must couple to gravity. We can also conclude that Dp-branes gravitate from the fact that Dp-branes interact with closed strings, for example by emitting a closed string, as in figure \ref{dbranefig}, and the fact that the massless mode of the closed string is the graviton. The gravitational physics of Dp-branes motivated the original statement of the AdS/CFT correspondence, to which we now turn.

\section{Statement of the Correspondence}
\label{statement}

In this section, we will review Maldacena's original D-brane construction and the resulting conjecture of the equivalence of two theories, $\N=4$ SYM and string theory on $AdS_5 \times S^5$ \cite{Maldacena:1997re}. Maldacena conjectured that two theories were equivalent, but he did not provide a dictionary between them, which we will need to compute anything. A more precise statement of the correspondence came shortly after Maldacena's conjecture \cite{Witten:1998qj,Gubser:1998bc}. We will review the precise statement of the correspondence in this section.

The argument for the original AdS/CFT conjecture begins with D3-branes. The logic is very simple: D3-branes have \textit{two different} low-energy descriptions, one in terms of open strings and one in terms of closed strings.

We know the open string description: $N_c$ D3-branes are described, at sufficiently low energy, by a $U(N_c)$ Yang-Mills theory formulated on the (3+1)-dimensional D3-brane worldvolume. To identify precisely \textit{which} Yang-Mills theory, we simply count supercharges. Type IIB string theory has 32 supercharges. Introducing the D3-branes breaks half of the supersymmetry, hence 16 supercharges remain, the maximal number in (3+1) dimensions. The theory must thus be the maximally supersymmetric (3+1)-dimensional Yang-Mills theory: $\N=4$ SYM. Recall that $\N=4$ SYM has six real scalar fields. These are in fact the six worldvolume scalars, describing low-energy fluctuations of the D3-brane in the six directions orthogonal to its worldvolume.

What about the closed-string description of D3-branes? The low-energy description of the closed string sector is supergravity, hence D3-branes must have a description in terms of the fields of supergravity. For D3-branes, the relevant fields are the metric and the RR 4-form, $A^{(4)}$. We can thus describe $N_c$ D3-branes as a geometry with some 4-form charge.

So what is the geometry of $N_c$ D3-branes? The answer must preserve the right symmetries. The D3-branes break the Lorentz group of $\mathbb{R}^{9,1}$, $SO(9,1)$, to Lorentz transformations along their worldvolume, $SO(3,1)$, and rotations in the remaining six directions, $SO(6)$. These constraints are in fact enough to determine the answer \cite{Horowitz:1991cd}:
\beq
\label{d3metric}
ds_{D3}^2 = Z(r_6)^{-1/2} \, \left( -dt^2 \, + \, d\vec{x}^2 \right) + Z(r_6)^{1/2} \, \left( dr_6^2 \, + \, r_6^2 \, ds^2_{S^5}\right),
\eeq
with $\vec{x} = (x_1,x_2,x_3)$, and where we have written the metric for the six transverse directions in spherical coordinates, with radial coordinate $r_6$, the distance to the D3-branes. Here $ds^2_{S^5}$ is the metric for a unit $S^5$. To preserve the symmetries, the warp factor $Z(r_6)$ can depend only on the radial coordinate $r_6$, as indicated. Explicitly, the warp factor that solves the supergravity equations is
\beq
Z(r_6) = 1 + \frac{L^4}{r_6^4}
\eeq
with the very important definition
\beq
\label{adsradius}
L^4 = 4 \pi \, g_s N_c \, \alpha'^2
\eeq
The explicit solution for $F^{(5)} = dA^{(4)}$ is
\beq
\label{fiveformsolution}
F^{(5)} = \left( 1 + \star \right) dt \wedge dx_1 \wedge dx_2 \wedge dx_3 \wedge d(Z(r_6)^{-1})
\eeq
with $\star$ the (9+1)-dimensional Hodge star. Notice that $F^{(5)}$ is self-dual, $F^{(5)} = \star \, F^{(5)}$, as it must be. In the six transverse directions, the D3-branes look like a point charge producing $A^{(4)}$ flux. We can compute the flux of D3-brane charge by integrating $F^{(5)}$ over the $S^5$ that surrounds the D3-branes: we find $\int_{S^5} F^{(5)} = N_c$, as of course we must.

To understand the D3-brane geometry, eq. (\ref{d3metric}), we take limits. Suppose we are far away from the D3-branes, so $r_6 \gg L$. We then drop the $L^4/r_6^4$ term in $Z(r_6)$. The metric then becomes that of (9+1)-dimensional Minkowski space. That makes sense: space may be warped near the D3-branes, but should approach flat space asymptotically far away.

What happens as we move close to the D3-branes? That is, what if we take $r_6 \ll L$? Now we drop the ``1'' in $Z(r_6)$, and the metric becomes
\beq
\label{nearhorizond3metric}
ds_{D3}^2 = L^2 \, \frac{dr_6^2}{r_6^2} \, + \, \frac{r_6^2}{L^2}  \left( -dt^2 + d\vec{x}^2 \right) \, + L^2 \, ds^2_{S^5}.
\eeq
Comparing to eq. (\ref{adsmetric}), we immediately identify the metric of $AdS_5$, with radius of curvature $L$, as well as the metric of $S^5$, with the same radius of curvature. The geometry near the D3-branes is thus $AdS_5 \times S^5$.

Roughly speaking, the geometry is divided into three regions: the asymptotic region, which is flat, the so-called ``near-horizon'' region, which is $AdS_5 \times S^5$, and the interpolating region, called the ``throat.'' The throat acts as a gravitational potential well. Suppose an excitation in the near-horizon region travels up the throat and escapes into the asymptotic region. An observer in the asymptotic region will measure an energy lower than what the excitation initially had, that is, a \textit{red-shift} will occur.

The question we now ask is, what low-energy physics will the observer in the asymptotically flat region see? Here again, low-energy means far below the string scale, $\ell_s^{-1}$. In the asymptotic region, the low-energy closed-string physics is just supergravity in flat space, which is a free theory in the infrared. Peering down into the throat, however, the observer will see all the modes of the string theory: even very high-energy string modes, if sufficiently far down the throat, will be very highly redshifted, hence the observer cannot discard them when writing a low-energy effective theory. The non-trivial low-energy dynamics is thus that of type IIB string theory at the bottom of the throat. In other words, the low-energy closed string description of D3-brane physics is type IIB string theory on $AdS_5 \times S^5$.

We thus have two theories, $\N=4$ SYM and type IIB string theory on $AdS_5 \times S^5$, which are two different descriptions of the same physics, namely the low-energy physics of a stack of $N_c$ D3-branes. Maldacena then conjectured that these two theories, which \textit{a priori} look so different, must in fact be \textit{equivalent}.

A quick and easy check of Maldacena's claim (that we actually already did at the end of section \ref{adsspace}) is to compare symmetries, which must be identical if the theories are equivalent. The isometry group of $AdS_5$, $SO(4,2)$, is identical to the (3+1)-dimensional conformal group. The isometry of the internal space, the $S^5$, is of course $SO(6)$, which is isomorphic to the $SU(4)_R$ part of the R-symmetry of $\N=4$ SYM.

Notice that the \textit{global symmetries} of the gauge theory are dual to \textit{gauge invariances} (local transformations) of the gravity theory: the conformal group and R-symmetry group were dual to isometries of the metric. Indeed, this is a generic feature of gauge-gravity duality, and arises because gravity admits no global symmetries\footnote{The statement that gravity admits no global symmetries comes from black hole physics. A particle carrying some charge can fall into a black hole. The charge must be destroyed in this process: a black hole has no hair, so no observer outside the hole could know that a charge fell into it. Notice we are not discussing electric charge, since this is not associated with a global symmetry. Electrically charged black hole solutions do exist in gravitational theories, and if such a black hole ate an electrically charged particle, then the black hole's charge would change by one unit. Arguments also exist showing that string theory admits no global symmetries (see for example section 4.1 of ref. \cite{ArkaniHamed:2006dz}).}.

Matching symmetries is not a statement about dynamics, however, and thus is far from a proof of Maldacena's conjecture. Nevertheless, as explained in the Introduction, a great deal of evidence suggests that the correspondence is true.

An important question, though, is where exactly can we \textit{trust} our solution for the metric and five-form? Remember, these were solutions of classical supergravity! We should not trust our solution if the curvature reaches the string scale, where our effective description in terms of supergravity will break down (the supergravity action will receive stringy corrections). In other words, we want the radius of curvature to be bigger than the string length, $L^4 / \ell_s^4 \gg 1$. We can rewrite eq. (\ref{adsradius}) as
\beq
\frac{L^4}{\ell_s^4} \, = \, 4 \pi \, g_s N_c. 
\eeq
We understand perturbative string theory best, so we want $g_s \ll 1$. If $N_c$ is finite, that would mean $L^4 / \ell_s^4 \ll 1$, which is the opposite of what we want. We should thus take $N_c \ra \infty$ as we take $g_s \ra 0$ in such a fashion that $4 \pi g_s N_c$ remains finite, and then chose $4 \pi g_s N_c$ to be large, so that $L^4 / \ell_s^4 \gg 1$.

From eq. (\ref{gym}), with $p=3$, we can idenfity $g_{YM}^2 \, = \, 2\pi \, g_s$. What do the limits above translate into in the field theory? The first limit was $g_s \ra 0$ and $N_c \ra \infty$ with $g_s N_c$ fixed. Defining the 't Hooft coupling $\lambda \equiv 2 \, g_{YM}^2 N_c$, we can see that this limit is the 't Hooft limit, $g_{YM}^2 \ra 0$ and $N_c \ra \infty$ with $\lambda$ fixed. The next limit is then to take $\lambda \ra \infty$. The limits where the dual description as a classical theory of gravity is valid become the 't Hooft limit, with large 't Hooft coupling.

As we have repeated a few times now, AdS/CFT gives us a new tool for studying strongly-coupled gauge theory: classical gravity. Maldacena conjectured that the two theories are equivalent for \textit{all} values of $g_s$ and $N_c$, however, which is usually called the ``strong form'' of the conjecture. The ``weak form'' of the conjecture, which is perhaps on more solid ground, equates type IIB supergravity on $AdS_5 \times S^5$ with $\N=4$ SYM theory in the 't Hooft limit, at large 't Hooft coupling.

How do we actually use gravity to compute observables in the field theory, though? We need a more precise statement of the correspondence. Roughly speaking, the idea is that two theories being ``dual'' means their path integrals should be identical. The degrees of freedom over which we integrate will be different, but upon performing the path integral the result will be the same, hence the theories are physically equivalent. From this perspective, a duality transformation is just a change of variables, which of course cannot change the result for the integral.

If we believe the strong form of the conjecture, we should imagine equating the path integral of $\N=4$ SYM theory with some kind of path integral for type IIB string theory. The problem is that no one knows how to define such a path integral for string theory, primarily because no one knows what the appropriate degrees of freedom would be.

If we limit ourselves to the weak form of the conjecture, we can argue that, whatever the string path integral should be, supergravity should be a good saddle point approximation to it. We can then write the precise statement of the correspondence. Let $\phi$ be some supergravity field, which could be a scalar, a vector, the metric, etc. (we will ignore all indices). Let $\overline{\phi}$ be a particular solution of the supergravity equations of motion, with leading asymptotic behavior $\overline{\phi}_0$ near the boundary. The AdS/CFT correspondence is the statement that \cite{Witten:1998qj,Gubser:1998bc}
\beq
\label{partitionfunctions}
\left < e^{i \, \int d^4x \, \O \, \overline{\phi}_0 } \right>_{SYM} \, = \, e^{i \, S_{SUGRA}[\overline{\phi} \, \ra \, \overline{\phi}_0]}
\eeq
On the left hand side, we have the path integral of the $\N=4$ SYM theory, or more precisely, a generating functional: $\O$ is some operator, and $\overline{\phi}_0$ is a source for that operator. We compute correlators of $\O$ by taking functional derivatives with respect to $\overline{\phi}_0$. On the right-hand side is an exponential of the supergravity action, evaluated on the solution $\overline{\phi}$, also known as the ``on-shell'' action, representing a saddle-point approximation to the full string theory path integral.

To compute a correlator for $\O$ in $\N=4$ SYM, then, the AdS/CFT recipe is:

\smallskip

1.) Determine which field $\phi$ is dual to $\O$.

2.) Solve the supergravity equations for $\phi$.

3.) Plug the solution $\overline{\phi}$ into the supergravity action and exponentiate.

4.) Take variational derivatives with respect to the leading asymptotic value $\overline{\phi}_0$.

\smallskip

Notice that we can compute the \textit{connected} correlators of $\O$ by taking the logarithm of both sides of eq. (\ref{partitionfunctions}) and taking functional derivatives of the on-shell supergravity action itself. We will define the notation for connected correlator  
\beq
\langle \O \rangle \, = \, \frac{\langle \O \rangle_{SYM}}{\langle 1 \rangle_{SYM}} \, = \, \frac{\delta}{\delta \overline{\phi}_0} S_{SUGRA}
\eeq
where in the first equality, the denominator is the expectation value of the identity operator (also known as the path integral itself), and in the second equality we have invoked the correspondence to translate to supergravity quantities.

The first step in the AdS/CFT recipe can actually be very difficult. Given an operator, how do we determine the dual field, and vice versa? No sure-fire recipe exists for how to do this, so we usually have some detective work to do. Most of the time, though, the dimension and symmetries of the operator are enough to identify the dual field.

The precise statement of the correspondence itself, eq. (\ref{partitionfunctions}), also has a big problem: the quantities on both sides are divergent! On the left-hand side, the generating functional has ultraviolet divergences. On the right-hand side, the on-shell supergravity action generically has divergences due to the infinite volume of AdS space. To make eq. (\ref{partitionfunctions}) truly precise, we must introduce some consistent regularization and renormalization scheme. Which scheme is best often depends on the problem.

For example, suppose we have a metric that satisfies Einstein's equation and asymptotically approaches AdS. If we evaluate the Einstein-Hilbert action on such a solution, we will find a divergence coming from integration over the infinite distance to the boundary. A sensible way to regulate the divergence is to introduce a cutoff: using Fefferman-Graham coordinates, we would integrate only to $z=\e$ rather than $z=0$, with plans to send $\e \ra 0$ in the end. We could then renormalize by subtracting the on-shell action for pure AdS, similarly regulated. The two actions will have the same $\e \ra 0$ divergences, which will cancel for finite $\e$. The $\e \ra 0$ limit of the \textit{difference} is therefore finite.

This method is fairly simple, but requires an explicit ``background'' solution to cancel divergences. Such a background solution may not always be available. A more general method does exist, however, that makes no reference to background solutions, called ``holographic renormalization'' \cite{Henningson:1998gx,Balasubramanian:1999re,deHaro:2000xn,Skenderis:2000in,Bianchi:2001kw,Skenderis:2002wp}. In holographic renormalization, we again regulate by integrating only to $z=\e$. We then add counterterms on the $z=\e$ hypersurface to cancel those divergences coming from integration. These counterterms can preserve all symmetries, such as general covariance and gauge invariance (when both are restricted to the $z=\e$ hypersurface), and make no reference to any background solution.

An extremely important point is that the \textit{infrared} (long-distance) divergences on the supergravity side, from integrating over the infinite distance to the AdS boundary, translate into the \textit{ultraviolet} (short-distance) divergences of the field theory. We have found one manifestation of a more general property of the correspondence, the so-called ``UV-IR relation'' \cite{Susskind:1998dq,Peet:1998wn}. Heuristically, the UV-IR relation states that ultraviolet physics of the field theory is dual to physics ``near the boundary,'' while infrared physics in the field theory is dual to physics deep inside AdS. In what follows, we will make no distinction between ``near the boundary'' and ``the UV,'' or between ``deep inside AdS'' and ``the IR.''

More precisely, the radial coordinate can in fact be identified as the renormalization scale of the field theory \cite{Balasubramanian:1998de,Balasubramanian:1999jd,deBoer:1999xf}. A good mnemonic device to recall this is to perform a scale transformation, which is an isometry of the AdS metric. We take $x_{\mu} \ra \lambda \, x_{\mu}$, with $\lambda$ some positive real number (not the 't Hooft coupling). For the AdS metric in eq. (\ref{nearhorizond3metric}) to remain invariant, we must also take $r_6 \ra \lambda^{-1} \, r_6$, which is the way an energy must transform under a scale transformation. A more rigorous derivation also exists, in which the field theory renormalization-group equations are computed from supergravity in AdS \cite{Balasubramanian:1999jd,deBoer:1999xf}. The role of the radial coordinate as the renormalization scale is then explicit.

The UV-IR relation and the role of an extra dimension as an energy scale are believed to be general features of gauge-gravity duality. In other words, the way some gauge theories ``know'' about higher-dimensional gravity is through their renormalization group flow. The simplest flow is none at all: a conformal theory. Conformal theories are always dual to AdS geometries, simply because of symmetry (the conformal group). Many examples of non-trivial renormalization flows are also known, usually for theories that are conformal in the UV but flow to non-trivial IR fixed points. Holographically, such a flow appears as a geometry that is asymptotically AdS but deforms as we move into the bulk (a few examples appear in refs. \cite{Constable:1999ch,Pilch:2000ue}). Examples of theories exhibiting confinement in the IR are also known (as a start, see refs. \cite{Sakai:2004cn,Girardello:1999bd,Polchinski:2000uf,Klebanov:2000hb}). Of course, the dual of QCD is not known (yet)!

We will illustrate the mechanics of the correspondence in greater detail in the next chapter, when we discuss probe D-branes. There we will have D-branes living in $AdS_5 \times S^5$, with worldvolume scalars and gauge fields. We will identify the field theory operators dual to these fields, perform holographic renormalization for the DBI action, and explicitly compute one-point functions for field theory operators from the on-shell DBI action.

\section{The Correspondence at Finite Temperature}
\label{finiteTcorrespondence}

The generalization of the AdS/CFT correspondence to finite temperature is fairly intuitive. The geometry dual to $\N=4$ SYM at finite temperature is AdS-Schwarzschild (times $S^5$) \cite{Witten:1998zw}. We identify the Hawking temperature and the entropy of the AdS black hole with the temperature and entropy of the field theory \cite{Gubser:1996de,Witten:1998zw}. Additionally, if we are interested in thermodynamics\footnote{In the following chapters, we will also be interested in finite-temperature \textit{real-time} dynamics, which requires Lorentzian signature.}, we must also Wick rotate to Euclidean time and compactify the time direction into a circle of circumfrence $1/T$, as is standard for thermal field theory.

As mentioned above, if the only scale we introduce into a conformal field theory is the temperature, no thermal phase transitions can occur. Once we have characterized the state of the theory at any one temperature, we are done with thermodynamics: we know the state of the system at any other temperature. With only one dimensionful scale, the dependence of thermodynamic functions, like the free energy, entropy, etc., is completely determined by dimensional analysis to be some power of the temperature. For example, for $\N=4$ SYM at large $N_c$ but \textit{zero} 't Hooft coupling, the entropy density is $s \, = \, \frac{\pi^2}{2} \, N_c^2 \, T^3$ \cite{Gubser:1996de}.

The factor of $N_c^2$ is suggestive. Large-$N_c$ pure Yang-Mills theory has a first-order deconfinement transition, characterized for example by a sudden jump in the entropy density from order $N_c^0$ to order $N_c^2$. The heuristic way to understand this is is to think of $s$ as counting degrees of freedom. At low temperatures, the appropriate degrees of freedom should be color singlets, \textit{i.e.} glueballs, which provide the order $N_c^0$ degrees of freedom, while at sufficiently high temperature the appropriate degrees of freedom should be deconfined gluons, which provide the order $N_c^2$ degrees of freedom. If we \textit{define} ``deconfinement'' as the leading-order behavior of the entropy density at large $N_c$ (and ignore, for example, any statements about the potential between two test charges), then we may declare the \textit{free}, large-$N_c$ $\N=4$ SYM theory to be deconfined at any temperature. While imprecise, this can be a useful way to think about the physics of finite-temperature $\N=4$ SYM. 

Returning to eq. (\ref{partitionfunctions}), and upon Wick rotation to Euclidean time, we can re-label the left-hand side as $Z[\overline{\phi}_0]$, and call this the thermodynamic partition function in the presence of the source $\overline{\phi}_0$. In thermodynamics, we define the free energy $F$ by $Z[\overline{\phi}_0] = e^{-F/T}$. We can thus identify
\beq
F \, = \, T \, S_{SUGRA}[\overline{\phi} \, \ra \, \overline{\phi}_0]
\eeq
Up to a factor of $T$, the Euclidean on-shell supergravity action \textit{is} the free energy of the field theory at large-$N_c$ and large 't Hooft coupling.

What about finite-density thermodynamics? In other words, what happens if we introduce a chemical potential in the field theory? First of all, we know that with two dimensionful scales, interesting phase transitions become possible. Second, we have a choice of ensembles, canonical or grand canonical. Ultimately, the physics does not depend on our choice of ensemble, but which ensemble does the on-shell supergravity action describe? For charged AdS black holes, the answer is known: the on-shell action is the grand canonical potential, also known as the Gibbs free energy \cite{Chamblin:1999tk,Chamblin:1999hg}. We will denote this as $\Omega$. The potential in the canonical ensemble is the Helmholtz potential, which we have denoted as $F$. The two are related, as usual, by a Legendre transform.

Inserting the AdS-Schwarzschild metric (times $S^5$) into the Einstein-Hilbert action, and cancelling divergences by subtracting the action evaluated on $AdS_5 \times S^5$, we can compute the free energy for $\N=4$ SYM at large-$N_c$ and large 't Hooft coupling. From the free energy, we can compute the entropy density $s$. Surprisingly, the supergravity result for $s$ is precisely $3/4$ the free-field value \cite{Gubser:1996de}! Indeed, some have speculated that the entropy density smoothly interpolates from the free-field value to $3/4$ the free-field value as the coupling increases \cite{Gubser:1998nz,Fotopoulos:1998es}.

We can actually extract an important lesson for QCD from the supergravity result for $\N=4$ SYM, of the ``quantitative universal'' variety. In QCD, asymptotic freedom suggests that at asymptotically high temperatures, the value of $s$ should be that of an ideal gas of quarks and gluons. Lattice studies then showed that the value of $s$ at lower temperatures, including temperatures explored at RHIC, was about $80\%$ the ideal gas value \cite{Karsch:2003jg}. Many people therefore hoped that perturbative calculations performed for asymptotically high temperatures could be extrapolated to RHIC temperatures. The lesson from $\N=4$ SYM is this: a value of $s$ close to the free-field value, even $75\%$ of it, is definitely not a sign of weak coupling. Indeed, the results from RHIC, for example the large ``elliptic flow,'' indicate that the QCD plasma created there is very strongly coupled \cite{Shuryak:2003xe,Shuryak:2004cy}.

%
%
%
%
%
%
%
%
%
%
%
%
%
%
%
\chapter {Adding Flavor to AdS/CFT}
\label{flavor}

All of the fields of the $\N=4$ SYM theory transform in the adjoint representation of the gauge group. In QCD, the gluons are in the adjoint representation, but the quarks are in the fundamental representation. To make $\N=4$ SYM look a little bit more like QCD, then, we want to introduce additional fields that transform in the fundamental representation, \textit{i.e.} flavor fields. In this section we will describe how this is done in AdS/CFT. Put another way, we will address a question of the ``qualitative universal'' variety: ``What do flavor fields look like holographically?''

\smallskip

\textbf{IMPORTANT NOTE ABOUT UNITS}: Starting now, we measure all lengths in units of the $AdS_5$ radius, $L$. In other words, we use units in which $L\equiv1$. An important conversion from supergravity quantities to field theory quantities is then $\alpha'^{-2} = \lambda$.

\section{Adding D7-Branes}

We start at the start, with a stack of $N_c$ D3-branes in flat space. As described in section \ref{dbranes}, we know that the open string description leads to a gauge theory on the D3-brane worldvolume, at sufficiently low energies. If D3-branes are the only D-branes present, then both ends of an open string must end on the D3-branes. One endpoint acts as a point charge in the fundamental representation while the other endpoint acts as a point charge in the anti-fundamental representation. The fundamental and anti-fundamental together form the adjoint representation. In the situation we are describing, nothing forbids the string from collapsing to zero length, forming a massless excitation in the adjoint representation of $SU(N_c)$. We want fields in the fundamental representation, so we therefore need open strings with only one endpoint on the D3-branes. The other end must go somewhere, so we must introduce some other D-branes.

What D-branes should we introduce, and where should we put them? As mentioned in section \ref{dbranes}, one way to produce massive excitations in the fundamental representation of the gauge group is to separate some number of D3-branes from the initial stack, so that we then have one stack of, say, $n$ D3-branes and another of $N_c - n$ D3-branes. We will then have open strings stretched from the $n$ D3-branes to the $N_c - n$ D3-branes, and hence massive fundamental-representation excitations. The physics of the worldvolume theory is a little different from what we want, however. This construction is interpreted in the field theory as moving to a non-trivial point on the moduli space of $\N=4$ SYM, where the gauge group is broken from $SU(N_c)$ to $SU(N_c-n)$. The massive fundamental-representation fields are W-bosons, \textit{i.e.} massive \textit{vector} fields. These do have fermionic superpartners, which we could call quarks, but we would like to introduce flavor fields that are fermions or scalars, not vectors, and we would like to do so without having to move onto the moduli space of the theory and reducing the gauge group. How can we do this?

Let us jump to the answer \cite{Karch:2002sh}, and then explain why it's a good choice. We introduce a number $N_f$ of D7-branes, oriented relative to the D3-branes as described by the following array, with $X_{\mu}$ the coordinates of $\mathbb{R}^{9,1}$:
\begin{equation}
\begin{array}{ccccccccccc}
&X_0 & X_1 & X_2 & X_3 & X_4 & X_5 & X_6 & X_7 & X_8 & X_9\\
\mbox{D3} & \times & \times & \times & \times & & &  &  & & \\
\mbox{D7} & \times & \times & \times & \times & \times  & \times
& \times & \times &  &   \\
\end{array}
\end{equation}
The D7-branes overlap the D3-branes in all of the D3-brane's directions, but also extend into four more directions orthogonal to the D3-branes. Notice that two directions, $X_8$ and $X_9$, are orthogonal to both stacks of D-branes.

Two things motivate this choice. The first thing is supersymmetry. We know that adding D-branes will break at least half the supersymmetry, and possibly more. We want to preserve as much supersymmetry as possible, to guarantee stability and retain as much control over the system as possible. General analyses of intersecting D-branes reveal that precisely half the supersymmetries will be preserved if we have \textit{four} or \textit{eight} directions in which one D-brane is extended but the other is not \cite{Polchinski:1998rq,Johnson:2003gi,Becker:2007zj}. In terms of an array like the one above, we need four or eight directions in which only one row has an ``$\times$.'' We thus introduce the D7-branes as above, so that in the directions $X_4$ to $X_7$, one D-brane is extended while the other is not.

The second thing that motivates our choice is the issue of dimensionality. If our only demand was to preserve half the supersymmetries, we could have introduced D5-branes, arranged as
\begin{equation}
\begin{array}{ccccccccccc}
&X_0 & X_1 & X_2 & X_3 & X_4 & X_5 & X_6 & X_7 & X_8 & X_9\\
\mbox{D3} & \times & \times & \times & \times & & &  &  & & \\
\mbox{D5} & \times & \times & \times & & \times  & \times
& \times & &  &   \\
\end{array}
\end{equation}
Such an array again has four directions with only one D-brane, $X_3$, $X_4$, $X_5$, and $X_6$. Imagine an open string stretched between the D3-branes and the D5-branes, however. The endpoint on the D3-branes cannot propagate throughout the entire D3-brane worldvolume, but only along the (2+1)-dimensional intersection with the D5-branes, \textit{i.e.} in the $X_0$, $X_1$, and $X_2$ directions. We would like our flavor fields to propagate in the same 3+1 dimensions as the adjoint fields, just like the quarks in QCD. That, together with supersymmetry, uniquely determines the choice of D7-branes above.

By introducing the D7-branes, we have introduced open strings whose endpoints end on the D3-branes, and at low energies on the D3-brane worldvolume will behave as point charges in the fundamental representation of $SU(N_c)$. We can separate the two stacks of D-branes, so that these open strings will have nonzero length, and in the D3-brane's worldvolume theory the point charges will have nonzero mass. We can only separate the D-branes in directions orthogonal to both stacks, which in this case are the $X_8$ and $X_9$ directions. In the $X_8$-$X_9$ plane, both stacks of D-branes look like points. Without loss of generality we can separate these points in the $X_8$ direction. Doing so does not break any more supersymmetries.

\section{D3-Brane Worldvolume Theory}
\label{d3braneworldvolumetheory}

At low energy, what is the field theory living on the D3-branes? We still have the fields of $\N=4$ SYM, but we will now also have flavor fields, preserving only $\N=2$ supersymmetry and transforming in the fundamental representation of the gauge group. The matter multiplet of $\N=2$ supersymmetry is called the hypermultiplet, which is comprised of two complex scalars and two Weyl spinors of opposite chirality, or equivalently, one Dirac spinor. One scalar and one Weyl spinor transform in the $N_c$ of $SU(N_c)$, and the other scalar and spinor transform in the conjugate representation, $\overline{N}_c$. We will ignore this subtlety and refer to the fields as if they are all in the fundamental representation. In particular, we will refer to the hypermultiplet fermions as ``quarks,'' in analogy with QCD. A big difference from QCD, however, is the presence of scalars in the fundamental representation, which, being the superpartners of the quarks, we will call ``squarks.'' All of the fields in the hypermultiplet will of course have the same mass, $m$. With $\N=2$ supersymmetry, the Lagrangian mass and the physical mass are actually identical, \textit{i.e.} the Lagrangian mass is not renormalized.

The Lagrangian is that of $\N=4$ SYM theory plus terms for the hypermultiplet fields. Again, we will not reproduce the Lagrangian here. It is written explicitly in ref. \cite{Chesler:2006gr}. It will of course include the kinetic terms for the flavor fields, with gauge-covariant derivatives providing couplings to the gauge fields, as well as mass terms for the hypermultiplet fields, and quartic couplings for the hypermultiplet scalars. That is not everything, but to explain the remainder we need some notation. Let $\Phi$ and $\Psi$ represent a complex scalar and Dirac fermion (respectively) of the $\N=4$ multiplet, and let $\phi$ and $\psi$ represent a complex scalar and the Dirac fermion of the hypermultiplet. Speaking very schematically, the Lagrangian contains Yukawa couplings of the form $\bar{\psi} \Psi \phi$ and $\bar{\psi} \Phi \psi$, quartic couplings of the form $\phi^{\dagger} \Phi^{\dagger} \Phi \phi$, and a cubic coupling of the form $m \phi^{\dagger} \Phi \phi$.

What are the global symmetries of $\N=4$ SYM coupled to $\N=2$ hypermultiplets? We begin with the fermionic generators. We have already suggested that, of the sixteen supercharges of $\N=4$, only the eight of $\N=2$ supersymmetry remain. As for the sixteen superconformal generators, if $m=0$, then only eight are broken, while if $m$ is nonzero, then all sixteen are broken.

What about the R-symmetry? The flavor fields break the $SO(6)_R$ part of the R-symmetry to $SO(4)_R \times SO(2)_R$. The $SO(4)_R$ includes the expected $SU(2)_R$ and an $SU(2)$ that rotates two of the complex scalars in the $\N=4$ multiplet as a doublet, namely the two scalars that do not participate in the $m \phi^{\dagger} \Phi \phi$ coupling. In the D-brane picture, the breaking of $SO(6)_R$ to $SO(4)_R \times SO(2)_R$ is easy to see from the array above. Before we introduce the D7-branes, the six directions orthogonal to the D3-branes have an $SO(6)$ rotation symmetry. The D7-branes clearly break this to the $SO(4)$ rotations in the $(X_4,X_5,X_6,X_7)$ directions and $SO(2)$ rotations in the $X_8$-$X_9$ plane.

This $SO(2)_R \sim U(1)_R$ symmetry acts as a chiral symmetry: the left- and right-handed Weyl fermions in the hypermultiplet carry opposite charges under the $U(1)_R$. Of course, the $\N=4$ fields are also charged under this $U(1)_R$: one complex scalar has charge $+2$ (by convention), two Weyl fermions have charge $+1$ and two have charge $-1$. The scalars in the hypermultiplet are neutral. The crucial point is that this $U(1)_R$ is analogous to the axial symmetry of QCD, and in particular is anomalous when $N_c$ is finite, for the same reasons as in QCD. (To see the anomaly in the D-brane description requires introducing the dilaton-axion, which we will not discuss.) Also as in QCD, a finite quark mass explicitly breaks the axial symmetry. Again, this is easily seen in the D-brane picture. At zero mass, the D3-branes and D7-branes are at the same point, right on top of each other in the $X_8$-$X_9$ plane, so that $SO(2)$ rotations about that point are a good symmetry (again, ignoring the axion-dilaton). Once we separate the D3-branes and the D7-branes, however, $SO(2)$ rotations about the D3-branes are explicitly broken by the presence of the D7-branes.

The flavor fields will not spoil the Poincar\'{e} invariance of the theory, but what about the conformal symmetries? A finite mass $m$ will obviously break scale invariance, so the only place where we could hope to have conformal symmetry is the massless case, $m=0$. Even with $m=0$, if $N_f$ is on the order of $N_c$, then conformal invariance is lost: the theory dynamically generates a scale. More specifically, quantum effects due to the flavor fields cause the coupling to run, so that the coupling of the theory grows stronger in the UV (like QED, unlike QCD), suggesting that the theory requires some UV completion, that is, some new physics at short distances. Even at zero mass, then, the conformal group is broken to the Poincar\'{e} group (in the quantum theory, when $N_f$ is on the order of $N_c$).

If all $N_f$ hypermultiplets have the same mass $m$, that is, if the mass matrix in flavor space is proportional to the identity, then the theory also has a global, bosonic $U(N_f)$ symmetry, just like the vector symmetry of QCD (with mass-degenerate quarks). Notice that, in analogy with QCD, we can identify the diagonal $U(1)$ as baryon number. More generally, we could introduce a different mass for each flavor, in which case the flavor symmetry would be broken to $U(1)^{N_f}$. For simplicity, though, we will only study the case in which all flavors have the same mass $m$. In the D-brane picture, this is the statement that in the $X_8$-$X_9$ plane, we will keep all $N_f$ D7-branes right on top of each other (a single point) rather than separating them (many points) and placing them at different distances from the D3-branes.

Looking ahead a bit, we know that we will eventually take $N_c \ra \infty$. When we do so, we will always keep $N_f$ finite, so that $N_c \gg N_f$, and we will never go beyond order $N_f \, N_c$ in anything we compute (recall that generically the leading order is $N_c^2$). We call this the ``probe limit.'' From a field theory point of view, the probe limit is a simplification because the quantum effects of the flavor fields are highly suppressed, and if we only work to order $N_f N_c$, their effects are not apparent at all. In the language of perturbation theory, this would be the statement that no diagrams with quark (or squark) loops contribute at the order to which we work. Two nice things result from this.

The first nice thing is that, if $m=0$, conformal invariance is restored! More precisely, in the probe limit we neglect precisely those quantum effects, due to the flavor fields, that cause the coupling to run. In this limit, then, no dynamically generated scale will be apparent, and the theory will appear to be conformally invariant. Of course, a nonzero mass will explicitly break conformal invariance, but nevertheless, the probe limit is nice because the only scales in any problem will be the ones we introduce by hand.

The second nice thing that results from the probe limit is that we also neglect the quantum effects that produce the $U(1)_R$ anomaly, so that, when $m=0$, the $U(1)_R$ will appear to be a genuine symmetry of the theory. Again, however, a nonzero mass will explicitly break the $U(1)_R$.

As a final note, we have been describing the low-energy physics on the D3-branes' worldvolume, but we could just as well ask about the low-energy physics of the D7-branes' worldvolume. The difference is just one of perspective. The low-energy theory on the D7-branes will be a (7+1)-dimensional Yang-Mills theory. What was a global symmetry on the D3-branes, $U(N_f)$, is now the gauge invariance of the D7-brane worldvolume theory. The theory will be supersymmetric and contain adjoint fields, from open strings with both ends on the D7-branes, as well as fields in the fundamental representation of $U(N_f)$, from open strings stretched between the D3-branes and D7-branes. These flavor fields will be restricted to propagate along the (3+1)-dimensional intersection with the D3-branes, and will have a $U(N_c)$ global symmetry.

\section{D7-Branes in $AdS_5 \times S^5$}

\subsection{The Probe Limit}

We now want to repeat Maldacena's arguments, but with the extra ingredient of D7-branes. That is, we want to know the low-energy \textit{closed string} description of the system, \textit{i.e.} the supergravity description. In supergravity, the D3-branes will again act as a source for the metric and self-dual five-form, $F^{(5)}$. The D7-branes will also source the metric, and in addition the axion-dilaton, which is a (pseudo-)scalar field.

The D7-branes actually complicate the picture enough that the supergravity solution for the D3/D7 intersection is not known for all values of $N_c$ and $N_f$. Many problems arise when $N_f$ is on the order of $N_c$, for example, the curvature diverges close to the D-branes, while the dilaton diverges far away from them\footnote{More details about the supergravity solution of the D3/D7 intersection appear in chapter 4 ref. \cite{Erdmenger:2007cm}, and references therein.}.

Here is where the probe limit takes the spotlight. If we take $N_c \gg N_f$, so that the D3-branes vastly outnumber the D7-branes, then we expect the effect of the D7-branes to be negligible. In particular, if we have $N_c \ra \infty$ D3-branes and, say, just a single D7-brane, then we expect the D3-branes to warp the geometry much more than the D7-branes. We may then study a single D7-brane ``cleanly probing'' the geometry produced by the D3-branes: the D7-branes will be embedded in the geometry but will not deform it.

These statements can of course be made precise by examining the supergravity equations of motion, although we will not do so explicitly. The key point is that the D3-branes source the metric and five-form with a strength $g_s N_c$, while the D7-branes source the metric and axion-dilaton with a strength $g_s N_f$. As before, if we take $g_s \ra 0$ and $N_c \ra \infty$ such that $g_s N_c$ is fixed, then if $N_f$ remains finite we will have $g_s N_f \ra 0$, so we may safely ignore the axion-dilaton and the D7-branes' effect on the metric.

The background is the one we saw before, in section \ref{statement}, where the geometry near the D3-branes is $AdS_5 \times S^5$, and the five-form solution is still eq. (\ref{fiveformsolution}). The six coordinates transverse to the D3-branes became the radial coordinate of $AdS_5$ and the directions of the $S^5$. The D7-brane is extended in four of those directions, and hence will be extended along the radial direction and along an $S^3$ inside $S^5$. We thus expect the D7-brane to be extended along $AdS_5 \times S^3$ in the near-D3-brane geometry.

Lastly, taking the limit in which $g_s N_c \sim L^4 / \ell_s^4 \gg 1$, we arrive at an expanded version of the AdS/CFT correspondence, one that includes flavor fields: $\N=4$ SYM, at large-$N_c$ and large 't Hooft coupling, is dual to supergravity in $AdS_5 \times S^5$, and a number $N_f \ll N_c$ of $\N=2$ hypermultiplet fields are dual to a number $N_f$ of D7-branes extended along $AdS_5 \times S^3$.

\subsection{D7-brane Embeddings}
\label{d7braneembeddings}

We now want to make all of this precise, using equations. In particular, we will derive explicit, analytic solutions for the embedding of the D7-brane in the background geometry \cite{Karch:2002sh,Kruczenski:2003be}. Given these solutions, we will explain how to extract field theory quantities, such as the mass $m$ of the hypermultiplets.

The action that we will need for the probe D7-branes is just the DBI action. We will only need the Abelian version,
\beq
S_{D7} \, = \, - N_f \, T_{D7} \, \int d^8 \xi \, \sqrt{-\det(g^{D7}_{\mu \nu} + (2\pi\alpha')F_{\mu \nu})}
\eeq
We have $N_f$ D7-branes, so the action is just $N_f$ copies of the action for a single D7-brane, hence the overall factor of $N_f$. We can translate the D7-brane tension to field theory quantities as
\beq
T_{D7} \, = \, (2\pi)^{-7} \, g_s^{-1} \, \alpha'^{-4} \, = \, \frac{\lambda \, N_c}{2^5 \, \pi^6}.
\eeq
The D7-brane action is thus order $\lambda N_f N_c$.

To describe probe D7-brane embeddings, we will use two different coordinate systems, those of eq. (\ref{nearhorizond3metric}) and the Fefferman-Graham coordinates of eq. (\ref{adsbhmetric}). The latter are especially convenient for holographic renormalization. In this section we will study the worldvolume scalars of the D7-brane, not the gauge fields, which we leave for later sections. We will also focus only on zero temperature embeddings, leaving the finite-temperature case for future chapters.

We begin by rewriting the metric of eq. (\ref{nearhorizond3metric}) slightly, in a fashion that makes the symmetries of the D7-brane manifest,
\beq
ds_{D3}^2 \, = \, Z(r_6) \, \left( -dt^2 + d\vec{x}^2 \right) \, + \, Z(r_6)^{-1} \, \left( dr^2 + r^2 \, ds^2_{S^3} + dy^2 + y^2 \, ds^2_{S^1} \right).
\label{adsmetricrcoordinate}
\eeq
The warp factor is $Z(r_6) = r_6^2$. We have divided the six directions orthogonal to the D3-branes into the four in which the D7-brane is extended and the two in which neither kind of D-brane is extended. In the four directions parallel to the D7-branes, we have introduced spherical coordinates, with radial coordinate $r$ and the metric of $S^3$ denoted $ds_{S^3}^2$. In the remaining two directions, which in the asymptotic geometry form the $X_8$-$X_9$ plane, we have introduced polar coordinates, with radial coordinate $y$ and the metric of $S^1$ denoted $ds_{S^1}^2$. Notice, then, that $r_6^2 = r^2 + y^2$.

The D7-brane is extended along all of the Minkowski directions, as well as $r$ and the $S^3$ directions, and can fluctuate in directions orthogonal to itself, namely $y$ and the $S^1$ direction. We will not consider motion in the $S^1$ direction. We are very interested in motion in $y$, however: we know that a nonzero separation in the asymptotic $X_8$-$X_9$ plane corresponds to a nonzero hypermultiplet mass in the dual field theory.

We thus consider only motion in $y$. As a D7-brane worldvolume scalar, $y$ can in principle depend on any of the worldvolume coordinates: $y(\xi^{\mu})$. To simplify our search for solutions, we will demand that the D7-brane embeddings preserve certain symmetries. For example, we want to preserve the Lorentz invariance of the dual field theory, so we will not allow $y$ to depend on the Minkowski coordinates. For simplicity, we will also assume the embedding preserves the $SO(4)$ isometry of the $S^3$, hence $y$ will not depend on the $S^3$ coordinates. The only remaining possibility, then, is to write an ansatz $y(r)$.

As an aside, we mention that all of these assumptions may be relaxed, and indeed, more general embeddings contain a lot of interesting physics, about D-branes and about the field theory.

With the ansatz $y(r)$, we can use eq. (\ref{dbranemetric}) to write the induced metric of the D7-brane,
\beq
\label{inducedd7branemetric}
ds_{D7}^2 \, = \, Z(r_6) \, \left( -dt^2 + d\vec{x}^2 \right) + \, Z(r_6)^{-1} \, \left( dr^2 \, (1 + y'(r)^2) \, + \, r^2 \, ds^2_{S^3} \right).
\eeq
The new ingredient is the $y'(r)$ term (primes denote $\frac{\partial}{\partial r}$). Notice that this metric is only $AdS_5 \times S^3$ if $y(r)=0$, the meaning of which we will discuss shortly. Taking minus the determinant of the D7-brane metric and simplifying, the DBI action becomes
\beq
\label{yaction}
S_{D7} \, = \, - N_f \, T_{D7} \, V_{\mathbb{R}^{3,1}} \, V_{S^3} \int dr \, r^3 \, \sqrt{1+y'(r)^2}
\eeq
We have integrated over the ``spectator'' directions, namely the Minkowski and $S^3$ directions, producing a factor of $V_{\mathbb{R}^{3,1}}$, the (divergent) volume of Minkowski space, as well as $V_{S^3} = 2 \pi^2$, the volume of a unit-radius $S^3$. In what follows, we will always divide by $V_{\mathbb{R}^{3,1}}$, and work with action \textit{densities}, unless stated otherwise. We thus define the notation $S \equiv S_{D7} / V_{\mathbb{R}^{3,1}}$ for the action density, which we will henceforth refer to as the action. We will also define a more compact notation for the remaining constants in front,
\beq
\N \, \equiv \, N_f \, T_{D7} \, V_{S^3} \, = \, \frac{\lambda}{(2\pi)^4} \, N_f N_c,
\eeq
not to be confused with the $\N$ of supersymmetry (the difference should always be clear from the context). We have also written above the translation of $\N$ to field theory quantities.

The Lagrangian only depends on $y'(r)$, so the system has a constant of motion, which is given by the equation of motion for $y(r)$,
\beq
\label{yeom}
\partial_r \, \left ( r^3 \, \frac{y'(r)}{\sqrt{1+y'(r)^2}} \right)= 0,
\eeq
where the quantity in parentheses is ``conserved.'' Clearly, setting $y'(r)=0$ is a solution, that is, a \textit{constant} $y(r)$ is a solution. We could also consider a more general solution, by setting the quantity in parentheses equal to a constant, solving for $y'(r)$, and integrating to find $y(r)$. These solutions actually have larger action than the $y'(r)=0$ solutions, as we will show in Chapter $4$, so for now we focus on the $y'(r)=0$ solutions.

We have found that $y(r)=c$, for any constant $c$, describes the embedding of a D7-brane. We would like to interpret this solution, on both the supergravity side and the field theory side of the correspondence.

To understand what $y(r)=c$ means in the supergravity description, we return to the induced D7-brane metric, eq. (\ref{inducedd7branemetric}), setting $y'(r) = 0$ and $y(r) = c$. Notice that for the D7-brane, this means $r_6^2 = r^2 + c^2$, so that the metric becomes
\beq
\label{onshellinducedd7branemetric}
ds_{D7}^2 \, = \, \left( r^2 + c^2 \right) \, \left( -dt^2 + d\vec{x}^2 \right) + \, \frac{1}{r^2 + c^2} \, \left( dr^2 \, + \, r^2 \, ds^2_{S^3} \right).
\eeq
As we decrease $r$, from $r=\infty$ down to $r=0$, we find that $r_6$ decreases from $r_6 = \infty$ to $r_6 = c$, so for nonzero $c$ the D7-brane only reaches some minimum value of $r_6$. As we decrease $r$, from the metric factor in front of $ds_{S^3}^2$, we can also see that the volume of the $S^3$ descreases. Indeed, at $r=0$, the volume of the $S^3$ is zero! Recalling that $r_6$ is the radial coordinate of $AdS_5$, we have found that, for finite $c$, the D7-brane extends from the boundary at $r_6 = \infty$ down into the bulk of $AdS_5$, with the $S^3$ shrinking as it goes, until at $r_6 = c$ the $S^3$ has collapsed completely\footnote{The $S^3$ is a topologically trivial cycle of the $S^5$, which makes this possible.} and the D7-brane simply terminates, extending no further in $r_6$.

\begin{figure}
{\centering
\includegraphics[width=0.50\textwidth]{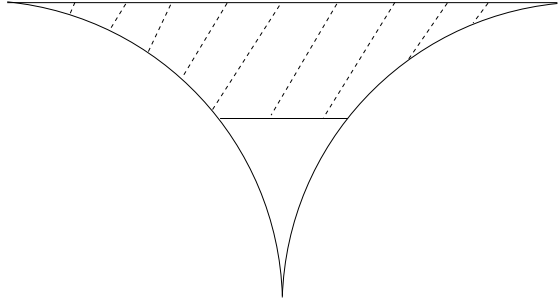}
\caption{\label{d7fig} Cartoon of D7-brane embedded in AdS space.}
}
\end{figure}

The embedding of the D7-brane is depicted (crudely) in figure \ref{d7fig}, which is a side-on view of AdS as depicted in figure \ref{adsfig}. The radial direction $r_6$ runs vertically, while one of the Minkowski spatial directions runs horizontally. The boundary at $r_6 = \infty$ is at the top while the Poincar\'{e} horizon $r_6=0$ is at the bottom. The D7-brane extends throughout the region of dashed lines, but ends at some $r_6 = c$, and does not extend to smaller values of $r_6$.

What is the field theory interpretation of $y(r)=c$? These solutions actually describe massive hypermultiplet fields. To see this, recall that if the D3-branes and D7-branes are separated in the $X_8$-$X_9$ plane (in the asymptotic geometry) then the stretched open strings between them represent massive flavor fields in the D3-brane theory. Recall also that, in the asymptotic geometry, $y$ is precisely the separation in the $X_8$-$X_9$ plane. Given $y(r)$, we find the asymptotic separation by taking by $\lim_{r \ra \infty} y(r)$, which is of course $c$. The mass of the flavor fields is just the string tension times this separation: $m = \frac{c}{2\pi\alpha'}$.

We have thus discovered what massive flavor fields look like holographically: D7-branes that ``end'' somewhere in $AdS_5$. Such a description has a nice interpretation in the field theory. Far enough in the IR, we expect any massive fields to be absent from the dynamics. Indeed, the D7-brane is absent for $r_6 < c$. Conversely, far in the UV we expect a finite mass to be unimportant. As $r \ra \infty$, we can drop the $c^2$ part of $r_6^2 = r^2 + c^2$, so that the induced D7-brane metric becomes pure $AdS_5 \times S^3$. The $AdS_5$ factor indicates that in the UV the conformal symmetry is restored: the finite mass is not apparent.

If $m$ precisely equals zero, then $r_6=r$ and the D7-brane extends all the way to $r_6=0$, so that the D7-brane geometry is exactly $AdS_5 \times S^3$. In the field theory, conformal symmetry is completely restored, and not just approximate in the UV.

\begin{figure}
{\centering
\includegraphics[width=0.50\textwidth]{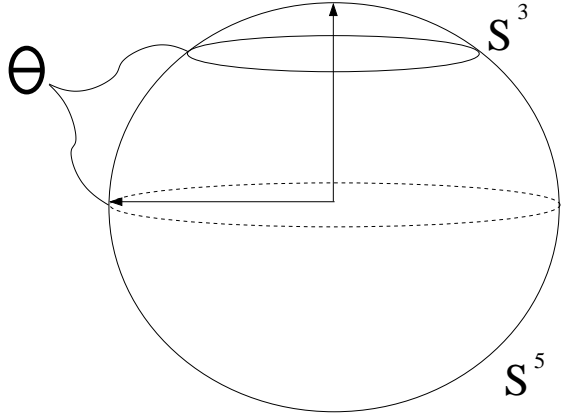}
\caption{\label{s5fig} Cartoon of $S^5$, with D7-brane wrapping $S^3$, and the definition of the angle $\theta$.}
}
\end{figure}

We end this section by re-writing the above action and solutions, making two changes: first, we will use Fefferman-Graham coordinates, and second, we will use a different parameterization of the worldvolume scalar. We rewrite the $AdS_5 \times S^5$ metric as
\beq
ds_{D3}^2 \, = \, \frac{1}{z^2} \, \left( dz^2 - dt^2 + d\vec{x}^2 \right) \, + \, d\theta^2 \, + \, \sin^2 \theta \, ds^2_{S^1} \, + \, \cos^2 \theta \, ds^2_{S^3}
\eeq
These coordinates are related to the previous ones as: $z = 1 / r_6$, $r = r_6 \cos \theta$, and $y = r_6 \sin \theta$. The worldvolume scalar is now $\theta$. Using the same symmetry arguments as before, we consider an ansatz in which $\theta$ depends only on the radial coordinate, $z$. In figure \ref{s5fig} we have depicted the $S^5$ as a sphere, and the $S^3$ that the D7-brane wraps as a circle, with the angle $\theta$ indicated. $\theta$ runs from zero, where the D7-brane wraps the maximum-volume equatorial $S^3$, to $\pi/2$, where the $S^3$ has collapsed to zero volume (at the ``North pole'').

The induced D7-brane metric is now
\beq
\label{feffgrahaminducedd7branemetric}
ds_{D7}^2 \, = \, \left( \frac{1}{z^2} + \theta'(z) \right) dz^2 \, + \, \frac{1}{z^2} \left( - dt^2 + d\vec{x}^2 \right) \, + \, \cos^2 \theta(z) \, ds^2_{S^3}
\eeq
where the prime on $\theta'(z)$ denotes $d/dz$. The D7-brane action becomes
\beq
\label{thetaaction}
S \, = \, - \N \, \int dz \, \frac{1}{z^5} \, \cos^3 \theta(z) \, \sqrt{1 + z^2 \, \theta'(z)^2}
\eeq
The equation of motion for $\theta(z)$ is not particularly illuminating, so we will not reproduce it. Fortunately, however, we already know that $y(r) = c$ is a solution, so we can just translate that solution into these coordinates. Using $y = r_6 \sin \theta$, we have $\theta(z) = \arcsin (c z)$.

We can translate what we know about these solutions into our new notation. The boundary is now at $z=0$, where the $\arcsin$ is zero, hence $\theta(z) = 0$. Looking at the D7-brane metric, we can see that the $S^3$ has maximum volume ($\cos \theta(z) = 1$) there. As we move into the bulk, increasing $z$, we see that $\theta(z)$ increases, until we reach $z = 1 / c$, at which point $\theta(z)$ reaches its maximum, $\theta(z) = \pi/2$, where the volume of the $S^3$ is zero ($\cos \theta(z) = 0$). The asymptotic form of the solution is $\theta(z) = c \, z + O(z^3)$, so in these coordinates we identify the mass from the coefficient of the term linear in $z$.

\section{Holographic Renormalization}
\label{holorg}

\subsection{Regulator and Counterterms}

Recalling the precise statement of the AdS/CFT correspondence, eq. (\ref{partitionfunctions}), we know that eventually we will need to compute the on-shell D7-brane action and take functional derivatives with respect to the leading asymptotic values of the D7-brane worldvolume fields. Let us try to do so na\"{i}vely. Inserting the solution $y(r)=c$ into the action eq. (\ref{yaction}), we find
\beq
S \, = \, - \N \, \int_0^{\infty} dr \, r^3
\eeq
where we have now explicitly written the endpoints of the $r$ integration. The integral is plainly divergent! In this section we will explain how to regulate and renormalize the D7-brane action, using the method of holographic renormalization \cite{Henningson:1998gx,Balasubramanian:1999re,deHaro:2000xn,Skenderis:2000in,Bianchi:2001kw,Skenderis:2002wp}, as first done in ref. \cite{Karch:2005ms}.

We will work in Fefferman-Graham coordinates throughout this section. We will thus use eq. (\ref{thetaaction}) as the form of the D7-brane action. The main reason for doing so is that the scalar $\theta(z)$ explicitly depends on the $AdS_5$ radial coordinate, and hence looks like a scalar field in $AdS_5$, as opposed to $y$, which depends on $r$ rather than $r_6$ (or $z=1/r_6$). 

Holographic renormalization proceeds in two steps (the same steps as in any textbook treatment of renormalization): introduce a regulator, and then introduce counterterms to cancel divergences as the regulator is removed. The procedure is best illustrated by example, so we will describe it as we go.

The action diverges because of the integration over the radial direction: $AdS_5$ has infinite volume, heuristically because ``the boundary is infinitely far away.'' We thus introduce a regulator simply by integrating, not all the way to the boundary at $z=0$, but only to some $z=\e$, with the intention of sending $\e \ra 0$ in the end. We thus write a regulated action,
\beq
\label{regulatedaction}
S_{reg} \, = \, - \N \, \int_{\e}^{\infty} dz \, \frac{1}{z^5} \, \cos^3 \theta(z) \, \sqrt{1 + z^2 \, \theta'(z)^2}
\eeq
We next need to determine the divergences. To do so, we need the asymptotic form of $\theta(z)$ (the form near the boundary). We can use $\theta(z)$'s equation of motion to show that it must have the asymptotic form
\beq
\label{asymptoticsolution}
\theta(z) \, = \, z \left( \, \theta_{(0)} \, + \, \theta_{(2)} \, z^2 \, + \, O(z^4) \, \right).
\eeq
In the $\theta(z) = \arcsin (c z)$ example, we have $\theta_{(0)} = c$ and $\theta_{(2)} = \frac{1}{6} \, c^3$, but in fact the above asymptotic form will be the same in any background geometry that is asymptotically $AdS_5$, as we will discuss later (around eq. (\ref{asymptoticallyadsmetric})). We will therefore leave $\theta_{(0)}$ and $\theta_{(2)}$ as generic asymptotic coefficients, only plugging in $\theta_{(0)} = c$ and $\theta_{(2)} = \frac{1}{6} \, c^3$ in explicit examples.

The equation of motion actually determines all the higher-order coefficients in terms of these two, that is, all the higher order coefficients are just some functions of $\theta_{(0)}$ and $\theta_{(2)}$ that we can compute by demanding that the equation of motion vanish order-by-order in powers of $z$. The equation of motion does not determine $\theta_{(0)}$ or $\theta_{(2)}$, however. We know that the first is (related to) the mass $m$ of the hypermultiplets, which we should be free to vary. What determines $\theta_{(2)}$? The equation of motion is (in this case) just some ordinary differential equation. To solve it, we must specify boundary conditions. One boundary condition we specify is the leading asymptotic value, that is, the value of $\theta_{(0)}$. We then imagine ``integrating in'' the solution into the bulk. Somewhere in the bulk (the IR), we require another boundary condition, such as smoothness of the solution $\theta(z)$. The entire solution is thus specified, and in particular, the value of $\theta_{(2)}$ is fixed. In short, $\theta_{(2)}$ is fixed by $\theta_{(0)}$, but we will not see that in the \textit{asymptotic} expansion of the equation of motion, since the way it is fixed requires information about the boundary conditions in the \textit{bulk} of $AdS_5$. These statements are in fact true generally for fields in $AdS_5$: in the asymptotic expansion, the leading and sub-leading coefficients determine all higher coefficients, and the sub-leading term is fixed by the leading term via some boundary condition in the bulk \cite{Balasubramanian:1998sn,Balasubramanian:1998de}.

In a little while, we will provide a field theory interpretation of these statements. Before we can do so, we need to know how to translate $\theta_{(2)}$ into field theory quantities. We will do that in the next section.

We can now plug the asymptotic solution into the regulated action, Taylor expand in $z$, and integrate to determine the $\e \ra 0$ divergences,
\beq
\label{integratedregulatedaction}
S_{reg} \, = \, -\N \, \int_{\e} dz \left( \, \frac{1}{z^5} \, - \, \theta_{(0)}^2 \, \frac{1}{z^3} \, + \, O(z) \, \right) \, = \, -\N \, \left( \, \frac{1}{4} \frac{1}{\e^4} \, - \frac{1}{2} \, \theta_{(0)}^2 \frac{1}{\e^2} \, + \, O(\e^2) \, \right)
\eeq
Notice that we have not specified the upper endpoint of integration, since that may vary from solution to solution. For example, in the $\theta(z) = \arcsin (c z)$ solution, we should only integrate to $z = 1 / c$.

We now want to introduce some kind of counterterms to cancel the $\e \ra 0$ divergences. These counterterms will be localized on the hypersurface $z=\e$. To understand why, we invoke the UV/IR relation: we know that these IR divergences in supergravity translate into UV divergences in the field theory. In the field theory, the cancellation of UV divergences does not involve IR physics. On the supergravity side, then, we should be able to cancel the divergences just with counterterms\footnote{On the gravity side, the counterterms are often also necessary for variational problems to be well-posed. See for example ref. \cite{Papadimitriou:2005ii}.} ``near the boundary,'' at $z=\e$.

In the field theory we know that the regulator and counterterms will preserve all global symmetries, except for anomalous symmetries. On the supergravity side, we thus expect the counterterms to respect all the gauge invariances of the theory, and in particular, coordinate transformations, or at least coordinate transformations on the $z=\e$ hypersurface.

We define the induced metric on the regulator hypersurface as $\g_{ij}$. Written explicitly for $AdS_5$ in Fefferman-Graham coordinates, $\g_{ij}$ is
\beq
ds^2_{\e} \, = \, \g_{ij} \, dx^i dx^j \, = \, \frac{1}{\e^2} \left( -dt^2 + d\vec{x}^2 \right) 
\eeq
We define $\gamma \equiv \det \g_{ij}$, which for $AdS_5$ is simply $\gamma = - 1/\e^8$. We can now write the counterterms we need to cancel the $\e \ra 0$ divergences of the D7-brane action \cite{Karch:2005ms}:
\beq
L_1 \, = \, + \frac{1}{4} \, \N \, \sqrt{-\gamma} \qquad \, L_2 \, = \, - \frac{1}{2} \, \N \, \sqrt{-\gamma} \, \theta(\e)^2
\eeq
These have an $\e \ra 0$ expansion
\beq
L_1 = + \N \, \frac{1}{4} \, \frac{1}{\e^4}, \qquad L_2 = - \N \, \frac{1}{2} \, \theta_{(0)}^2 \, \frac{1}{\e^2} \, - \, \N \, \theta_{(0)} \, \theta_{(2)} \, + \, O(\e^2)
\eeq
which, by construction, cancel the $\e \ra 0$ divergences in eq. (\ref{integratedregulatedaction}).

Notice that the leading divergence, and the first counterterm, $L_1$, are independent of $\theta(z)$. The leading divergence is just the divergence from the infinite volume of $AdS_5$, and will be present for any solution $\theta(z)$. Also notice that the second counterterm, $L_2$, includes a finite contribution of the form $\theta_{(0)} \, \theta_{(2)}$, which will be important shortly.

We construct the renormalized action by summing the regulated action and counterterms, and then removing the regulator, that is, taking $\e \ra 0$. We will call the sum of the regulated action and counterterms the ``subtracted action,''
\beq
\label{subtractedaction}
S_{sub} \, = \, S_{reg} \, + \, \sum_{i} L_i
\eeq
because we have subtracted the terms that diverge as $\e \ra 0$, and the ``renormalized action'' as the $\e \ra 0$ limit of the subtracted action,
\beq
\label{renormalizedaction}
S_{ren} \, = \, \lim_{\e \ra 0} \,  S_{sub}.
\eeq

Actually, one ambiguity remains: we can introduce a \textit{finite} counterterm:
\beq
L_f \, = \, C \, \N \, \sqrt{-\g} \, \theta(\e)^4 \, = \, C \, \N \, \theta_{(0)}^4 \, + \, O(\e^2)
\eeq
where $C$ is, at the moment, an undetermined coefficient. How can we fix $C$?

We can determine $C$ as follows. We know that the solution $\theta(z) = \arcsin (c z)$ is supersymmetric\footnote{Recall that this solution describes hypermultiplets in the field theory with an $\N=2$ supersymmetry-\textit{preserving} mass $m$.}. Supersymmetry demands that the action evaluated on the solution, \textit{i.e.} the energy of the D7-brane, be zero. We will adjust $C$ to make this so. Plugging $\theta(z) = \arcsin (c z)$ into the action, we find
\bea
\label{arcsinregulatedaction}
\left . S_{reg} \right|_{sol} & = & - \N \, \int_{\e}^{1/c} dz \, \frac{1}{z^5} \, \left( \, 1 \, - \, c^2 \, z^2 \, \right) \nonumber \\ & = & -\N \, \left( \frac{1}{4} \frac{1}{\e^4} \, - \frac{1}{2} \, c^2 \, \frac{1}{\e^2} \, + \, \frac{1}{4} \, c^4 \right)
\eea
Notice that the upper endpoint of the $z$ integration is $1/c$. Recalling that $\theta_{(0)} = c$, we see that the divergences have the expected form shown in eq. (\ref{integratedregulatedaction}). Adding $L_1$ and $L_2$ and taking $\e \ra 0$, we find that the renormalized action is
\beq
\label{arcsinrenormalizedaction}
S_{ren} \, = \, -\N \, \frac{5}{12} \, c^4
\eeq
which is clearly nonzero. To get zero, we fix the coefficient $C$ of the finite counterterm:
\beq
L_f \, = \, + \frac{5}{12} \, \N \, \sqrt{-\g} \, \theta(\e)^4
\eeq
Starting now, the renormalized action will always include $L_f$ as well as $L_1$ and $L_2$.

The method we have outlined is actually valid for a wide class of asymptotically AdS geometries. To be specific, consider a metric of the form
\beq
\label{asymptoticallyadsmetric}
ds^2 \, = \, \frac{dz^2}{z^2} \, + \, \frac{1}{z^2} \, g_{ij}(x^i,z) \, dx^i dx^j.
\eeq
From Einstein's equation, we can show that $g_{ij}(x^i,z)$ has an asymptotic expansion
\beq
g_{ij}(x^i,z) \, = \, g^{(0)}_{ij}(x^i) \, + \, g^{(2)}_{ij}(x^i) \, z^2 \, + \, O(z^4),
\eeq
For the moment, let us assume that $g^{(0)}_{ij}$ is Ricci-flat. In such cases, a straightforward exercise reveals that in fact $g^{(2)}_{ij} = 0$, and the first sub-leading term enters at order $z^4$ \cite{deHaro:2000xn}.

Any such metric has the same near-boundary behavior as pure $AdS_5$, that is, the spatial slices diverge as $1/z^2$ when $z\ra0$, hence anything that depended only on the asymptotic behavior in $z$ will be unchanged. In particular, the asymptotic form of $\theta(z)$, eq. (\ref{asymptoticsolution}), will be unchanged, as will the divergences of the on-shell D7-brane action. Some of the intermediate equations may look different, for example, factors of $\sqrt{-\det g^{(0)}}$ will appear, and $\g_{ij}$ will have a different form, but the counterterms are written covariantly, so their form is unchanged. As long as $g^{(0)}$ is Ricci-flat, $L_1$, $L_2$ and $L_f$ will suffice.

As an example, the AdS-Schwarzschild metric in eq. (\ref{adsbhmetric}) has the form of eq. (\ref{asymptoticallyadsmetric}) with Ricci-flat $g^{(0)}_{ij}$, so the counterterms $L_1$, $L_2$ and $L_f$ will be sufficient. This agrees nicely with our field theory expectations. A general statement in thermal field theory is that introducing a finite temperature does not introduce new UV divergences. In other words, in finite-temperature field theory, the zero-temperature counterterms are sufficient.

If $g^{(0)}_{ij}$ is not Ricci-flat, additional counterterms are required. To be covariant, these counterterms must be built from the Ricci tensor (with indices properly contracted) and Ricci scalar of $\g_{ij}$. Introducing a $g^{(0)}_{ij}$ with nonzero curvature corresponds to studying a field theory on a curved space. Many interesting physical questions arise when we study $\N=4$ SYM on a curved manifold, simply because the manifold introduces a new scale into the theory: the curvature. We will only be interested in field theories in flat space, though, so we will not need the additional counterterms. They are written explicitly in ref. \cite{Karch:2005ms}.

$L_1$ and $L_2$ are actually not unique to the D7-brane. $L_1$ is present merely to renormalize the infinite volume of $AdS_5$, and hence will be present in any action that involves integrating over $AdS_5$, or at least over the region near the boundary. $L_2$ is present for a free scalar in $AdS_5$, with a particular mass. Both of these are not surprising: the D7-brane fills $AdS_5$ and possesses a worldvolume scalar. Indeed, suppose we Taylor expand the action in powers of $\theta(z)$. The result will have the schematic form
\beq
S \, \sim \, \int dz \, \sqrt{-g} \, \left( 1 \, + \, \frac{1}{2} \left( g^{\mu \nu}\partial_\mu \theta \partial_{\nu} \theta \, + \, M^2 \theta^2 \right) \, + \, O(\theta(z)^4) \right)
\eeq
with $g_{\mu \nu}$ the metric of $AdS_5$ and $g$ its determinant. Clearly, the first term is just the volume of $AdS_5$ while the quadratic term is that of a free scalar with mass $M^2$. The remainder looks like an infinite series of interaction terms.

The mass-squared of the worldvolume scalar turns out to be $M^2 = -3$ (in units of the $AdS_5$ radius), which looks worrisome at first, being negative. Fortunately, in AdS space, scalar fields can have a slightly negative $M^2$ and still be stable: due to the curvature of the space, the total energy remains finite as long as $M^2 \geq - 4$. This is called the Breitenlohner-Freedman bound \cite{Breitenlohner:1982jf,Breitenlohner:1982bm}. The worldvolume scalar's mass is clearly in the safe region.

Only $L_f$ is really special to the D7-brane. In fact, we will see in the next section that $L_f$ must be present to extract the correct answer for the one-point function of the operator dual to $\theta(z)$.

\subsection{The Dual Operator and Its One-Point Function}

In this section we identify the field theory operator dual to $\theta(z)$ and use holographic renormalization to compute its expectation value (one-point function) from the renormalized on-shell D7-brane action.

What field theory operator is dual to the D7-brane worldvolume scalar? We know that the leading asymptotic value of the worldvolume scalar is, up to some constants, the hypermultiplet mass $m$. Recalling the precise definition of the AdS/CFT correspondence, eq. (\ref{partitionfunctions}), we know that the leading value of the bulk field acts as a source for the dual operator. The dual operator must therefore be given by taking $\partial / \partial m$ of the field theory Lagrangian, and must therefore include at least the mass operator of the fermions in the hypermultiplet and $m$ times the mass operator for the scalar fields in the hypermultiplet. Notice the operator will thus explicitly depend on $m$. We will denote the operator as $\Om$.

A precise derivation of $\Om$ was performed in ref. \cite{Kobayashi:2006sb}. Due to supersymmetry, $\Om$ actually includes more than just the mass operators of the hypermultiplet fields. It also includes an interaction term with an adjoint field, from the $m \phi^{\dagger} \Phi \phi$ coupling mentioned in section \ref{d3braneworldvolumetheory}. We will just write the operator schematically. Let $\phi_a$ represent the two complex scalars in the hypermultiplet ($a=1,2$) and $\psi$ represent the Dirac fermion in the hypermultiplet, and let $\Phi$ represent one of the complex scalars in the $\N=4$ multiplet. $\Om$ has the form
\beq
\O_m \, = \, \bar{\psi} \psi \, + \, \sum_a \left( m \, \phi_a^{\dagger} \phi_a \, + \, \sqrt{2} \, \phi_a^{\dagger} \Phi \phi_a \right) \, + \, h.c.
\eeq
For most practical purposes, however, we can think of $\Om$ as being just the hypermultiplet mass operator.

We should be able to compute correlation functions of $\Om$ from the on-shell D7-brane action. We will illustrate the procedure by computing the one-point function for $\Om$. According to the precise statement of the correspondence, the expectation value of $\Om$ is given by taking the derivative of the on-shell D7-brane action with respect to the leading asymptotic value of $\theta(z)$:
\beq
\Omv \, = \, \frac{\delta S_{ren}}{\delta m} \, = \, (2\pi\alpha') \, \frac{\delta S_{ren}}{\delta \theta_{(0)}}.
\eeq
After introducing the regulator, $\Omv$ can actually be re-written as
\beq 
\Omv \, = \, (2\pi\alpha') \, \lim_{\e \ra 0} \, \frac{1}{\e^3} \, \frac{1}{\sqrt{-\gamma}} \, \frac{\delta S_{sub} }{\delta \theta(\e)}. 
\eeq
Essentially, all we have done here is introduce some factors of $\e$ so that we can write $\Omv$ as a variational derivative with respect to $\theta(\e)$, which is easier to compute than the derivative with respect to $\theta_{(0)}$. The contribution from the regulated action is
\beq 
\frac{\delta S_{reg}}{\delta \theta(\e)} \, = \, \left . \frac{ \delta L}{\delta \theta'(z)} \right |_{\e} \, = \, +
\N \, \frac{1}{\e^5} \, \cos^{3}\theta(\e) \, \frac{\e^2 \theta'(\e)}{\sqrt{ 1 + \e^2 \theta'(\e)^2}}.
\eeq
where we have defined the Lagrangian via $S_{reg} \, = \, \int_{\e} dz \, L$ and in the first equality we have used the equation of motion,
\beq
\frac{\delta L}{\delta \theta(z)} \, = \, \frac{d}{dz} \, \frac{ \delta L}{\delta \theta'(z)}
\eeq
and then performed the $z$ integration. Using the asymptotic form of $\theta(z)$, we find
\beq
\frac{1}{\e^3} \, \frac{1}{\sqrt{-\gamma}} \, \frac{\delta S_{reg}}{\delta \theta(\e)} \, = \, \N \, \frac{\theta_{(0)}}{\e^2} \, + \, \N \, \left( 3 \, \theta_{(2)} \, - \, 2 \, \theta_{(0)}^3 \right) \, + \, O(\e^2)
\eeq

The contribution from the counterterms is simple. $L_1$ is independent of $\theta(z)$ and hence contributes nothing. For the remaining counterterms we find
\beq
\frac{\delta L_2}{\delta \theta(\e)} \, = \, - \N \, \sqrt{-\g} \, \theta(\e), \qquad \frac{\delta L_f}{\delta \theta(\e)} \, = \, + \frac{5}{3} \, \N \, \sqrt{-\g} \, \theta(\e)^3
\eeq
so that these contribute
\begin{subequations}
\beq
\frac{1}{\e^3} \, \frac{1}{\sqrt{-\gamma}} \, \frac{\delta L_2}{\delta \theta(\e)} \, = \, - \N \, \frac{\theta_{(0)}}{\e^2} \, -\N \, \theta_{(2)} \, + O(\e)^2
\eeq
\beq
\frac{1}{\e^3} \, \frac{1}{\sqrt{-\gamma}} \, \frac{\delta L_f}{\delta \theta(\e)} \, = \, +\frac{5}{3} \, \N \, \theta_{(0)}^3 \, + \, O(\e^2)
\eeq
\end{subequations}
Summing everything and taking $\e \ra 0$, we find
\beq
\label{omvformula}
\Omv \, = \, \N (2\pi\alpha') \, \left( 2 \, \theta_{(2)} \, - \, \frac{1}{3} \, \theta_{(0)}^3 \right)
\eeq
Given a solution $\theta(z)$, we can compute the associated value of $\Omv$ by extracting the asymptotic coefficients $\theta_{(0)}$ and $\theta_{(2)}$, and combining them as indicated in eq. (\ref{omvformula}). We now also understand the meaning of $\theta_{(2)}$ in the field theory: its value determines $\Omv$.

In fact, this is a generic feature in gauge-gravity duality: given some solution of some field, in the asymptotic expansion the coefficient of the leading term acts as a source for the dual operator while the coefficient of the sub-leading term determines the expectation value of the dual operator \cite{Balasubramanian:1998sn,Balasubramanian:1998de}. The leading term, being the lower power in $z$, will be non-normalizable, meaning that if we insert just that term into the action and attempt to integrate all the way to the boundary, we find a divergence. The sub-leading term will be normalizable, meaning that in the same process the result will be finite.

We can now return to a point we mentioned above. From the supergravity point of view, we have a field with some equation of motion, in this case $\theta(z)$, whose equation of motion is just an ordinary differential equation in $z$. We specify a boundary condition at the AdS boundary, in our case the value of $\theta_{(0)}$, and integrate the solution into the bulk, where we impose some other condition, for example regularity of the solution. The rest of the solution is then determined. In particular, our $\theta_{(2)}$ will be determined.

We are now ready to interpret these statements in the field theory. Choosing the value of $\theta_{(0)}$ means choosing $m$, a parameter in the Lagrangian. From a field theory point of view, we then expect the dynamics of the theory, including IR physics, to determine $\Omv$. That is what we see happening in the supergravity description when we see the boundary condition in the bulk (the IR) fixing the value of $\theta_{(2)}$.

We end with an example: what value of $\Omv$ does the solution $\theta(z) = \arcsin (c z)$ produce? Recalling that 
$\theta_{(0)} = c$ and $\theta_{(2)} = \frac{1}{6} \, c^3$, we find precisely $\Omv = 0$. Indeed, supersymmetry demands that $\Omv = 0$ \cite{Babington:2003vm}. A fermion bilinear, such as the mass operator of the hypermultiplet fermions, may be written as the supersymmetric variation of another operator, which must vanish. Notice that the finite counterterm $L_f$, which was required by supersymmetry, was essential to get $\Omv = 0$.

If we break supersymmetry, however, we may find a nonzero value for $\Omv$. We can break supersymmetry by introducing a finite temperature, density, or electric and magnetic fields. The effect of doing so is the subject of the next two chapters, although most of our attention will be on the on-shell value of the D7-brane action, rather than the value of $\Omv$.

%
%
%
%
%
%
%
%
%
%
%
%
%
%
%
\chapter {Thermodynamics of Flavor Fields}\label{thermo}

We are now ready to study thermodynamics of flavor fields using the holographic description. We will add one ingredient at a time: temperature, chemical potential, electric fields and magnetic fields. The emphasis will be on the physics and the qualitative form of the phase diagram. We will only present explicit results in two cases, both at finite chemical potential: finite temperature and zero mass ($T>0$,$\, m=0$), and zero temperature and finite mass ($T=0$,$\, m>0$). In each case, we will present exact solutions for D7-brane worldvolume fields, and using these and the on-shell D7-brane action, draw conclusions about the thermodynamics of the field theory. In the former case, we will find no phase transitions, while in the latter case we will find a second-order transition. We will describe how these results fit into the ``big picture'' of the full phase diagram as we go.

\section{Finite Temperature}

\subsection{Thermal Phase Transition}
\label{thermaltransition}

In $\N=4$ SYM, in the 't Hooft limit and with large 't Hooft coupling, a state in thermodynamic equilibrium at temperature $T$ is dual holographically to supergravity on the (4+1)-dimensional AdS-Schwarzschild geometry times $S^5$. The temperature and entropy of the field theory are identified with the temperature and entropy of the AdS-Schwarzschild black hole. The AdS-Schwarzschild metric is written explicitly in eqs. (\ref{originaladsbhmetric}) and (\ref{adsbhmetric}). We will work in Euclidean signature, with compact time direction, until stated otherwise.

We introduce flavor fields as before, with a holographic description as probe D7-branes. The on-shell action of the D7-branes now corresponds (up to a factor of $T$) to their contribution to the free energy, which will be order $\lambda \, N_f N_c$ because their action is order $\lambda \, N_f N_c$.

In the AdS-Schwarzschild background, two types of D7-brane embedding are possible \cite{Babington:2003vm,Kirsch:2004km,Ghoroku:2005tf,Mateos:2006nu}. As before, we will have D7-brane embeddings in which the D7-brane ends at some radial position, where the $S^3$ collapses to a zero volume, outside of the black hole horizon. Starting now, we will follow convention and call these ``Minkowski embeddings.'' The presence of the horizon admits a second class of embeddings, however, the so-called ``black hole embeddings.'' These are D7-branes for which the $S^3$ shrinks but never fully collapses as the D7-brane extends into the bulk: the D7-brane exends all the way to, and intersects, the black hole horizon. Intuitively, we can think of the D7-brane as extending into, and ending somewhere inside, the black hole. The black hole interior is not part of the geometry, however, so we should not take this too seriously. An observer in AdS-Schwarzschild will simply see the D7-brane intersect the horizon.

The two classes of embeddings have distinct topology. For a Minkowski embedding, the $S^3 \subset S^5$ collapses to zero volume, while for a black hole embedding it does not. Instead, for a black hole embedding, at the horizon the $g_{tt}$ component of the metric vanishes, indicating that the Euclidean time circle collapses to zero size there.

Invoking the UV-IR relation and recalling that the radial direction acts as an energy scale in the field theory, the position of the horizon corresponds to the scale $T$, while the position where the D7-brane ends corresponds to the scale $m$. Roughly speaking, the Minkowski embeddings correspond to large $m$, compared to $T$, while the black hole embeddings correspond to small $m$. For example, the zero-mass solution, $y(r)=0$, will obviously intersect the horizon: the $S^3$ sits at the equator of the $S^5$ for all values of $r$.

In the field theory, we now have two scales, $T$ and $m$, so we have the possibility of a phase transition. Suppose we fix $m$ to be some large value and then change the temperature. We must measure $T$ in units of $m$, so we imagine increasing $T/m$ and asking whether a phase transition occurs.

The supergravity analysis reveals that, indeed, a phase transition does occur \cite{Babington:2003vm,Kirsch:2004km,Ghoroku:2005tf,Mateos:2006nu,Albash:2006ew,Karch:2006bv,Mateos:2007vn}. In the holographic picture, we are starting with a Minkowski D7-brane and increasing $T$, so the horizon begins moving up towards the D7-brane. Eventually, the horizon will reach the D7-brane's endpoint. Following convention, we will call this the ``critical embedding'' of the D7-brane. As we continue increasing $T$, the D7-brane will intersect the horizon, becoming a black hole embedding. The topology of the D7-brane changes in this process, so we expect some kind of discontinuous transition.

By computing the value of the on-shell D7-brane action for these embeddings, a transition has been discovered: the minimum of the D7-brane's action does jump from a Minkowski embedding to a black hole embedding. In fact, this occurs before the horizon reaches the Minkowski D7-brane's endpoint. In other words, the D7-brane ``jumps into'' the black hole \textit{before} reaching the critical embedding.

\begin{figure}
{\centering
$\begin{array}{cc}
\includegraphics[width=0.45\textwidth]{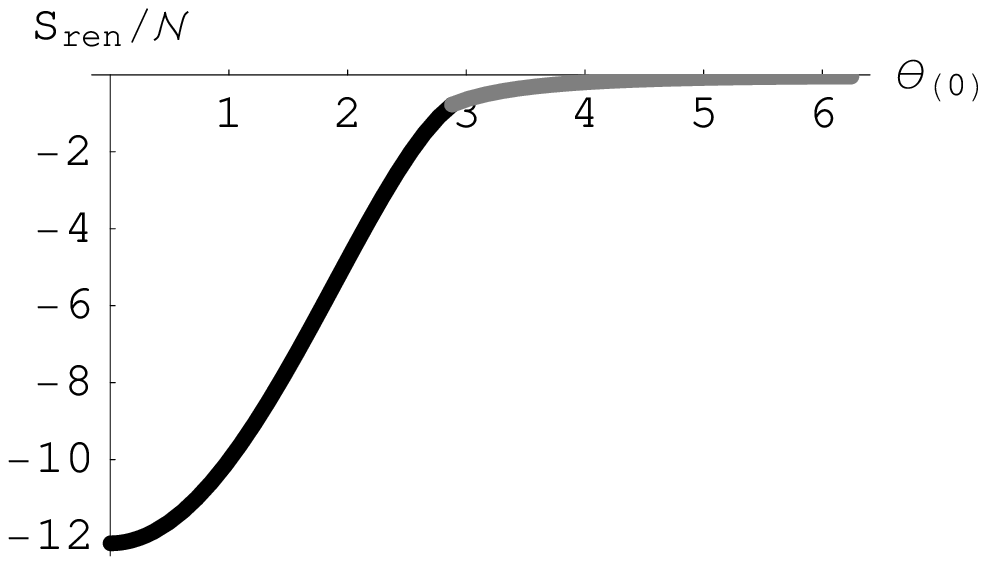} & \includegraphics[width=0.45\textwidth]{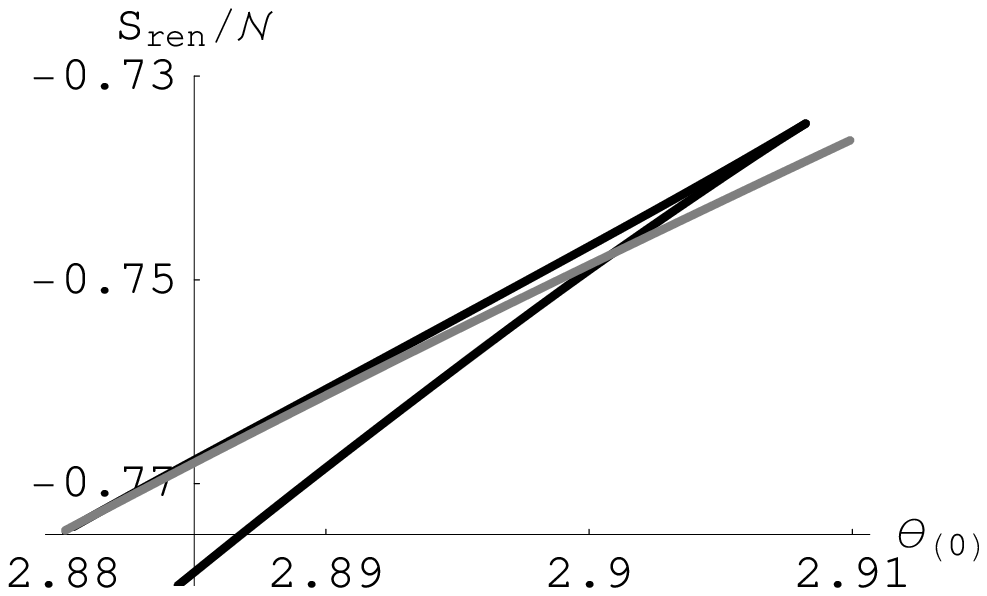} \\ (a.) & (b.)
\end{array}$
\caption{\label{d7actionfig} Value of $S_{ren}/\N$ versus $\theta_{(0)}$ for $T=1$, in units of the 
$AdS_5$ radius.}
}
\end{figure}

To show this explicitly, we show in figure \ref{d7actionfig} the results of the numerical supergravity calculation. To perform the calculation, we use Fefferman-Graham coordinates and the worldvolume scalar field $\theta(z)$, with the counterterms for the action written in section \ref{holorg}. In the plots, the vertical axis is the value of the renormalized on-shell D7-brane action, $S_{ren}$, divided by the normalization factor $\N$. On the horizontal axis is the leading asymptotic coefficient, $\theta_{(0)}$, in units of the $AdS_5$ radius. The black curves come from black hole embeddings while the gray curves come from Minkowski embeddings. The critical embedding is thus where the black and gray curves meet.

Recall that the hypermultiplet mass is given by $m = \frac{\theta_{(0)}}{2\pi \alpha'}$, and conversely $\theta_{(0)} = 2 \pi \sqrt{\lambda} \, m$, again in units of the $AdS_5$ radius. Our calculations are done with $T=1$ in units of the $AdS_5$ radius. We may thus write $\theta_{(0)} = 2 \pi \sqrt{\lambda} \, m/T$. 

In figure \ref{d7actionfig} (a.) we show the large-scale structure. Notice in particular that as $\theta_{(0)}$ grows, the action approaches zero. This conforms to our field theory expectation that as $m \ra \infty$, the flavor fields' contribution to the free energy should vanish, as they decouple from the dynamics. Notice that, on these scales, the action appears to be single-valued.

In figure \ref{d7actionfig} (b.) we show a close-up near the critical embedding, which reveals that the action is not single-valued. The ground state will be the state with the smallest value of the on-shell action (in field theory language: the lowest free energy). Suppose we fix $m$ and increase $T$, which appears in this plot as moving from the right to the left, since $\theta_{(0)} \sim m/T$. As we enter figure \ref{d7actionfig} (b.) from the right, the smallest values are on the gray curve, \textit{i.e.} the Minkowski embeddings. We can see that two other branches of solutions also appear, corresponding to black hole embeddings, with larger action. Once we reach the critical solution at $\theta_{(0)} \approx 2.90$, however, the minimum values of the action are now on the black curve, \textit{i.e.} the black hole embeddings. As we continue moving to the left, the minimum action solutions remain the black hole embeddings. The system never accesses the critical embedding, at $\theta_{(0)} \approx 2.88$. The action/free energy is thus continuous, but has a discontinuous first derivative, a ``kink,'' at the phase transition point, $\theta_{(0)} \approx 2.90$. 

Translating to field theory language, we conclude that the flavor fields experience a first-order phase transition at $T \approx 2.90 \, / \,2 \pi \sqrt{\lambda} \, m$. We then expect any observable associated with the flavor fields to exhibit a discontinuous change at that point. A simple observable to compute on the supergravity side is $\Omv$, which is the first derivative of the free energy with respect to $m$, and so of course displays a discontinuity at the transition point \cite{Babington:2003vm,Kirsch:2004km,Ghoroku:2005tf,Mateos:2006nu,Albash:2006ew,Karch:2006bv,Mateos:2007vn}.

A more detailed analysis of near-critical D7-brane solutions revealed that, in fact, the solutions exhibit a self-similar structure. Such self-similarity of the supergravity solutions naturally suggests that the phase transition in the field theory will be first order \cite{Mateos:2006nu,Mateos:2007vn}.

\subsection{Meson ``Melting''}

The simplest way to characterize the phase transition is in terms of the meson spectrum in the field theory. We have not yet described the meson spectrum in the field theory, or how to compute it from supergravity, so let us briefly review the ``qualitative universal'' question, ``How do we compute a meson spectrum holographically?'' We begin at zero temperature.

$\N=4$ SYM plus $\N=2$ hypermultiplets, in the 't Hooft limit, with large 't Hooft coupling and $N_f \ll N_c$, is not a confining theory\footnote{The potential between quarks (and/or squarks) is known in some limits, see for example ref. \cite{Kruczenski:2003be}.}, and will not have mesons in the same sense as QCD, \textit{i.e.} a quark and anti-quark connected by a flux tube. The theory may have quark-anti-quark bound states, however. At zero temperature, we know from dimensional analysis that the mass of any such meson must be proportional to $m$.

Holographically, a meson is represented by an open string with both ends attached to the D7-brane: one endpoint represents the quark and the other endpoint represents the anti-quark. In flat space (not $AdS_5 \times S^5$), we know that such strings will have massless excitations, in which the string collapses to zero length, and massive excitations, with masses on the order of $\ell_s^{-1}$. The former give rise to the massless worldvolume fields of the D7-brane, which appear in the DBI action. Now embedding the D7-brane and open strings in $AdS_5 \times S^5$, in the limit where the string length is much smaller than the $AdS_5$ radius, we expect the masses of all states to acquire a correction on the order of the inverse $AdS_5$ radius. The massless excitations thus acquire masses of order the inverse $AdS_5$ radius. The spectrum of excited string states will be largely unchanged, but will be far heavier, by at least a factor of $\ell_s^{-1}$, than the lightest states. In the field theory, we want to know the lightest degrees of freedom, as these dominate the low-energy physics, so we will focus only on the lightest string excitations.

The lightest states excite the D7-brane worldvolume fields, which we know how to describe using the DBI action. Our question thus becomes: what is the spectrum of excitations of the D7-brane worldvolume fields? In practical terms, we first pick a particular D7-brane embedding and then compute the action for fluctuations about that solution. As we only want the spectrum, we only need the action for such fluctuations to quadratic order. We will then need to solve the linearized equation of motion of the fluctuations, which in general is a complicated partial differential equation.

Remarkably, this calculation has been done \textit{analytically} for the probe D7-brane in $AdS_5 \times S^5$, in ref. \cite{Kruczenski:2003be}. In other words, at zero temperature, \textit{exact} solutions for the fluctuations of the D7-brane fields have been computed, and hence the exact meson spectrum is known. We will skip the details. The three salient features are: the spectrum is \textit{gapped}, \textit{discrete}, and the meson masses scale as $m/\sqrt{\lambda}$. The 't Hooft coupling, $\lambda$, is very big, so these mesons are actually much lighter than the fundamental quarks and squarks. Put another way, these mesons are very deeply bound.

That the spectrum is discrete is easy to understand just from basic facts about differential equations. If $m$ is nonzero, then the D7-brane ends at some value of the radial coordinate. A fluctuation will then have either Neumann or Dirichlet boundary conditions at the endpoint. If we imagine the fluctuation as a wave traveling down the D7-brane, then at the endpoint the wave will be reflected, giving rise to a discrete set of eigenfrequencies.

A special case is zero mass. With no endpoint to ``bounce off of,'' the fluctuation will simply continue traveling all the way to the Poincar\'{e} horizon. The spectrum of excitations, and hence the spectrum of mesons in the field theory, will be \textit{gapless} and \textit{continuous}.

Now we turn to finite temperature, where we have two classes of embeddings. For Minkowski embeddings, the spectrum resembles the zero-temperature spectrum: it is gapped, discrete, and the masses are proportional to $m/\sqrt{\lambda}$ \cite{Hoyos:2006gb}. Again, we can understand this just from the boundary conditions on fluctuations at the D7-brane's endpoint.

The story for black hole embeddings, however, resembles the story from zero temperature and zero mass. Again, picture a wave traveling down the D7-brane. Such a wave can be absorbed, at least partially, into the black hole. In other words, we must impose a boundary condition that any such wave is purely outgoing (from the bulk, into the black hole). The spectrum will thus be not only gapless and continuous, but also contain eigenfrequencies with nonzero \textit{imaginary} part \cite{Hoyos:2006gb}.

Translating to the field theory language: at high enough temperature, the flavor fields undergo a first-order transition, and in the high-temperature phase the mesons acquire some \textit{width}, which in fact increases as the temperature rises, suggesting that the mesons ``melt'' into the plasma \cite{Hoyos:2006gb}. (A broadening of quasi-particle peaks is also visible in spectral functions \cite{Myers:2007we}.) The first-order thermal phase transition thus leaves its most dramatic imprint on the meson spectrum. To emphasize this, starting now we will refer to the transition as a ``meson melting'' transition \cite{Hoyos:2006gb}.

In the following sections, we will introduce additional thermodynamic variables, namely a density and electric or magnetic fields. We will characterize phases of the field theory by their meson spectrum. If the phase is described holographically by a Minkowski embedding, we will refer to the phase as a phase with stable mesons. If the phase is described by a black hole embedding, we will refer to is as a phase with melted mesons.

\section{Finite Density}

\subsection{Only Black Hole Embeddings are Allowed}
\label{bhembeddingsallowed}

As mentioned in section \ref{d3braneworldvolumetheory}, the $N_f$ flavors of hypermultiplet possess a global $U(1)$ symmetry that we may identify as baryon number, and which we will denote as $U(1)_B$. We will thus have a conserved current, $J^{\mu}$, for baryon number. Again, let $\psi$ represent the Dirac fermion in the hypermultiplet and $\phi_a$ the two complex scalars. $J^{\mu}$ then has the schematic form
\beq
J^{\mu} \, = \, i \bar{\psi} \psi \, + \, \sum_{a=1}^2 i \left( \phi^{\dagger}_a D^{\mu} \phi_a \, - (D^{\mu} \phi_a)^{\dagger} \phi_a  \, \right) \, + \, h.c.
\eeq
with $D^{\mu}$ the gauge-covariant derivative. We want to study the field theory with a finite baryon number density. In other words, we want to study the theory in a state with $\langle J^t \rangle \neq 0$. Here we arrive at another question of the ``universal qualitative'' variety: what does a finite baryon number density look like holographically?

To answer this question, we need only recall one of the basic facts about the AdS/CFT correspondence from section \ref{statement}: a global symmetry in the field theory will be dual to a gauge invariance of the gravity theory. In this case, the global symmetry is the $U(1)_B$ associated with the flavor fields, which must be dual to a $U(1)$ gauge invariance associated with the D7-branes, which must be the D7-brane worldvolume gauge field. More explicitly, the field theory operator $J^{\mu}$ will be dual to the D7-brane's gauge field $A_{\mu}$, hence $J^t$ will be dual to $A_t$, so if we want a state with nonzero $\langle J^t \rangle$, then we need a nontrivial $A_t$.

$A_t$ can in principle depend on all of the D7-brane worldvolume coordinates. What ansatz for $A_t$ should we write? As we did for the worldvolume scalar, we make symmetry arguments. We of course want to preserve the rotation and time translation invariance of the field theory, so $A_t$ should not depend on the field theory directions. For simplicity, we do not want $A_t$ to depend on the $S^3$ directions. The remaining option is for $A_t$ to depend on the AdS radial direction. We thus want $A_t(r)$ (here we are using the radial coordinate of eq. \ref{adsmetricrcoordinate}). We will always work in a gauge where the component of $A_{\mu}$ in the radial direction vanishes: $A_r = 0$. We then have an electric field, $F_{rt} = \partial_r A_t$, on the D7-brane, pointing in the radial direction.

Suppose we are working at finite temperature, so that we have both Minkowski and black hole embeddings. We can see that a problem will arise if we ``turn on'' $A_t(r)$ for a Minkowski embedding. We have electric field lines pointing in the radial direction. What happens at the D7-brane's endpoint? The field lines cannot simply terminate: they must end on something. To turn the question around, we must ask what \textit{source} is producing these field lines.

The most natural source is a finite density of strings\footnote{Another possible source is a baryon, which in AdS/CFT appears as a D5-brane wrapping the $S^5$ factor of $AdS_5 \times S^5$. Solutions for the D5/D7 bound state are not yet known \cite{Kruczenski:2003be}, however, so we will not consider this possibility.}. Recall again that a single string endpoint holographically represents a quark in the field theory. We want to study a finite density of quarks (and squarks), which means that in the D7-brane picture we want to study a finite density of strings, each with one endpoint on the D7-brane and the other endpoint piercing the horizon. The string endpoints on the D7-brane of course act as a source for the gauge field on the D7-brane, producing $A_t(r)$.

A straightforward analysis reveals, however, that the force such a density of strings exerts on the D7-brane is greater than the tension of the D7-brane \cite{Kobayashi:2006sb}. We thus expect the strings to pull the D7-brane into the horizon, so that the electric field lines then end safely on the horizon. Actually showing that this occurs is difficult, however. Doing so requires computing a time-dependent solution of the D7-brane's evolution. What can be shown explicitly is that, with nonzero $A_t(r)$, black hole embeddings are sufficient to cover the entire range of masses, from $m=0$ to $m = \infty$ \cite{Kobayashi:2006sb}. Our intuitive, physical picture thus seems to be consistent, so we conclude that when we introduce $A_t(r)$, we need only include black hole embeddings in our analysis.

An interesting question, though, is what large-mass solutions look like. A large-mass Minkowski embedding is a D7-brane that ends near the boundary. When we turn on $A_t(r)$, however, we are claiming that the D7-brane must somehow reach all the way to the horizon. How can this be? An analysis of the D7-brane action reveals that in such a limit, the D7-brane develops a ``spike,'' in which the $S^3$ \textit{almost} collapses to zero volume somewhere near the boundary, but then remains finite-volume all the way to the horizon \cite{Kobayashi:2006sb}. In fact, the action for the spike, as derived from the D7-brane action, is identical in form to the action of a bundle of strings (that is, a number $\< J^t \>$ of strings) \cite{Kobayashi:2006sb}.

Crucially, black hole embeddings force upon us a particular boundary condition for $A_t(r)$. At the horizon, the Euclidean time circle collapses to zero size. For $A_{\mu}$ to be well-defined as a one-form there\footnote{The precise statement is: the event horizon of the geometry is a Killing horizon which contains a bifurcation surface where the Killing vector $\partial_t$ vanishes. For the gauge field to be regular as a one-form, we must have $A_t(r_h)=0$ \cite{Kobayashi:2006sb}.}, $A_t$ must vanish: $A_t(r_h) = 0$. Notice that this boundary condition, forced upon us by the geometry of the space, restricts gauge transformations, as only the subset of gauge transformations that preserve $A_t(r_h)=0$ are allowed.

\subsection{D7-brane Action and Gauge Field Solution}

We are now ready to write the D7-brane action, including the worldvolume electric field $F_{r_6 t} = \partial_{r_6} A_t(r_6) = A_t'(r_6)$, for which we will solve. We will use the bulk metric
\beq
ds^2 \, = \, \frac{dr_6^2}{h(r_6)} - h(r_6)\,dt^2 + r_6^2 \, d\vec{x}^2 + d\theta^2 + \sin^2\theta \, ds_{S^1}^2 + \cos^2 \theta \, ds_{S^3}^2
\eeq
where $h(r_6) = r_6^2 - \frac{r_h^4}{r_6^2}$ and $r_h$ is the black hole horizon, related to the temperature by $r_h = \pi T$. Notice also that we have chosen Lorentzian signature.

We will describe the D7-brane embedding by $\theta(r_6)$. Introducing the worldvolume gauge field, the D7-brane action becomes
\beq
\label{finiteTdbi}
S \, = \, -\N \, \int \, dr_6 \, r_6^3 \, \cos^3 \theta \, \sqrt{1 + h(r_6) \, \theta'^2 - (2\p\a')^2 \, A_t'^2}
\eeq
The action only depends on the derivative of $A_t$, so we have a constant of motion, which for the moment we will call $D$,
\beq
D \, \equiv \, \N \, r_6^3 \, \cos^3 \theta \, \frac{(2\p\a')^2 \, A_t'}{\sqrt{1 + h(r_6) \, \theta'^2 - (2\p\a')^2 \, A_t'^2}}
\eeq
We can now solve for $A_t'(r_6)$,
\beq
\label{finiteTgaugesol}
A_t'(r_6) \, = \, \frac{D}{(2\p\a')^2} \, \sqrt{\frac{1 + h(r_6) \, \theta'^2}{\N^2 \, r_6^6 \, \cos^6 \theta + \frac{D^2}{(2\p\a')^2}}}.
\eeq
Inserting this solution back into the action, we find
\beq
\label{finiteTonshellaction}
S \, = \, - \N^2 \, \int dr_6 \, r_6^6 \, \cos^6 \theta \, \sqrt{\frac{1 + h(r_6) \, \theta'^2}{\N^2 \, r_6^6 \, \cos^6 \theta + \frac{D^2}{(2\p\a')^2}}}
\eeq
We may obtain the equation of motion for $\theta(r_6)$ in two ways. We may either derive it from eq. (\ref{finiteTdbi}) and then plug in the gauge field eq. (\ref{finiteTgaugesol}), or we may eliminate the gauge field at the level of the action via a Legendre transform and then derive $\theta(r_6)$'s equation of motion. The Legendre-transformed action, which we denote $\hat{S}$, is
\beq
\label{finiteTlteffaction}
\hat{S} \, = \, S \, - \, \int dr_6 \, F_{r_6 t} \, \frac{\delta S}{\delta F_{r_6 t}} \, = \, - \int dr_6 \, \sqrt{1 + h(r_6) \, \theta'^2} \sqrt{\N^2 \, r_6^6 \, \cos^6 \theta + \frac{D^2}{(2\p\a')^2}}.
\eeq

We would like to translate into field theory quantities. Recalling that $z=1/r_6$, we know from the analysis in section \ref{holorg} that $\theta(r_6)$ will have an asymptotic expansion with the leading, non-normalizable term going as $1/r_6$ and the subleading, normalizable term going as $1/r_6^3$. The coefficient of the leading term will be, up to a factor of $(2\pi\alpha')$, the hypermultiplet mass $m$, and the coefficient of the sub-leading term will determine $\Omv$. In particular, these statements do not change in the AdS-Schwarzschild geometry, or in the presence of $A_t(r_6)$, as can be confirmed explicitly from $\theta(r_6)$'s equation of motion.

We also know that we should interpret the on-shell action, eq. (\ref{finiteTonshellaction}), up to a Wick rotation to Euclidean signature and a factor of temperature, as the Gibbs free energy (grand canonical ensemble) and the Legendre transform, eq. (\ref{finiteTlteffaction}) as the Helmholtz free energy (canonical ensemble) \cite{Chamblin:1999tk,Chamblin:1999hg}.

How do we extract $\langle J^t \rangle$ and the chemical potential from the supergravity description? $A_t(r_6)$ is dual to $J^t$. The coefficient of the leading, non-normalizable term in $A_t(r_6)$'s asymptotic expansion should give the source for $J^t$, which is just the baryon number chemical potential, $\mu$. The coefficient of the sub-leading, normalizable term should determine the expectation value $\langle J^t \rangle$. Let us see how this works. From eq. (\ref{finiteTgaugesol}) we find $A_t'(r_6)$'s asymptotic expansion,
\beq
A_t'(r_6) \, = \, \frac{D}{\N (2\pi\alpha')^2} \, \frac{1}{r_6^3} \, + \, O\left(\frac{1}{r_6^5}\right)
\eeq
Integrating in $r_6$, we find
\beq
A_t(r_6) \, = \, \mu \, - \, \frac{1}{2} \, \frac{D}{\N (2\pi\alpha')^2} \, \frac{1}{r_6^2} \, + \, O\left(\frac{1}{r_6^4}\right)
\eeq
The leading behavior of $A_t(r_6)$ is a constant, which we identify as the chemical potential. Notice that, normally, such a constant would not be physical: we could make a gauge transformation to remove it. In our case, however, the boundary condition $A_t(r_h)=0$ forbids us from making such gauge transformations, so the constant is physical. In fact, we can write $\mu$ as the integral of a gauge-invariant quantity, the field strength:
\beq
\mu \, = \, A_t(\infty) \, = \, \, \int_{r_h}^{\infty} dr_6 \, A_t'(r_6)\, = \, \int^{r_h}_{\infty} dr_6 \, F_{t r_6}
\eeq
where the second equality relies on the fact $A_t(r_h)=0$. We can thus interpret $\mu$ as the work done to push a string endpoint against the radial electric field, $F_{t r_6}$, from the boundary (the UV) to the horizon (the IR). This is a nice holographic version of a field theory statement: the chemical potential is the energy needed to add a charge to the system.

The coefficient of the sub-leading term should determine the expectation value of the dual operator, $\langle J^t \rangle$. The quick way to determine the exact relation is to invoke holographic renormalization. (A longer, but more physical, argument appears in ref. \cite{Kobayashi:2006sb}.) We introduce a regulator by cutting off $r_6$ at some large value $\Lambda$. We can then write the expectation value for any current component $J^{\mu}$,
\beq
\langle J^{\m} \rangle \, = \, \lim_{\Lambda \ra \infty} \, \frac{\Lambda^4}{\sqrt{-\g}} \, \frac{\d S}{\d A_{\m}(\Lambda)}
\eeq
where $\g$ is the determinant of the induced metric on the $r_6 = \Lambda$ slice. Notice that $\sqrt{-\g} \sim \Lambda^4$. For $J^t$, we want the variation with respect to $A_t(r_6)$, with all other fields held fixed. Defining the Lagrangian via $S = \int dr_6 L$, we have
\beq
\d S \, = \, \int_{r_h}^{\Lambda} dr_6 \, \frac{\d L}{\d \partial_{r_6} A_t} \, \partial_{r_6} \d A_t \, = \, D \, \int_{r_h}^{\Lambda} dr_6 \, \partial_{r_6} \d A_t \, = \, D \left ( \d A_t(\Lambda) - \d A_t(r_h) \right ),
\eeq
Enforcing $\d A_t(r_h)=0$ we find $\frac{\d S}{\d A_t(\Lambda)}= D$ and hence $\langle J^t \rangle = D$.

\subsection{Finite Density Phase Diagram}
\label{finitedensityphasediagram}

Using the above holographic setup, or more precisely the Euclidean-signature version of the above setup, much of the phase diagram of $\N=4$ SYM coupled to massive $\N=2$ hypermultiplets at finite baryon density has been mapped out, in both the canonical and grand canonical ensembles \cite{Kobayashi:2006sb,Ghoroku:2007re,Karch:2007br,Mateos:2007vc}. In this section we will just present a qualitative sketch of the phase diagram, in its current form.

We begin in the canonical ensemble, in which the density is fixed, \textit{i.e.} is our thermodynamic control parameter, and hence an \textit{input}, and the chemical potential is an \textit{output}. By continuity we expect the transition observed at zero density (discussed in section \ref{thermaltransition}) to persist to finite density, at least for small density. In other words, in our phase diagram, which is now two-dimensional, with axes $\langle J^t \rangle/m^3$ and $T/m$, we expect a line of phase transitions to emerge from the transition point on the $\langle J^t \rangle /m^3 = 0$ axis.

\begin{figure}
{\centering
$\begin{array}{cc}
\< J^t \> /m^3 \, \, \, \, vs. \, \, \, \, T/m & \mu/m \, \, \, \, vs. \, \, \, \, T/m \\
\includegraphics[width=0.45\textwidth]{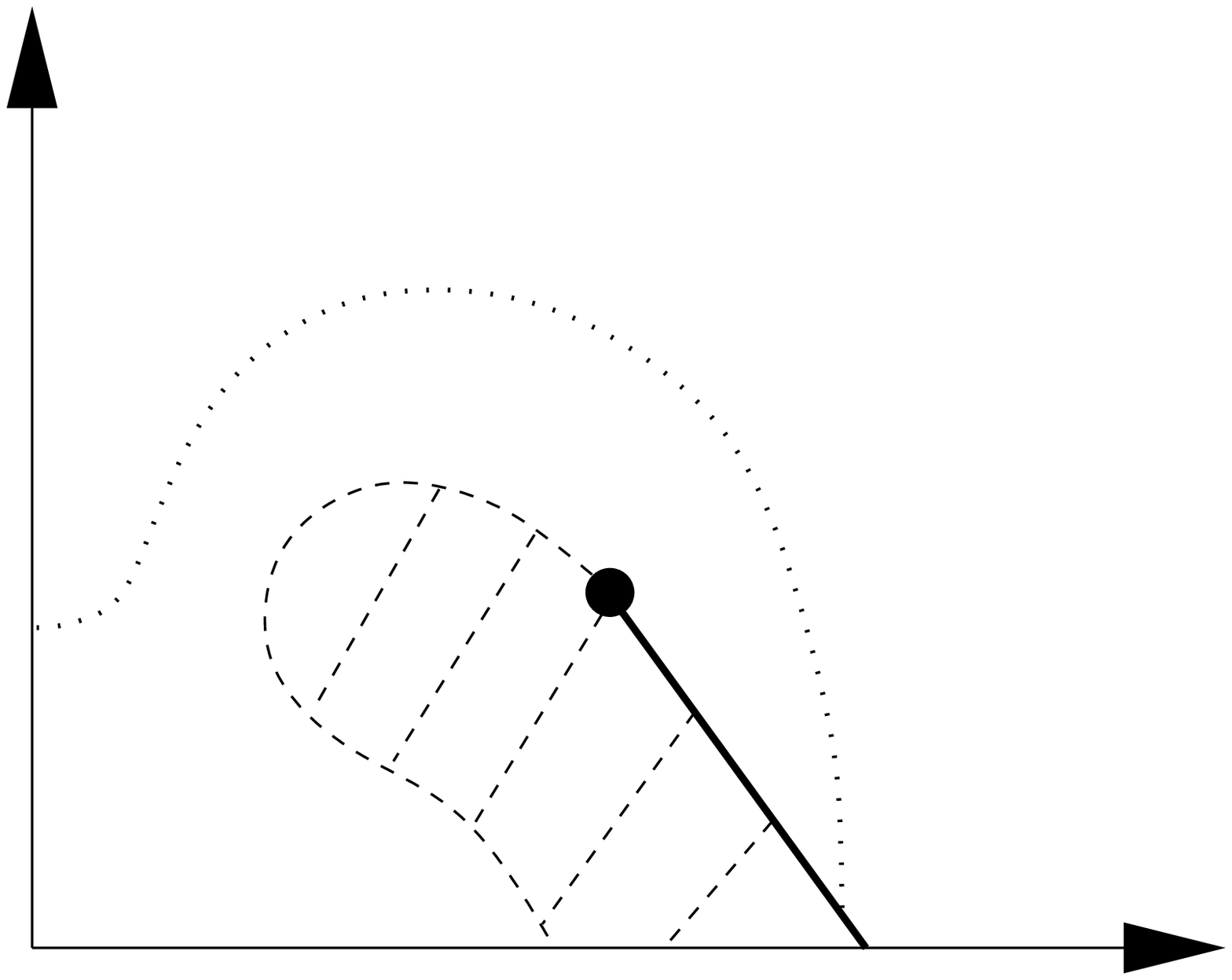} & \includegraphics[width=0.45\textwidth]{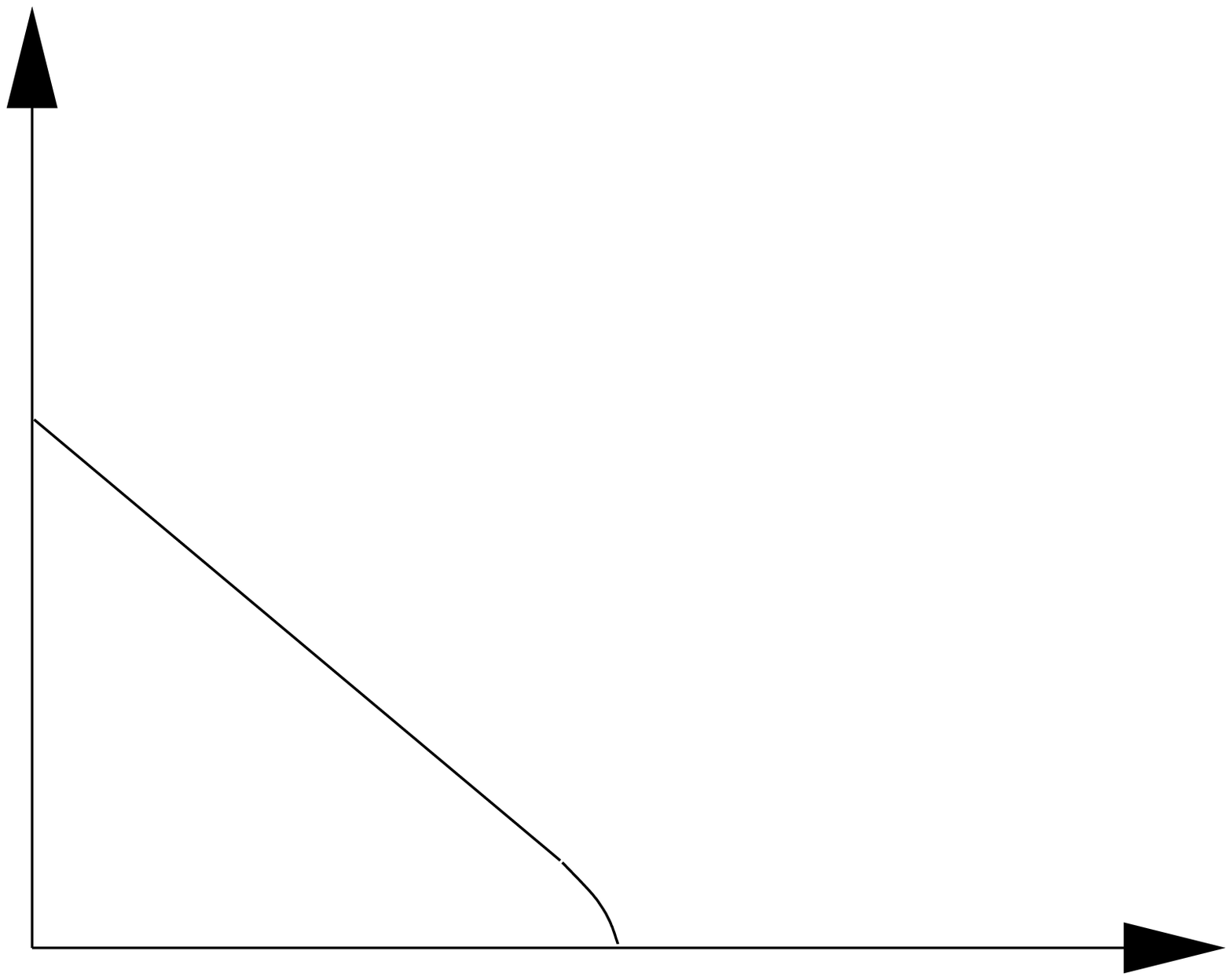} \\ (a.) & (b.)
\end{array}$
\caption{\label{phasediagramfig} Phase diagrams in (a.) the canonical ensemble and (b.) the grand canonical ensemble, as presently understood.}
}
\end{figure}

The analysis of ref. \cite{Kobayashi:2006sb} revealed that a line of transitions does emerge from the $\langle J^t \rangle/m^3 = 0$ axis, but eventually ends in a critical point. This is depicted schematically in figure \ref{phasediagramfig} (a.). In the figure, the vertical axis is $\< J^t \> /m^3$ and the horizontal axis is $T/m$. The solid line represent the line of first-order phase transitions that emerge from the transition point on the horizontal axis. The black dot represents the critical point.

Immediately to the left of the line of transitions is a region of thermodynamic instability, discovered in ref. \cite{Kobayashi:2006sb} and indicated in the plot with dashed lines (distinct from the dotted line, which we will discuss shortly). In this region, the condition $\frac{\partial \mu}{\partial \langle J^t \rangle} > 0$, for stability of the system against fluctuations in the density, is violated. We must then ask what the true ground state is in this region. The grand canonical ensemble will offer some insight into this question.

An important lesson emerges from the line of first-order transitions at finite density. Recall that at zero density, the first-order transition appeared in the bulk as a topology change: the D7-brane transition was from a Minkowski to a black hole embedding. Here, however, as we only have black hole embeddings, the transition is between a black hole embedding and another black hole embedding, so the topology does not change. Nevertheless, the transition is discontinuous.

What about the grand canonical ensemble, where $\mu$ is the control parameter and $\langle J^t \rangle$ is an output of the dynamics? Here we learn an important lesson. To describe a finite density in the field theory, we must use black hole embeddings of the D7-brane. The field theory includes states, however, with finite $\mu$ but \textit{zero density}. In a theory with a mass gap, chemical potentials of magnitude less than the mass gap do not lead to any finite density: if the lightest charge carriers have mass $m$, then a chemical potential $\mu<m$ cannot produce a nonzero density. In the canonical ensemble they never appear, since zero density corresponds to $\mu = m$. In the grand canonical ensemble we should of course be able to dial $\mu$ all the way down to zero, but the physics for $\mu < m$ is still expected to be trivial in this range, with $\< J^t\>=0$ and the free energy being independent of $\mu$. 

What D7-brane embeddings describe states with $\mu > 0$ but $\langle J^t \rangle = 0$? The answer is: Minkowski embeddings \cite{Ghoroku:2007re,Karch:2007br,Mateos:2007vc}. How can this be? The D7-brane action only depends on dertivatives of the gauge potential, in this case, derivatives of $A_t(r_6)$. A valid solution of $A_t(r_6)$'s equation of motion, in the absence of sources, is a constant, $A_t(r_6) = \mu$. Our nice physical intuition of $\mu$ as the work needed to bring a string endpoint from the boundary to the horizon fails here, but $\mu$ is nevertheless the leading asymptotic value of $A_t(r_6)$, and hence the chemical potential. Most importantly, $D=0$ for such a solution. We can thus have a Minkowski embedding with $A_t(r_6)=\mu$, describing a state in the field theory with finite $\mu$ but zero density. Notice also that the value of the on-shell action will be the same as the $A_t(r_6)=0$ solution, indicating that in the field theory the free energy is independent of $\mu$, as expected. In the grand canonical ensemble, we must include such $A_t(r_6)=\mu$ Minkowski embeddings if we wish to include all physically allowed values of $\mu$.

Including such embeddings, the phase diagram in the grand canonical ensemble was computed in refs. \cite{Ghoroku:2007re,Karch:2007br,Mateos:2007vc}, and is depicted schematically in figure \ref{phasediagramfig} (b.). The vertical axis is $\mu/m$ and the horizontal axis is $T/m$. The solid line indicates a line of first-order transitions emerging from the known transition on the horizontal axis. The solid line intersects the vertical axis precisely at $\mu/m=1$. In supergravity language, the transition across the line is from Minkowski embeddings with constant $A_t(r_6)$, below the line, to black hole embeddings with nontrivial $A_t(r_6)$, above the line. In field theory language, the transition is from a state with $\mu > 0$ but $\< J^t \> = 0$, with a meson spectrum that is gapped and discrete, to a state with $\mu > 0$ and $\< J^t \> > 0$, with a meson spectrum that is gapless and continuous. Notice, then, that the transition involves charge condensation, as $\< J^t \>$ jumps from zero to nonzero values across the transition line. Furthermore, the discontinuities in thermodynamic quantities across the phase transition line actually shrink as the system approaches the vertical axis \cite{Mateos:2007vc}, raising the possibility that the line of first-order transitions may end in a critical point when $T=0$ and $\mu = m$. In the next section, we will show that this is indeed the case.

The two ensembles should be equivalent, that is, should describe the same physics. To explain what this means in practical terms, consider the canonical ensemble. Here we choose values of $T$ and $\< J^t \>$, and the Helmholtz free energy then gives us the value of $\mu = \mu(T,\<J^t\>)$. If, in the grand canonical ensemble, we choose the same $T$ and the same value of $\mu$, the Gibbs free energy will return the value of $\<J^t\>$ that we started with. We can thus think of the two ensembles geometrically, as two equivalent parameterizations of a surface, the collection of points $(T, \mu(T,\<J^t\>),\<J^t\>)$ in the canonical ensemble or $(T,\mu,\< J^t \>(T,\mu))$ in the grand canonical ensemble. This surface is the manifold of equilibrium states in the parameter space. We should ask whether the phase diagrams in the two ensembles are consistent with one another.

If we map the line of first-order transitions in the grand canonical ensemble to the phase diagram in the canonical ensemble, we find the dotted line in figure \ref{phasediagramfig} (a.) \cite{Mateos:2007vc}. Below the line, Minkowski embeddings with constant $A_t(r_6)$ are thermodynamically preferred. These have $\<J^t\>=0$, so the grand canonical ensemble appears to have a gap. Notice also that the entire unstable region found in the canonical ensemble is within the region below the line.

The question is thus how the system accesses the states below the dotted line in figure \ref{phasediagramfig} (a.). The suggestion of ref. \cite{Mateos:2007vc} was that the ground state in this region may actually be an inhomogeneous mixture of the two phases, in analogy with the liquid-to-gas transition of ordinary water. Such a thing is difficult to describe in supergravity, however. It would be something like an inhomogeneous mixture of Minkowski and black hole embeddings. The important lesson is that the simple homogenous states that are easy to explore using supergravity are almost certainly insufficient to describe the entire phase diagram.

We emphasize that most of the work that we have summarized was numerical: the phase diagrams were derived from numerical solutions for D7-brane embeddings. We will now present analytic solutions in two limits: zero mass and zero temperature. These solutions will nicely confirm the general picture outlined above. In particular, the zero-temperature analytic solution will reveal that the transition at $\mu =m$ is second order.

\subsection{Exact Solution at Finite Density: $T>0, \, m=0$}
\label{zeromsol}

Reviewing the phase diagrams in figures \ref{phasediagramfig} (a.) and (b.), we can see that the $m=0$ limit actually looks pretty boring: it is the extreme upper right-hand corner of each plot, where we expect no phase transitions. We will compute an exact solution for the D7-brane embedding scalar and gauge field in this limit, and explicitly confirm the absence of any interesting phase structure.

We hasten to add, however, that the current phase diagram is probably incomplete. The key observation is that a chemical potential acts as a negative mass-squared for scalars. Suppose for the moment that we studied our theory at zero temperature, zero mass, and zero 't Hooft coupling, but with finite chemical potential. The theory will have no ground state: the potential for the scalars in the hypermultiplet would be unbounded from below. Introducing a finite mass or temperature can stabilize the potential, as can the quartic self-interactions and Yukawa couplings. In such cases we generically expect Bose-Einstein condensation: the hypermultiplet scalars will develop a nonzero expectation value, break the $U(1)_B$ symmetry, etc. How to see this in the supergravity description remains an open, and very important, question, one that we will not address here. We will work only with the D7-brane configurations described above.

Our goal is thus to integrate the solution for $A_t'(r_6)$ in eq. (\ref{finiteTgaugesol}) and the on-shell action, eq. (\ref{finiteTonshellaction}). To do so in general requires an explicit solution for $\theta(r_6)$. We know the solution in the massless case: $\theta(r_6)=0$, which also describes $\Omv = 0$ in the field theory.

With $\theta(r_6)=0$, the integral for $A_t(r_6)$ becomes
\beq
A_t(r_6) \, = \, \int_{r_h}^{\infty} \, \frac{D}{(2\p\a')^2} \, \left ( \N^2 \, r_6^6 + \frac{D^2}{(2\p\a')^2} \right )^{-1/2}.
\eeq
This integral can be performed analytically using incomplete Beta functions. We have collected the details in an Appendix to this chapter. The result is
\beq
A_t(r_6) = \frac{1}{6} \frac{1}{2\pi\alpha'} \left( \frac{D}{\N (2\pi\alpha')}\right)^{1/3} \left( B\left( \frac{r_6^6}{r_6^6 + \frac{D^2}{\N^2 (2\p\a')^2}}; \frac{1}{6}, \frac{1}{3} \right) - B\left( \frac{r_h^6}{r_h^6 + \frac{D^2}{\N^2(2\pi\alpha')^2}}; \frac{1}{6}, \frac{1}{3} \right)  \right)
\eeq
Notice that $A_t(r_h)=0$, as it should be. The value of $\mu$ is given by taking $r_6 \ra \infty$, for which the first term in parentheses goes to $B\left( \frac{1}{6}, \frac{1}{3} \right) \approx 8.41$.

The integral for the on-shell action is:
\beq
S \, = \, - \N^2 \, \int dr_6 \, r_6^6 \, \left( \N^2 \, r_6^6 + \frac{D^2}{(2\p\a')^2} \right)^{-1/2}.
\eeq
The $r_6$ integral is of course divergent. We regulate the divergence by cutting off the $r_6$ integration at some large value $\Lambda$. Rather than use counterterms, we will subtract a background solution. The simplest solution is a D7-brane with $\theta(r_6) = A_t(r_6)=0$, for which we denote the regulated on-shell action as $S_0$:
\beq
S_0 = - \N \int^{\Lambda}_{r_h} dr_6 r_6^3 = - \frac{1}{4} \N \Lambda^4 + \frac{1}{4} \N r_h^4.
\eeq
We then define the renormalized action $S_{ren}$ as
\beq
S_{ren} \, = \, \lim_{\Lambda \ra \infty} \left( S - S_0 \right),
\eeq
with the result
\beq
S_{ren} = - \frac{1}{4} \N r_h^4 - \frac{1}{6} \N^{-1/3} \left(\frac{D}{2\pi\a'}\right)^{4/3} B \left ( \frac{\left(\frac{D^2}{\N^2 (2\pi\a')^2}\right)}{r_h^6 + \left( \frac{D^2}{\N^2 (2\pi\alpha')^2} \right)}; - \frac{2}{3}, \frac{7}{6} \right ).
\eeq
The result has no interesting features, \textit{i.e.} is completely monotonic, and its derivatives display no discontinuities, which is consistent with the numerical results computed in refs. \cite{Kobayashi:2006sb,Ghoroku:2007re,Karch:2007br,Mateos:2007vc} and displayed in figure \ref{phasediagramfig}.

\subsection{Exact Solution at Finite Density: $T=0, \, m>0$}
\label{zerotsol}

The main feature of the phase diagram of figure \ref{phasediagramfig} (b.) is the line of first-order transitions between a phase with a gapped, discrete spectrum of stable mesons and zero baryon density and a phase with a gapless, continuous spectrum of unstable mesons and finite baryon density. As the line of transitions approaches the $T/m=0$ (vertical) axis, the first order transition grows weaker: the discontinuities in the derivatives of thermodyanmic potentials shrink. A natural question is whether the transition eventually becomes continuous. In other words, does the line of first-order transitions end in a critical point when $T=0$ and $\mu = m$? In this section we will compute the thermodynamics precisely along the $T/m=0$ axis, and discover a second-order transition when $\mu=m$.

In supergravity language, our goal is to solve for the D7-brane fields, and use them to compute the D7-brane's on-shell action. Even when $T=0$, finding a solution for $\theta(r_6)$ and then performing the integrals to find $A_t(r_6)$ and the on-shell action is very difficult.

The key to finding exact solutions is to change coordinates. We switch to the $AdS_5 \times S^5$ metric
\beq
ds^2 \, = \, Z(r_6) \, \left( -dt^2 + d\vec{x}^2 \right) \, + \, Z(r_6)^{-1} \, \left( dr^2 + r^2 \, ds^2_{S^3} + dy^2 + y^2 \, ds^2_{S^1} \right),
\eeq
with $Z(r_6)=r_6^2=r^2+y^2$. The D7-brane scalar is now $y(r)$ and the gauge field will be $A_t(r)$. With this ansatz for the fields, the D7-brane action becomes
\beq
S \, = \, -\N \, \int dr \, r^3 \, \sqrt{1+ y'^2 - (2\p\a')^2 A_t'^2}.
\eeq
Remarkably, the action depends only on derivtiaves of \textit{both} $y(r)$ and $A_t(r)$. We thus have \textit{two} constants of motion. Defining the Lagrangian by $S = \int dr L$, we denote the two constants of motion as $c$ and $d$:
\beq
\frac{\delta L}{\delta y'} = - \N r^3 \frac{y'}{\sqrt{1+ y'^2 - (2\p\a')^2 A_t'^2}} \equiv -c, \qquad \frac{\delta L}{\delta A_t'} = \N r^3 \frac{(2\p\a')^2 A_t'}{\sqrt{1+ y'^2 - (2\p\a')^2 A_t'^2}} \equiv d.
\eeq
The ratio implies
\beq
\label{ratio}
A_t'^2 = \frac{d^2}{(2\p\a')^4 \, c^2} \, y'^2.
\eeq
so that if we can perform the integration for one field, we automatically find the answer for the other. We can solve algebraically for $y'(r)$ and $A_t'(r)$ in terms of the integration constants $c$ and $d$,
\beq
y' = \frac{c}{\sqrt{\N^2 r^6 + \frac{d^2}{(2\p\a')^2}-c^2}}, \qquad A_t' = \frac{d/(2\p\a')^2}{\sqrt{\N^2 r^6 +
\frac{d^2}{(2\p\a')^2}-c^2}},
\eeq
which can be integrated using incomplete Beta functions. The result depends on the sign of $\frac{d^2}{(2\p\a')^2}-c^2$. When $c=d=0$, we obtain the solution with $y'(r)=0$ and $A_t'(r)=0$, so $y(r)$ and $A_t(r)$ are constants. If $A_t(r)=0$, we recover the embeddings of section \ref{d7braneembeddings}, with $y(r)$ constant. We will investigate the other possibilities in what follows.

The D7-brane action evaluated on these solutions is
\beq
S \, = \, -\N \, \int^{\Lambda} dr \, r^3 \sqrt{\frac{\N^2 r^6}{\N^2 r^6 + \frac{d^2}{(2\p\a')^2} - c^2}}
\eeq
The lower endpoint of integration will depend on the sign of $\frac{d^2}{(2\p\a')^2} - c^2$. The integral diverges if we
integrate to $r = \infty$, of course, so we have regulated the integral with a cutoff at $r = \Lambda$. The divergence is already present for $c = d = 0$:
\beq
S_0 \, = \, -\N \, \int_0^{\Lambda} dr \, r^3 \, = \, - \frac{1}{4} \, \N \, \Lambda^4
\eeq
so we will again define the renormalized on-shell action, $S_{ren}$, as
\beq
S_{ren} = \lim_{\Lambda \ra \infty} (S - S_0)
\eeq

In the field theory, the thermodynamic potential $\Omega$ of the grand canonical ensemble is given by\footnote{In section \ref{finiteTcorrespondence}, when $T>0$ and $\mu=0$, we made the identification $F = T S_{SUGRA}$, and mentioned that if $\mu >0$ we should identify $\Omega = T S_{SUGRA}$. Here we have $T=0$ and $\mu>0$, however, for which the identification above is the correct one.} $\Omega = - S_{ren}$. We can thus determine whether solutions with nonzero $d$ and $c$ are thermodynamically favored relative to the $c=d=0$ case as follows: $S_{ren} >0$ means the configuration is \textit{favored} and $S_{ren} < 0$ means the configuration is \textit{disfavored}.

We Legendre transform to obtain the free energy density in the canonical ensemble, $F = \Omega + \mu \langle J^t \rangle$. Recall that at zero temperature, free energy and energy are the same, so at zero temperature $F$ is also the energy density.

The conserved charges $c$ and $d$ determine $\Omv$ and $\langle J^t \rangle$ as follows,
\beq
\Omv = \frac{\delta \Omega}{\delta m} = - (2\p\a') \frac{\delta S}{\delta y(\infty)}, \qquad \langle J^t \rangle = -
\frac{\delta \Omega}{\delta \mu} = \frac{\delta S}{\delta A_t(\infty)}
\eeq
where in each case when we vary one field we hold the other fixed. We can write
\beq
\delta S = \int dr \left ( \frac{\delta L}{\delta A_t'(r)} \partial_r \delta A_t(r) + \frac{\delta L}{\delta y'(r)} \partial_r \delta y(r) \right) = d \, \delta A_t(\infty) - c \, \delta y (\infty)
\eeq
where we demand that $\delta A_t(r)$ and $\delta y(r)$ are always zero at the lower endpoint of the $r$ integration. If we vary $A_t(r)$ while holding $y(r)$ fixed ($\delta y(r) = 0$), we find $\langle J^t \rangle = d$, and similarly we find $\Omv = (2\p\a') c$. Starting now, we will stick to the notation $c$ and $d$ and refer to these as the
condensate and density.

\begin{figure}
{\centering
\includegraphics[width=0.70\textwidth]{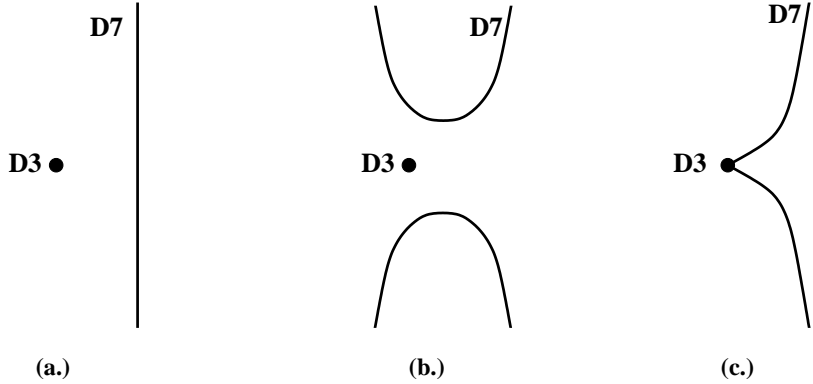}
\caption{\label{zeroTfig} Cartoons of D7-brane embeddings, for various values of $\frac{d^2}{(2\p\a')^2}- c^2$.}
}
\end{figure}

We now turn to the various possibilities for the values of $c$ and $d$, and in particular the sign of $\frac{d^2}{(2\p\a')^2}- c^2$. The possibilities are depicted, in a very schematic way, in figure \ref{zeroTfig}. In the figure, the vertical axis is one of the directions transverse to the D3-branes (in Maldacena's construction) but parallel to the D7-branes, while the horizontal axis is an overall transverse direction. Figure \ref{zeroTfig} (a.) depicts the $c=d=0$ case, in which $A_t(r)=0$ and $y(r)$ is a constant. Figure \ref{zeroTfig} (b.) depicts the $\frac{d^2}{(2\p\a')^2}- c^2 <0$ case, and figure \ref{zeroTfig} (c.) depicts the $\frac{d^2}{(2\p\a')^2}- c^2 > 0$ case, which we will now describe in turn.

We begin with $\frac{d^2}{(2\p\a')^2}- c^2 < 0$, and furthermore, we restrict to $d=0$. Such solutions were first studied in ref. \cite{Herzog:2007kh}. These are, in fact, precisely the solutions we mentioned in section \ref{d7braneembeddings}, below eq. (\ref{yeom}), when we first discussed D7-brane embeddings.

For such $d=0$ embeddings, $y'(r)$ diverges as $\frac{1}{\sqrt{r-r_0}}$ at a critical radius $r_0 = \N^{-1/3} c^{1/3}$. This behavior indicates that the D7-brane has a turn-around point where the D7-brane smoothly matches onto a second branch. This solution thus describes a D7/anti-D7-brane pair connected by a smooth neck, as depicted schematically in figure \ref{zeroTfig} (b.).

In the asymptotic region, we may specify the distances of the D7-brane and the anti-D7-brane to the D3-branes separately. In other words, we may associate two masses to such a configuration. We denote these as $m$ and $\overline{m}$. A D7-brane solution exists for any $m$ and $\Omv$. The mass (or asymptotic separation) of the anti-D7-brane is then fixed in terms of these two input parameters. From the field theory point of view, we would like to interpret the two masses $m$ and $\overline{m}$ as input parameters, so that the above construction dynamically determines the value of the condensate $\Omv$. From the supergravity picture, we can see that the condensate only depends on the difference of the masses, $m - \overline{m}$. As we will show (and as first noted in ref. \cite{Herzog:2007kh}), all such solutions have $S_{ren} < 0$, and hence will be meta-stable, at best. The true ground state of the system has vanishing $\Omv$.

Now we re-introduce nonzero $d$. To connect the two halves of the D7-brane smoothly we need the gauge field and its derivative to be continuous at the neck. In particular, the field strength on the two branches should be equal in magnitude but opposite in direction. The baryon number density is then equal and opposite at the two ends, so this configuration does not describe any net baryon number! The Lagrangian of the field theory has a global $U(1) \times U(1)$ symmetry, and, at least in the grand canonical ensemble, we can imagine turning on different chemical potentials for the two $U(1)$'s at the level of the Lagrangian, even though the symmetry is spontaneously broken to the diagonal $U(1)_B$ in the state represented by the connected D7-brane configuration. If we wanted to have a brane/anti-brane system with finite baryon density we would have to include an explicit source at the neck. Such additional sources will back-react and alter the configuration, so we will not discuss this option further.

In any case, these brane/anti-brane systems correspond to a different field theory Lagrangian than that for systems with a single D7-brane reaching the asymptotic region. The dual field theory for the brane/anti-brane is ${\cal N}=4$ SYM coupled to two hypermultiplets that preserve opposite ${\cal N}=2$ supersymmetries, so that the full theory is non-supersymmetric. When comparing all possible bulk configurations that correspond to hypermultiplets preserving $\N=2$ supersymmetry, the brane/anti-brane configurations never contribute.

For nonzero $d$, keeping in mind that for the brane/anti-brane system this does not correspond to a net baryon density, the D7-brane reaches the turn-around point
\beq
r_0 = \N^{-1/3} \left( c^2 - \frac{d^2}{(2\p\a')^2} \right)^{1/6}
\eeq
The integrals for $y(r)$, $A_t(r)$ and the regulated on-shell action are performed in the Appendix. We find
\beq
y(r) = c \frac{1}{6} \N^{-1/3} \left( c^2 - \frac{d^2}{(2\p\a')^2} \right)^{-1/3} \left ( B\left( \frac{1}{3},\frac{1}{2} \right) - B\left( \frac{r_0^6}{r^6}; \frac{1}{3},\frac{1}{2} \right) \right )
\eeq
and $A_t(r)$ is simply $\frac{d}{(2\p\a')^2} \frac{1}{c}$ times $y(r)$, as shown in eq. (\ref{ratio}). Notice $y(r_0)= A_t(r_0) = 0$. The mass and chemical potential are given by the asymptotic values at $r \ra \infty$, with $\lim_{r \ra \infty} B\left(\frac{r_0^6}{r^6}; \frac{1}{3},\frac{1}{2} \right) = 0$. The renormalized on-shell action for the D7-brane is
\beq
S_{ren} =  - \frac{1}{6} \N^{-1/3} \left( c^2 - \frac{d^2}{(2\p\a')^2} \right)^{2/3} B\left(-\frac{2}{3},\frac{1}{2}
\right)
\eeq
The on-shell action for the anti-D7-brane is identical, so the total action is twice $S_{ren}$. Notice that $S_{ren} < 0$, so these connected brane/anti-brane configurations always have a higher free energy than the $c=d=0$ case, and hence are thermodynamically disfavored.

At this point we should also consider $c$ and $d$ nonzero, but such that $\frac{d^2}{(2\pi\alpha')^2} - c^2 = 0$. This configuration shares a similar feature with the brane/anti-brane configurations: it does not satisfy the required UV boundary conditions for a theory with a single D7-brane added. When $\frac{d^2}{(2\pi\alpha')^2} -c^2 = 0$, we have $y'(r)
\sim\frac{1}{r^3}$, so the D7-brane misses the D3-branes at the origin and returns to the asymptotic region at $r \rightarrow \infty$. We will therefore not study the $c = \frac{d}{2\pi\alpha'}$ case further.

Now we come to the interesting case: $\frac{d^2}{(2\p\a')^2}- c^2 > 0$. We will show that these are the zero-temperature limit of black hole embeddings. Notice first that these solutions have no turn-around point, and describe D7-branes alone.

For a $\frac{d^2}{(2\p\a')^2}- c^2 >0$ solution, suppose we take $c=0$. The solution with constant $y(r)$ is still allowed.
Geometrically, this is the zero-temperature Minkowski embedding, but now with nonzero $A_t'(r)$. As mentioned before, the constant charge density along the D7-brane requires a source at $r=0$, so this embedding is not physical unless we add extra strings connecting the D3-branes and D7-branes, or some other source.

We can avoid this issue, however, by demanding that the D7-brane touch the horizon at $r=0$. Since the horizon is located at $0=r_6^2 = r^2 +y^2$, this immediately implies a boundary condition on $y$, namely $y(r=0) = 0$. The embedding satisfying this boundary condition is the zero temperature analogue of black hole embeddings. We will stick to this
name, even though at zero temperature ``horizon crossing'' would be more appropriate. This black hole embedding is displayed schematically in figure \ref{zeroTfig} (c.).

The integral for $y'(r)$ is done in the Appendix, with the result
\beq
y(r) = c \frac{1}{6} \N^{-1/3} \left( \frac{d^2}{(2\p\a')^2} - c^2 \right)^{-1/3} B\left( \frac{\N^2 r^6}{\N^2 r^6 + \frac{d^2}{(2\p\a')^2} - c^2}; \frac{1}{6},\frac{1}{3} \right)
\eeq
and again, $A_t(r)$ is $\frac{d}{(2\p\a')^2} \frac{1}{c}$ times $y(r)$. Notice that $y(0) = A_t(0)=0$. Identifying $m$ and $\mu$ from the asymptotic values $y(\infty)$ and $A_t(\infty)$, where as $r \ra \infty$ the incomplete Beta function becomes $B\left(\frac{1}{6},\frac{1}{3} \right)$, we find
\begin{subequations}
\label{relations}
\beq
c = \gamma \N (2\p\a')^3 \left( \m^2 - m^2 \right) m
\eeq
\beq
\frac{d}{2\p\a'} = \gamma \N (2\p\a')^3 \left( \m^2 -
m^2 \right) \mu
\eeq
\end{subequations}
where we have defined the constant
\beq
\gamma \equiv  \left( \frac{1}{6} B\left(\frac{1}{6},\frac{1}{3}\right)
\right)^{-3} \approx 0.363 .
\eeq
Notice that since these black hole embeddings were valid only when $\frac{d^2}{(2\p\a')^2}- c^2 >0$, we can immediately see that they only exist for $\mu > m$, as expected for a theory with a mass gap, in a state with nonzero density.

The renormalized on-shell action in this case is
\beq
\label{bhrenaction}
S_{ren} = \frac{1}{4} \gamma^{-1/3} \N^{-1/3} \left( \frac{d^2}{(2\p\a')^2} - c^2 \right)^{2/3} = \frac{1}{4} \gamma \N
(2\p\a')^4 \left(\mu^2-m^2 \right)^2
\eeq
Notice that $S_{ren} > 0$, so these embeddings are thermodynamically favored relative to the $c=d=0$ case. Notice that we observe no instability in $\Omega$. Stability requires $\frac{\partial d}{\partial\mu} \geq 0$, which is clearly satisfied in eq. (\ref{relations}). This is consistent with the numerical results of ref. \cite{Mateos:2007vc} in the zero-temperature limit.

The thermodynamic potential in the canonical ensemble is
\beq
F = \Omega + \mu \langle J^t \rangle = \frac{1}{4} \N^{-1/3} \gamma^{-1/3} \left( \frac{d^2}{(2\p\a')^2} - c^2 \right)^{-1/3} \left(3\frac{d^2}{(2\p\a')^2} + c^2 \right)
\eeq
so that $F > 0$. The black hole embedding still turns out to be the configuration with lowest $F$, however.

To see the phase transition between black hole and Minkowski embeddings, let us return to $c=0$ and constant $y(r)$. These are solutions everywhere away from $r=0$. These D7-branes do not obey the boundary condition $y(0)=0$. In other words, when $r=0$ the D7-branes are still a distance $y$ away from the D3-branes.

The integral for $A_t(r)$ will be unchanged, however. In particular, we still integrate from $r=0$ to $r=\infty$, and $A_t(0)=0$. From eq. (\ref{relations}) we read off the relation between $\mu$ and $d$,
\beq
d = \gamma \N (2\p\a')^4 \mu^3,
\eeq
The integral for the on-shell action is also unchanged, so we simply set $c=0$ in the first equality of eq. (\ref{bhrenaction}),
\beq
S_{ren}  = \frac{1}{4} \gamma^{-1/3} \N^{-1/3} (2\p\a')^{-4/3} d^{4/3} = \frac{1}{4} \gamma \N (2\p\a')^4 \mu^4
\eeq
and hence $\Omega = -S_{ren} \sim - \mu^4$. Naively this $\Omega$, for any nonzero $d$ (or $\mu > m$), is smaller (more
negative) than the one for the black hole embedding, eq. (\ref{bhrenaction}). Notice, however, that unlike the nice
interpretation we found for the relations among $c$, $d$, $\mu$ and $m$ for the black hole embeddings, this result for $\Omega$, a pure quartic in $\mu$, is completely independent of the mass and therefore appears to be unphysical. For example, in the decoupling limit $m \ra \infty$, $\Omega$ should go to zero, while this result clearly does not.

The problem of course is that the Minkowski embedding, as it stands, is not consistent: we completely ignored the contribution to $\Omega$ due to whatever object sources the D7-brane gauge field. The simplest source is a finite density $d$ of fundamental strings stretching from the D3-branes to the D7-brane. These D7-branes with strings attached are not solutions to the full equations of motion following from the combined DBI and Nambu-Goto actions, however. Naively, we might think that the string back-reaction on the D7-brane can be neglected in the large-$N_c$ and large-$\lambda$ limit. For a \textit{single} string this is certainly true. The prefactor of the Nambu-Goto action scales as $\sqrt{\lambda}$, while the prefactor $\N$ of the D7-brane action scales as $N_f N_c \lambda$. We want a finite density $d$ of strings, however. If we keep all geometric distances of order one in units of the AdS curvature radius, the mass will scale as $\sqrt{\lambda}$, so we want the chemical potential also to scale as $\sqrt{\lambda}$. From eq. (\ref{relations}), we can
see that the density $d$ will then scale as $N_f N_c \sqrt{\lambda}$. With this, $d$ times the Nambu-Goto action has the same $N_f N_c \lambda$ scaling as the D7-brane action, and so the back-reaction is order one.

To include the back-reaction we would have to re-solve the equations of motion including the extra source term. We already found the most general solution to the equations of motion with sources localized at $r=0$, so the back-reacted solution including any such sources must be within this class. The only well-behaved solution in this class which asymptotically becomes a single brane is the black hole embedding, so it must be the back-reacted solution.

To show that this is physically reasonable we can show that the brane-plus-strings configuration, which
could serve as a consistent initial data for a full time-dependent physical solution, has higher energy than the black hole embedding. We add a term $d m$ to the action representing the finite density $d$ of strings with length $ y = (2 \pi \alpha') m$. The full free energy including the D7-branes and strings in this case is
\beq
\frac{\Omega}{ \frac{1}{4} \gamma \N (2 \pi \alpha')^4} = - \mu^4 + 4 \mu^3 m = \mu^3 (4 m - \mu)
\eeq
which is positive at $\mu=m$ and is disfavored relative to the black hole embedding at all values of $\mu>m$. We can also allow some mixture, where the D7-brane embedding satisfies the boundary condition $y(r=0)=y_0$ for some $m > \frac{y_0}{2 \pi \alpha'} > 0$, with a density $d$ of strings extending from $y=0$ to $y=y_0$. In figure \ref{zeroTfig}, we imagine sliding the black hole embedding to the right a distance $y_0$, such that the total asymptotic separation remains $y$, and then introducing the strings. The free energy is then
\beq
\frac{\Omega}{\frac{1}{4} \gamma \N (2 \pi \alpha')^4} = - \left( \mu^2 - \left( m - \frac{y_0}{2 \pi \alpha'} \right)^2 \right)^2 + 4 \mu \left( \mu^2 - \left( m - \frac{y_0}{2 \pi \alpha'} \right)^2 \right) \frac{y_0}{2 \pi \alpha'}
\eeq
and we can easily check that in the relevant regime $\mu>m > \frac{y_0}{2 \pi \alpha'} > 0$ the derivative of this expression with respect to $y_0$ is strictly positive. The D7-brane with strings attached can continuously lower its energy until it turns into the black hole embedding with all strings dissolved in the D7-brane.

While none of the finite-density Minkowski embeddings solve the equations of motion, the trivial Minkwoski embeddings with constant $A_t(r)$ do. As only $F_{rt}$ enters the action and not $A_t(r)$ itself, these embeddings have $F=\Omega=d=0$, and $\mu$ is a free parameter. For $\mu<m$ these are the only allowed configurations, so they dominate the ensemble. The fact that they are indistinguishable from the vacuum is completely natural from a field theory point of view, since the theory has the mass gap $m$. When $\mu >m$ these solutions still exist, but in that regime the black hole embedding, with negative $\Omega$, dominates the ensemble.

At $\mu=m$ we hence have a second order phase transition between the trivial Minkowski embedding and the
 black hole embeddings. Both $\Omega$ and its first derivative with respect to $\mu$ are zero at that point, but the second derivative jumps from zero, for the trivial Minkowski embedding, to $-2 \gamma \N (2\p\a')^4 m^2$, for the black hole embedding. The critical exponents are given by their mean-field values, that is, for $\mu = m + \epsilon$ we have $\Omega \sim \epsilon^2$, $c \sim \epsilon$, $d \sim \epsilon$.

We have thus shown that, indeed, the line of first-order transitions in the phase diagram of figure \ref{phasediagramfig} ends in a critical point on the $T=0$ axis, explicitly confirming the numerical results of ref. \cite{Mateos:2007vc}.

\section{Finite Electric and Magnetic Fields}

In the next chapter we will study the transport properties of flavor fields using the D7-brane description. We will introduce external electric and magnetic fields and then compute the resulting charge currents, from which we can extract a conductivity tensor. Before studying transport, however, we must study thermodynamics: we must know what the equilibrium state is before we perturb around it. In this section we very briefly review what is currently known about the thermodynamics of the flavor fields in the presence of constant electric and magnetic fields \cite{Filev:2007gb,Filev:2007qu,Albash:2007bk,Erdmenger:2007bn,Albash:2007bq}. We stress that all work to date regarding the thermodynamics in the presence of external fields has been done with \textit{zero} baryon density: the phase diagram in the full parameter space ($m$, $T$, $\langle J^t \rangle$, $E$, and $B$) remains largely incomplete.

\subsection{Electric Field}
\label{electric}

How can we introduce an electric field in the field theory? The flavor fields have the baryon number symmetry $U(1)_B$. We can introduce an electric field that couples only to excitations that carry this $U(1)_B$ charge. Notice that we will not \textit{gauge} the $U(1)_B$: we will not have a dynamical ``photon.'' The electric field we introduce will be an external parameter, one that is only ``felt'' by fields carrying $U(1)_B$ charge.

The primary motivation for introducing an electric field in this fashion is because the supergravity description is relatively easy. In the field theory, we want a nonzero field strength $F^{tx} = E$, describing a constant electric field $E$ in the $x$ direction. Under the right circumstances, we expect a current in the $x$ direction, that is, an expectation value $\langle J^x \rangle$. The former we expect to arise as the leading, non-normalizable mode of some D7-brane field, while the latter we expect to arise from a sub-leading, normalizable mode. The $U(1)_B$ in question is described by the $U(1)$ gauge field on the D7-brane worldvolume. We can thus introduce $E$ and $\langle J^x \rangle$ by assuming an ansatz for the D7-brane gauge field that includes \cite{Karch:2007pd}
\beq
A_x(t,r_6) \, = \, - E \, t \, + \, f_x(r_6)
\eeq
with $f_x(r_6)$ an $r_6$-dependent function for which we must solve. The first term produces the $F^{tx}$ in the gauge theory, while we should be able to extract $\langle J^x \rangle$ as a coefficient of a sub-leading term in $f_x(r_6)$'s asymptotic expansion. We may thus study the field theory at finite temperature, baryon density, and finite $E$ by adding $A_x(t,r_6)$ to our above ansatz for the fields. Currently, however, the thermodynamics with a finite electric field has only been studied at zero density.

We can actually anticipate what will occur with a finite electric field, using our field theory intuition \cite{Albash:2007bq}. Suppose we are at zero density, and with a finite temperature below the meson melting transition. The low-energy degrees of freedom are then stable mesons. We expect that an electric field will tend to pull a meson apart: if the meson is a quark and anti-quark, the electric field will pull the quark one way and the anti-quark the other. In other words, the binding energy of the quark is reduced in the presence of an electric field. We should then not need to heat the system up as much to see mesons melt. Our expectation is thus that the meson melting temperature should decrease.

Once the mesons have been pulled apart and/or melted, we also expect a current, $\langle J^x \rangle$, since we will then have free charge carriers (quarks and squarks) in the presence of an electric field. The meson melting transition in this case may be called an ``insulator-metal'' transition, at least in the sense that charge carriers are liberated.

Overall, the supergravity analysis is consistent with the above expectations, though with some minor subtleties. From the supergravity point of view, the electric field on the D7-brane worldvolume tends to ``push the D7-brane into the horizon,'' facilitating the first-order transition from Minkowski to black hole embeddings. Some D7-brane embeddings were found (numerically) to be singular, however, meaning these embeddings, as they stand, are probably unphysical \cite{Erdmenger:2007bn,Albash:2007bq}. A small part of the phase diagram thus remains in question. The status of these embeddings is still unresolved.

\begin{figure}
{\centering
$\begin{array}{cc}
E /T^2 \, \, \, \, vs. \, \, \, \, T/m & B/T^2 \, \, \, \, vs. \, \, \, \, T/m \\
\includegraphics[width=0.45\textwidth]{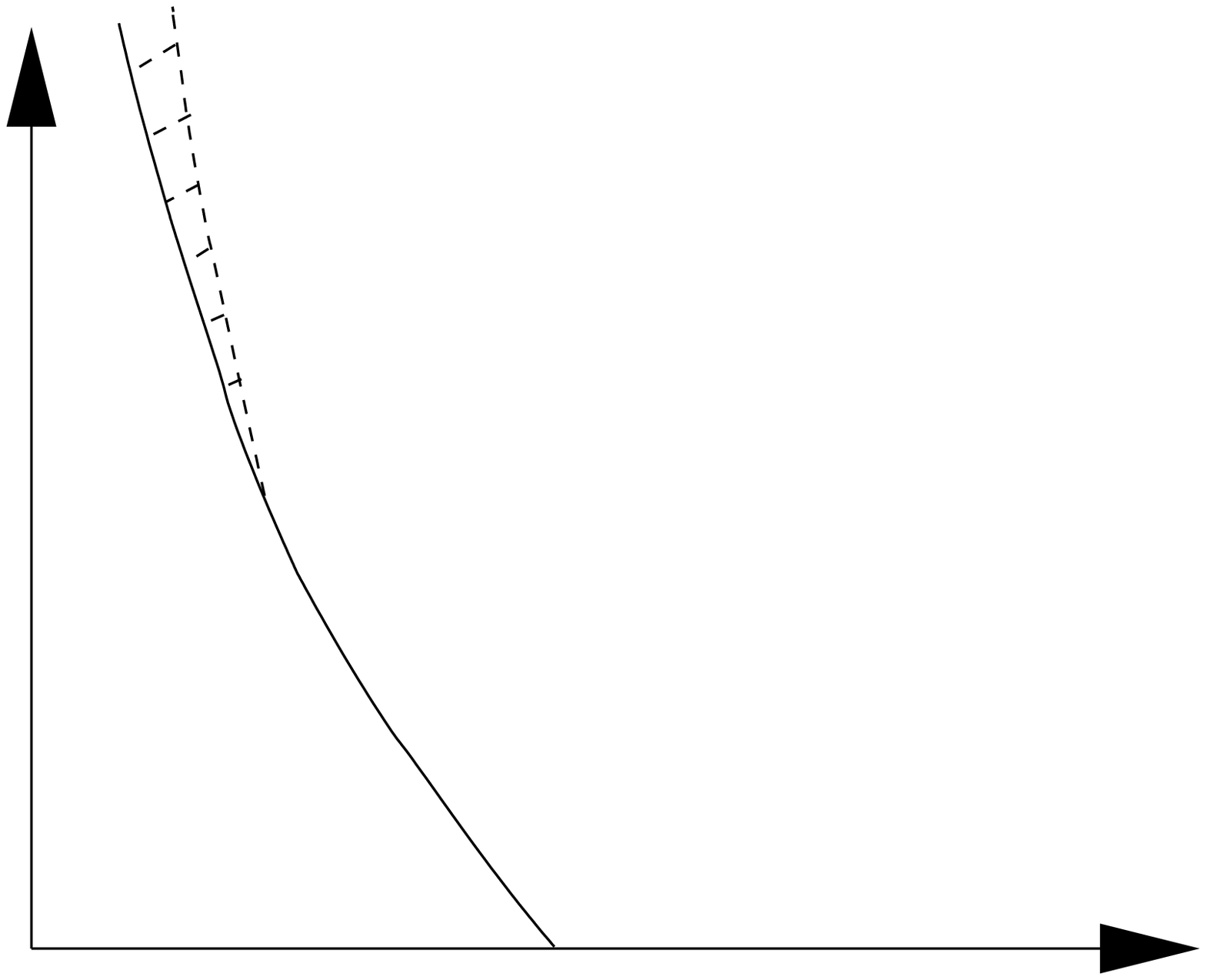} & \includegraphics[width=0.45\textwidth]{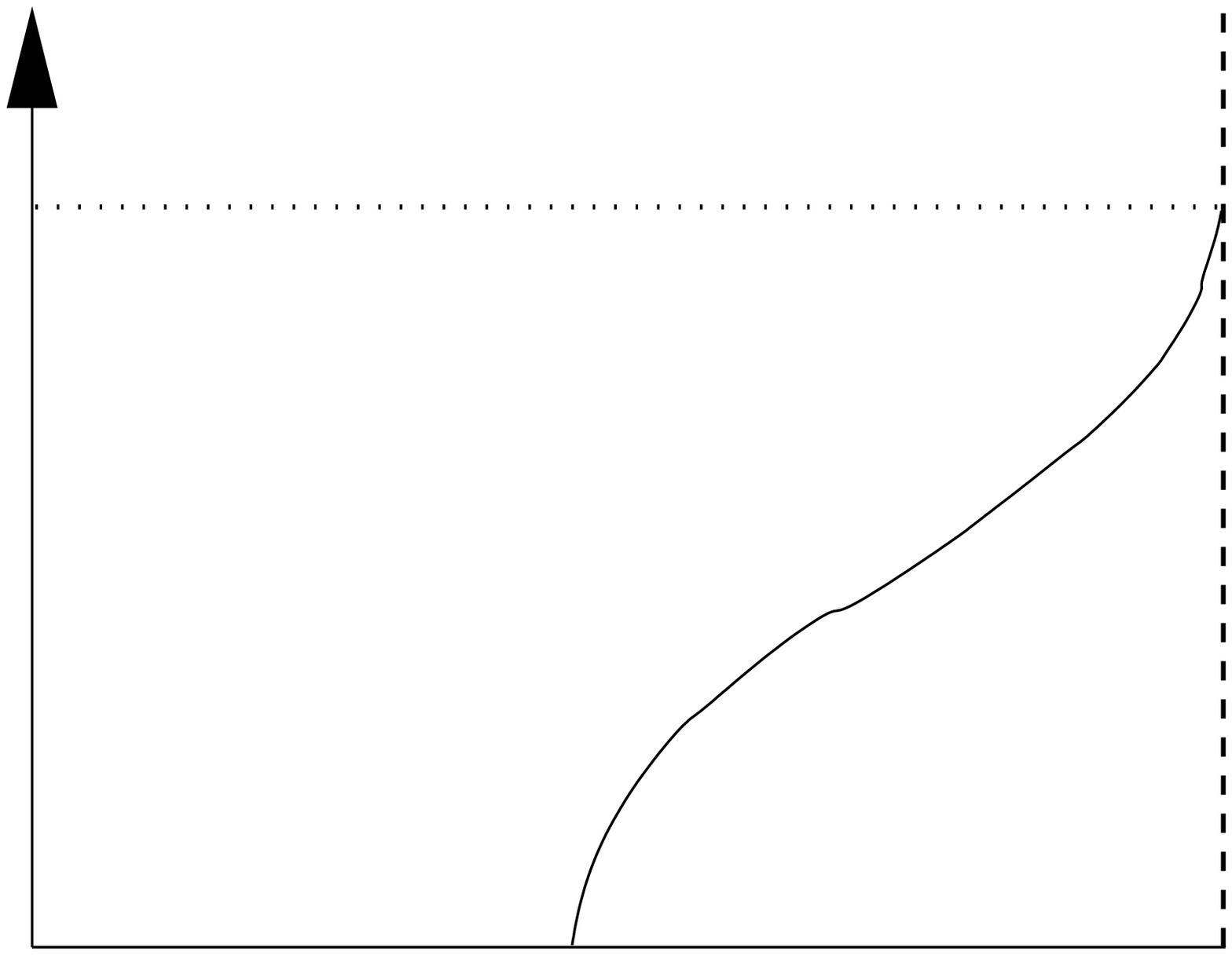} \\ (a.) & (b.)
\end{array}$
\caption{\label{eandbfieldsfig} Phase diagrams with (a.) nonzero electric field and (b.) nonzero magnetic field.}
}
\end{figure}

The result for the phase diagram, as derived from D7-brane solutions \cite{Erdmenger:2007bn,Albash:2007bq}, is depicted schematically in figure \ref{eandbfieldsfig} (a.). In the figure the vertical axis is $E/T^2$ while the horizontal axis is $T/m$. The solid line represents the phase transition from Minkowski to black hole embeddings: Minkowski on the left, black hole on the right. In field theory language, the region to the left of the solid line is a phase with stable mesons and no current $\langle J^x \rangle$, while to the right of the line is a phase with unstable mesons and a net current. The solid line emerges from the known transition point on the horizontal axis, where $E/T^2=0$. Clearly the transition temperature decreases as $E/T^2$ increases. The region marked with dashed lines corresponds to singular D7-brane embeddings. The ground state in this region is currently unknown.

\subsection{Magnetic Field}
\label{magnetic}

We introduce a magnetic field in the field theory is essentially the same way as we introduced the electric field: as a non-dynamical external field that couples to anything carrying $U(1)_B$ charge. We then want a nonzero $F^{xy} = B$. If we also introduce $E$ as above, we expect a current $\langle J^x \rangle$ but also now, via the Hall effect, $\langle J^y \rangle$. To our ansatz for D7-brane field we should thus add \cite{OBannon:2007in}
\beq
A_y(x,r_6) \, = \, B \, x \, + \, f_y(r_6)
\eeq
The leading term will produce $F^{xy}$ while $f_y(r_6)$ should produce $\langle J^y \rangle$. In this section, however, we consider only $E=0$, so that the currents are zero.

Two interesting things occur with nonzero $B$. The first occurs at zero temperature and zero hypermultiplet mass: chiral symmetry is spontaneously broken \cite{Filev:2007gb,Filev:2007qu}. We mean by this that, even with zero temperature and zero mass, nonzero $B$ produces a nonzero $\Omv$. As $\Omv$ is charged under the $U(1)$ that acts on the quarks as a chiral symmetry (dual to rotations in the $X_8$-$X_9$ plane), we call this chiral symmetry breaking. A Goldstone boson should then appear in the meson spectrum, which a supergravity analysis confirms \cite{Filev:2007gb,Filev:2007qu}. The Goldstone boson obeys the Gell-Mann-Oakes-Renner relation, as it should \cite{Filev:2007gb,Filev:2007qu}. Additionally, the spectrum of massive mesons exhibits Zeeman splitting \cite{Filev:2007gb,Filev:2007qu}.

The second interesting thing occurs at finite temperature: we find a critical value of the magnetic field above which mesons are always stable \cite{Albash:2007bk,Erdmenger:2007bn}. In heuristic terms, the magnetic field acts to ``hold mesons together.'' Suppose we fix $m$ and $T$ such that the system is in a phase with stable mesons. As $B$ increases, no transition will occur. If instead we choose $m$ and $T$ such that the system is in the phase with unstable mesons, then as $B$ increases we will eventually see a transition to the phase with stable mesons.

From a supergravity point of view, the black hole of the AdS-Schwarzschild background pulls on the D7-brane, while the worldvolume magnetic field acts to pull the D7-brane away from the black hole, so that the temperature of the transition from Minkowski to black hole embeddings increases. For fixed $m$ and $T$, Above a critical value of $B$ only Minkowski embeddings exist. These embeddings can then be shown to describe the entire range of $m$, and indeed, for $m=0$, the supergravity solution produces $\Omv \neq 0$, hence chiral symmetry is spontaneously broken.

The result for the phase diagram is depicted schematically in figure \ref{eandbfieldsfig} (b.). In the figure the vertical axis is $B/T^2$ and the horizontal axis is $T/m$. We have included in the plot the value $T/m = \infty$, indicated with the dashed line on the right. The solid line represents transitions from Minkowski embeddings, on the left, to black hole embeddings, on the right, or in field theory language, transitions from states with stable mesons to states with unstable mesons. The line emerges from the known transition point on the horizontal axis, where $B/T^2=0$.

Clearly, if we start on the $B/T^2=0$ axis, to the \textit{left} of the transition point, and then increase $B/T^2$, moving up in the figure, we encounter no transition. If we start to the right of the transition point, we can see that eventually we encounter a transition to the Minkowski/stable meson phase. Indeed, even when $T/m=\infty$, which we could achieve with $T$ finite and $m=0$, if we increase $B/T^2$ past a critical value, indicated in the figure by the dotted line, a transition occurs to the Minkowski/stable meson phase.

\section*{Appendix: Beta functions and Incomplete Beta functions}
\setcounter{equation}{0}
\renewcommand{\theequation}{\arabic{equation}}

The integrals we need for this chaper all take the form of Beta functions or incomplete Beta functions. Here we collect some of the important properties of these functions, and perform the integrals necessary for sections \ref{zeromsol} and \ref{zerotsol}. We use the notation of section \ref{zerotsol}, and make note of which integrals are relevant for section \ref{zeromsol}.

The Beta function is defined as
\beq
\label{bf}
B(a,b) = \frac{\G(a)\G(b)}{\G(a+b)} = \int_0^1 dt (1-t)^{b-1} t^{a-1} = \int_{0}^{\infty} du (1+u)^{-(a+b)} u^{a-1}
\eeq
and the incomplete Beta function as
\beq
B\left( x;a,b \right) = \int_0^{x} dt (1-t)^{b-1} t^{a-1} = \int_{0}^{x/(1-x)} du (1+u)^{-(a+b)} u^{a-1}.
\label{ibf}
\eeq
These satisfy the recursion relation
\beq
B(x;a,b) = B(a,b) - B(1-x;b,a)
\eeq
and are related to the hypergeometric function as
\beq
B(x;a,b) = a^{-1} x^a F(a,1-b;a+1;x).
\eeq

The $\Lambda^4$ divergence of $S$ in section \ref{zerotsol} appears in the expansion, for $a = -2/3$ about $x=0$,
\beq
B\left( x; - \frac{2}{3}, b \right) = - \frac{3}{2} x^{-2/3} + O(x^{1/3})
\eeq
where for $S$ we will have $x \sim \Lambda^{-6}$ (see for example eq. (\ref{sless}) below).

The integrals we need can easily be brought into the form of eq. (\ref{ibf}). For notational simplicity, we will replace $\frac{d^2}{(2\p\a')^2}$ with just $d^2$. We begin with the $d^2 - c^2 < 0$ case. The integral needed for $y(r)$ and $A_t(r)$ is
\beq
I_<(r) = \int_{r_0}^{r} dr \left ( \N^2 r^6 - (c^2 - d^2) \right)^{-1/2},
\eeq
with $r_0 = \N^{-1/3} \left( c^2 - d^2 \right)^{1/6}$. We change variables to $t = r_0^6 / r^6$,
\bea
I_<(r) & = & \frac{1}{6} \N^{-1/3} (c^2 - d^2)^{-1/3} \int_{t}^{1} dt (1-t)^{-1/2} t^{-2/3} \\ & = & \frac{1}{6} \N^{-1/3} (c^2 - d^2)^{-1/3} \left ( B\left( \frac{1}{3},\frac{1}{2} \right) -B\left( \frac{r_0^6}{r^6}; \frac{1}{3},\frac{1}{2} \right) \right ).\nonumber
\eea
The same change of variables works for the regulated on-shell action,
\bea
\label{sless}
S_< & = & - \frac{1}{6} \N r_0^4 \int_{r_0^6/\Lambda^6}^1 dt (1-t)^{-1/2} t^{-5/3} = - \frac{1}{6} \N r_0^4 \left ( B\left(-\frac{2}{3},\frac{1}{2} \right) - B\left(\frac{r_0^6}{\Lambda^6};-\frac{2}{3},\frac{1}{2}\right) \right ) \\ & = & - \frac{1}{6} \N r_0^4 B\left(-\frac{2}{3},\frac{1}{2} \right) - \frac{1}{4} \N \Lambda^4 + O(\Lambda^{-2}) \nonumber
\eea

Now we turn to the $d^2 - c^2 > 0$ case of section \ref{zerotsol}. The integrals here are also relevant for section \ref{zeromsol}. The integral needed for $y(r)$ and $A_t(r)$ (and of the same form as the integral for $A_t(r_6)$ in section \ref{zeromsol}) is
\beq
I_>(r) =\int_0^{r} dr \left ( \N^2 r^6 + d^2 - c^2 \right)^{-1/2}.
\eeq
We change variables to $u = \frac{\N^2 r^6}{d^2 - c^2}$,
\bea
I_>(r) & = & \frac{1}{6}\N^{-1/3} \left( d^2 - c^2 \right)^{-1/3} \int_{0}^{u} du (1+u)^{-1/2} u^{-5/6} \\ & = & \frac{1}{6} \N^{-1/3} \left( d^2 - c^2 \right)^{-1/3} B\left( \frac{\N^2 r^6}{\N^2 r^6 + d^2 - c^2}; \frac{1}{6},\frac{1}{3} \right). \nonumber
\eea
The integral for the regulated on-shell action (also relevant for the on-shell action in section \ref{zeromsol}) is, with $\Lambda ' = \frac{\N^2 \Lambda^6}{d^2 - c^2}$
\bea
S_> & = &- \frac{1}{6} \N^{-1/3} (d^2 - c^2)^{2/3} \int_0^{\Lambda'} du (1+u)^{-1/2} u^{1/6} \\ & = & - \frac{1}{6} \N^{-1/3} (d^2 - c^2)^{2/3} \left( B \left ( \frac{7}{6}, - \frac{2}{3} \right ) - B \left( \frac{1}{\Lambda' + 1}; - \frac{2}{3}, \frac{7}{6} \right ) \right ) \nonumber \\ & = & \nonumber - \frac{1}{6} \N^{-1/3} (d^2 - c^2)^{2/3} B \left ( \frac{7}{6}, - \frac{2}{3} \right ) - \frac{1}{4} \N \Lambda^4 +O(\Lambda^{-2}) \nonumber
\eea

%
%
%
%
%
%
%
%
%
%
%
%
%
%
%
\chapter {Transport of Flavor Fields}\label{transport}

Having seen some of the equilibrium thermodynamics of flavor fields, we are now ready to study the transport of flavor fields, that is, we are ready to study perturbations about equilibrium. Such perturbations are characterized by transport coefficients, such as the shear and bulk viscosity, diffusion constants, and conductivities. In this chapter we will use the supergravity description of flavor fields to compute their conductivity tensor. We first introduce electric and magnetic fields, and then use the D7-brane description to compute the resulting currents. From Ohm's law, we can then simply read off the conductivity. As an added bonus, we can compute the drag force experienced by the flavor fields as they are pushed through the $\N=4$ SYM plasma, or more accurately the drag coefficient, in the limit of large hypermultiplet mass $m$, where the flavor fields behave as quasi-particles.

Most of the material in this chapter comes from refs. \cite{Karch:2007pd,OBannon:2007in}.
\section{Preliminaries}
\label{prelims}
\setcounter{equation}{0}
\renewcommand{\theequation}{\thesection.\arabic{equation}}

To start, we recall some basic definitions. The conductivity tensor $\s_{ij}$ measures the response of a conducting medium to externally applied fields. It is defined by
\beq
\< J_i \> = \s_{ij} E_j \nonumber
\eeq
where $E_j$ are externally applied electric fields and $\< J_i \>$ are the currents induced in the medium. An external magnetic field $B$ produces off-diagonal elements in $\s_{ij}$: the induced current is perpendicular to both $E$ and $B$. This is the Hall effect. For a rotationally-invariant system with $E$ in the $x$ direction and $B$ perpendicular to the $xy$ plane, $\s_{xx} = \s_{yy}$ and $\s_{xy} = -\s_{yx}$. The component $\s_{xx}$ is called the Ohmic conductivity and $\s_{xy}$ the Hall conductivity.

We will now review two results from classical electromagnetism that we will reproduce from our supergravity calculation in appropriate limits. These will provide some nice checks of our supergravity answer for $\sigma_{ij}$.

Imagine filling the vacuum with a charge density $\< J^t \>$. In the lab frame we may introduce a magnetic field $\vec{B}$. In a frame moving with velocity $-\vec{v}$ relative to the lab frame we will find a current $\vec{J} = \< J^t \> \vec{v}$ and an electric field
\beq
\vec{E} = - \vec{v} \times \vec{B} = - \frac{1}{\< J^t \>} \vec{J} \times \vec{B}.
\eeq
If we take $\vec{B} = (0,0,B)$ we find the conductivity
\beq
\label{zerotempconductivity}
\s_{xx} = 0, \qquad \s_{xy} = \< J^t \> / B.
\eeq
Notice that this argument does not require that the charge density be comprised of quasi-particle charge carriers. Indeed, this argument relies only on Lorentz invariance.

Now imagine a density $\< J^t \>$ of massive quasi-particles propagating non-relativistically through an isotropic, homogeneous, neutral medium. In the rest frame of the medium we introduce an electric field $E$ in the $\hat{x}$ direction in addition to the magnetic field. The force on a quasi-particle is then
\beq
\frac{d \vec{p}}{dt} = \vec{E} + \vec{v} \times \vec{B} - \m_d \vec{p},
\eeq
where our quasi-particle has charge $+1$ and $\m_d$ is the drag coefficient (hence the subscript, to distinguish it from the chemical potential). We replace the momentum with the velocity using $\vec{p} = M \vec{v}$ for quasi-particle mass $M$. We then replace the velocity with the induced current using $\vec{v} = \< \vec{J} \>/\< J^t \>$. Imposing the steady-state condition $\frac{d \vec{p}}{dt} = 0$ and solving for $\< \vec{J}\>$ yields
\beq
\label{drude}
\s_{xx} = \frac{\s_0}{(B / \m_d M)^2 + 1}, \qquad \s_{xy} = \frac{\s_0 (B / \m_d M)}{(B / \m_d M)^2 +1}
\eeq
where $\s_0 = \< J^t \> / \m_d M$ is the conductivity when $B=0$, and is also known as the Drude form of the conductivity, although Drude would have called $\< J^t \>$ the number density of electrons and $M$ the electron mass.

\section{The Physics of Flavor Transport}
\label{transportphysics}

We want to study the transport properties of a number $N_f \ll N_c$ of $\N=2$ hypermultiplet fields propagating in an $\N=4$ SYM plasma, at large-$N_c$ and large 't Hooft coupling $\lambda$. We will introduce a finite density $\< J^t \>$ of flavor fields, and electric and magnetic fields $E$ and $B$ that couple to $U(1)_B$ charge. We will thus think of the density $\<J^t\>$ as a number density of \textit{charge carriers}. We will define the baryon charge of the flavor fields to be $+1$.

The physical picture of what will happen in the presence of nonzero $E$ is fairly intuitive. The $\N=4$ SYM fields will not ``feel'' the electric field since they do not carry $U(1)_B$ charge. The $\N=2$ hypermultiplet fields, however, will be pushed by $E$. They will still be interacting with the $\N=4$ SYM fields, of course, with coupling strength $\lambda$. The $\N=4$ SYM plasma will thus provide a \textit{drag force}, as the charge carriers must push through the plasma. We may hope that when the force due to the electric field balances the drag force we will find a steady-state current $\<J^x\>$. With nonzero $B$, we also expect a current $\< J^y \>$ from the Hall effect. Our goal is to compute these currents, using the supergravity description, and extract from them the components of the conductivity tensor.

Ultimately, the simple physical picture described above will turn out to be correct, but the reasons why this is so are actually quite subtle. To explain why, let us consider for a moment the same system, but with $N_f$ on the order of $N_c$. In this case, we do \textit{not} expect a finite answer for $\sigma_{ij}$. Why not? The two crucial features of the system are:

1.) The system has a finite charge density in a constant electric field, and

2.) The system is translationally invariant.

The first point tells us that the electric field will be doing \textit{net work} on the system. The system will thus be gaining energy at a constant rate as time passes. The second point tells us that momentum will be conserved in the system. As the electric field does work on the charge carriers, they will gain energy and momentum. If momentum is conserved, the charge carriers will transfer momentum to the $\N=4$ SYM plasma. Put more simply, the charge carriers will begin to drag the plasma along with them. Our constant electric field will continue to do work on the system, dumping in more and more energy, so that over time the entire system will move faster and faster, without bound. The end result is definitely \textit{not} a steady state with constant currents $\< J^x\>$ and $\<J^y\>$. Instead, we end up with a plasma moving infinitely fast! In a word, the problem is \textit{dissipation}, or rather, the lack thereof: our system seems to have no mechanism by which the charge carriers can dissipate momentum, which is necessary to produce a finite conductivity.

To address this problem, we could use an electric field with harmonic time dependence, $E(t) \sim e^{i \omega t}$, in which case the electric field will do no \textit{net} work on the system. Indeed, other gauge-gravity systems have been studied with harmonic fields, giving rise to finite, frequency-dependent conductivities \cite{Herzog:2007ij,Hartnoll:2007ai,Hartnoll:2007ih,Hartnoll:2007ip}. We will work with constant $E$, however.

We could also try to break translation invariance, as occurs in real materials. A crude picture of a metal, for example, is a gas of conduction electrons propagating through a lattice of ions, where the lattice clearly breaks translation invariance. The system then does not conserve momentum, and the conduction electrons can experience dissipation. Impurity potentials that break translation invariance have been introduced in gauge-gravity systems, giving rise to dissipation and hence finite conductivity, even in a constant external electric field \cite{Hartnoll:2008hs}. We will not break translation invariance, however.

How on earth can we find a finite conductivity, then? The answer comes from the probe limit, in which we can basically \textit{fake} dissipation. In the probe limit, which in particular means $N_f \ll N_c$, the stress-energy tensor of the system will cleanly separate into two terms, one of order $N_c^2$, coming from the adjoint fields, and one of order $N_f N_c$, coming from the flavor fields. We depict this schematically as
\beq
T_{\mu \nu} \, \sim \, \O(N_c^2)_{\mu \nu} \, + \, \O(N_f N_c)_{\mu \nu}.
\eeq
The flavor fields may then transfer momentum to the $\N=4$ SYM fields at a constant rate for a \textit{time} that is parametrically large in $N_c$, that is, a time on the order of $N_c$. At earlier times, the velocity of the $\N=4$ SYM plasma will be negligible. Only at very \textit{late} times will the flavor fields have transferred enough momentum to the plasma for the velocity of the plasma to be non-negligible. For any finite time, then, the $\N=4$ SYM plasma basically acts as an energy-momentum reservoir for the flavor fields. In simple terms, all we are saying is that because the system has \textit{so few} flavor fields, we can \textit{neglect} their effect on the plasma, at least for a very long time.

In the end, then, we will find a finite conductivity, even with finite charge density, constant electric field, and translation invariance, because we work in the probe limit. We should always bear in mind, however, that our result will not be valid for all time.

If we take $m \ra \infty$, making the flavor fields very heavy, then we expect the charge carriers to behave as quasi-particles: as $m \ra \infty$, the Compton wavelength of the flavor excitations will become arbitraily shorter than their mean free path, so that they will behave essentially as isolated particles. In other words, the charge carriers will be very heavy probes lumbering through the plasma, their dynamics essentially classical.

In the $m \ra \infty$ limit, then, the quasi-particles will obey a classical force law,
\beq
\label{classicalforcelaw}
\frac{d\vec{p}}{dt} \, = \, -\mu_d \, \vec{p} \, + \, \vec{E} \, + \, \vec{v} \times \vec{B}
\eeq
where $\mu_d$ is a drag coefficient and $\vec{p}$ is the momentum of the quasi-particles. Here we see explicitly the drag force due to the $\N=4$ SYM plasma and the Lorentz force due to the external fields.

We must actually be more precise with what we mean by ``$m \ra \infty$.'' To explain why, let us consider a single quark, rather than a density $\<J^t \>$ of quarks (and squarks). A single quark is represented in the supergravity description as a single string stretched from the endpoint of a Minkowski-embedded D7-brane to the AdS-Schwarzschild horizon. The rest mass of the quark is given by the length of the string times the string tension. The rest mass thus differs from the Lagrangian mass $m$. Defining $\Delta m \equiv \frac{1}{2} \, \sqrt{\lambda}T$, the rest mass $M_{rest}$ is, for $m \gg \Delta m$ \cite{Herzog:2006gh},
\beq
M_{rest} = m - \Delta m + O\left( \Delta m^2/m \right)
\eeq
For arbitrary $m$, the rest mass also differs from the kinetic mass, $M_{kin}$, which will be the mass that appears in the force law above. More specifically, $M_{kin}$ is the mass we must use if we write $\vec{p}$ in terms of $\vec{v}$. (Notice we used the kinetic mass $M$ in section \ref{prelims}, around eq. (\ref{drude}).) If $m \gg \Delta m$, then the kinetic mass and thermal rest mass only differ by \cite{Herzog:2006gh}
\beq
M_{kin} = M_{rest} + O(\Delta m^2 / m)
\eeq
Our point is that the large-mass limit $m \ra \infty$ actually means not just $m \gg T$, but $m \gg \Delta m$. In this limit, we will drop all corrections and equate $M_{kin} = M_{rest} = m$.

Analysis of single-string solutions \cite{Herzog:2006gh} revealed that in this limit the correct relation between velocity and momentum is relativistic, that is,
\beq
\label{relativistic}
\vec{p} = \g M_{kin} \vec{v} = \frac{m \vec{v}}{\sqrt{1-v^2}}
\eeq
with $\g$ the usual relativistic factor $\g = \frac{1}{\sqrt{1-v^2}}$. In the second equality we have invoked the $m \ra \infty$ limit and replaced $M_{kin}$ with $m$. For a single string moving under the influence of the D7-brane electric field, in the $m \gg \Delta m$ limit, the result for the drag coeffcient is \cite{Herzog:2006gh,Gubser:2006bz}
\beq
\mu_d \, m = \frac{\pi}{2} \sqrt{\lambda} \, T^2
\eeq

We will study a finite density $\< J^t \>$ of strings, with $\<J^t \>$ on the order of $N_f N_c$, in the presence of external $E$ and $B$ fields, and calculate $\mu_d \, m$ in the $m \ra \infty$ limit. Using the relativistic relation eq. (\ref{relativistic}), we will find \textit{exactly the same} result as the single-string calculation. The reason why we find the same answer comes from the probe limit.

In the probe limit, the plasma contains order $N_c^2$ adjoint degrees of freedom and order $N_f N_c \ll N_c^2$ flavor degrees of freedom. The flavor excitations are thus dilute in the large-$N_c$ limit. In a perturbative analysis, the flavor excitations will be more likely to scatter off of adjoint degrees of freedom than off of other flavor excitations. Scatterings with adjoint degrees of freedom will thus be the flavor excitations' primary mechanism for the microscopic energy loss that results in the macroscopic drag force. Introducing a density $\<J^t\>$ of order $N_f N_c$ will not change this to leading order in large-$N_c$. Increasing the stength of the coupling muddies the picture of isolated scatterings but does not affect the argument, which relies only on large-$N_c$ counting. Taking $m \ra \infty$, and in particular $m \gg \m$ (the chemical potential, not the drag coefficient), serves only to dilute the charge carriers further. We therefore \textit{expect} to recover the zero-density result at leading order in the $N_f \ll N_c$ limit.

The $E$ and $B$ independence follows from this, simply because the zero-density result $\frac{\p}{2} \sqrt{\lam} T^2$ was already, curiously, independent of the quasi-particle momentum, or equivalently of $m$ and $v$ \cite{Herzog:2006gh,Gubser:2006bz}. As $v$ is determined by $E$ and $B$, and is the only place where $E$ and $B$ could appear in the answer, we expect the answer to be independent of $E$ and $B$.

The $m \ra \infty$ limit is also where we will reproduce the classical result eq. (\ref{drude}), which we derived assuming the charge carriers were quasi-particles. 

We now turn to the details of the supergravity calculation.

\section{The D7-brane Solution}
\label{eandbd7branesol}

In this section, we will present the solution for the D7-brane fields describing a field theory state with nonzero $m$, $T$, $\< J^t \>$, $E$ and $B$.  

We will use an AdS-Schwarzschild metric in Fefferman-Graham coordinates, as in eq. (\ref{adsbhmetric}),
\beq
\label{transportmetric}
ds^2 = \frac{dz^2}{z^2} - \frac{1}{z^2} \frac{(1 - z^4 / z_H^4)^2}{1+z^4/ z_H^4} dt^2 + \frac{1}{z^2} (1+z^4 / z_H^4) d\vec{x}^2
\eeq
Recall that the boundary is at $z = 0$ and the black hole horizon is at $z = z_H$ with $z_H^{-1} = \frac{\p}{\sqrt{2}} T$. We will not actually need an explicit form for the metric in most of what follows, since most of our arguments are based on general properties of AdS-Schwarzschild, in particular that $g_{tt}$ vanishes at the horizon and that the metric has a second-order pole at the boundary. We will use an $S^5$ metric
\beq
ds^2_{S^5} \, = \, d\theta^2 \, + \, \sin^{2}\theta \, ds_{S^1}^2 \, + \, \cos^{2}\theta \, ds_{S^3}^2.
\eeq
The embedding of the D7-brane will be described by $\theta(z)$. We will not explicitly solve for $\th(z)$, but we will consider limits. We will be considering only black hole embeddings (as we want finite density in the field theory), so $m=0$ is represented by $\th(z)=0$. When $m \ra \infty$ we expect the D7-brane to form a ``spike,'' as explained in section \ref{bhembeddingsallowed}. In this case, we may approximate $\th(z) \approx \pi/2$ or $\cos \th(z) \approx 0$, at least for $z$ along the spike. In particular, we will make this approximation for $z$ near the horizon.

In the field theory, we want a density $\< J^t \>$, external fields $E$ and $B$, and the resulting induced currents $\< J^x \>$ and $\< J^y \>$. As explained in sections \ref{electric} and \ref{magnetic}, we thus introduce worldvolume gauge field components $A_t(z)$ and
\beq
A_x(z,t) = - E t + f_x(z), \qquad A_y(z,x) = B x + f_y(z)
\eeq
so that at the boundary we have electric and magnetic fields $F^{tx} = E$ and $F^{xy} = B$. As part of our gauge choice we take $A_z = 0$.

As our gauge fields only depend on $(z,t,x,y)$, the D7-brane action is simply a (3+1)-dimensional Born-Infeld action, with some ``extra'' factors in front from the $S^3$ and the extra spatial direction, which may be written as
\beq
S = - \N \int d^4x \cos^3\th g_{xx}^{1/2} \sqrt{- g - (2 \p \a')^2 \frac{1}{2} g F^2 - (2 \p \a')^4 \frac{1}{4} \left(  F \wedge F \right)^2}
\label{originaldbi}
\eeq
In contrast to our previous definition of $S$, here we define $S$ as $S_{D7}$ divided by a single factor of the volume of $\mathbb{R}$. We have also defined $d^4x = dz\,dt\,dx\,dy$, and defined $g=g_{zz} \, g_{tt} \, g_{xx}^2$ as the determinant of the induced metric in the $(z,t,x,y)$ subspace, with $g_{zz} = 1/z^2 + \th'(z)^2$. Writing $F^2 = F^{\m\n} F_{\m\n}$, where Greek indices run over $(z,t,x,y)$, and $\tilde{F}^{\m\n} = \frac{1}{2} \e^{\m\n\a\b} F_{\a \b}$ for totally antisymmetric $\e^{\m\n\a\b}$ with $\e^{ztxy} = +1$, we have explicitly
\begin{subequations}
\label{gaugeadef}
\beq
\frac{1}{2} g F^2 = g_{xx}^2 A_t'^2 + g_{tt} g_{xx} A_x'^2 + g_{tt} g_{xx} A_y'^2 + g_{zz} g_{xx} \dot{A}_x^2 + g_{zz} g_{tt} \bar{A}_y^2
\label{a2def}
\eeq
\beq
\frac{1}{4} \left( F \wedge F\right)^2 = \left( \frac{1}{4} \tilde{F}^{\m\n} F_{\m\n} \right)^2 = \bar{A}_y^2 A_t'^2 + \dot{A}_x^2 A_y'^2 + 2 \bar{A}_y A_t' \dot{A}_x A_y'.
\label{a4def}
\eeq
\end{subequations}
where dots, $\dot{A}$, denote derivatives with respect to $t$, primes, $A'$, denote derivatives with respect to $z$, and bars, $\bar{A}$, denote derivatives with respect to $x$.

The action only depends on the derivatives of $A_t(z)$, $f_x(z)$ and $f_y(z)$, so we will have three conserved charges. In the Appendix to this chapter we identify these as $\< J^t \>$, $\<J^x\>$ and $\<J^y \>$,
\begin{subequations}
\label{jdef}
\beq
\N (2 \p \a')^2 g_{xx}^{1/2} \cos^3 \th \frac{-g_{xx}^2 A_t' -(2 \p \a')^2 (\bar{A}_y^2 A_t' + \bar{A}_y \dot{A}_x A_y')}{ \sqrt{- g - (2 \p \a')^2 \frac{1}{2} g F^2 - (2 \p \a')^4 \frac{1}{4} \left( F \wedge F \right)^2} } = \< J^t \>
\label{jtdef}
\eeq
\beq
\N (2 \p \a')^2 g_{xx}^{1/2} \cos^3 \th \frac{|g_{tt}| g_{xx} A_x'}{\sqrt{- g - (2 \p \a')^2 \frac{1}{2} g F^2 - (2 \p \a')^4 \frac{1}{4} \left( F \wedge F \right)^2}} = \< J^x \>
\label{jxdef}
\eeq
\beq
\N (2 \p \a')^2 g_{xx}^{1/2} \cos^3 \th \frac{|g_{tt}| g_{xx} A_y' - (2\p\a')^2 (\dot{A}_x^2 A_y' + \bar{A}_y \dot{A}_x A_t')}{\sqrt{- g - (2 \p \a')^2 \frac{1}{2} g F^2 - (2 \p \a')^4 \frac{1}{4} \left( F \wedge F \right)^2}} = \< J^y \>
\label{jydef}
\eeq
\end{subequations}

With a little algebra we solve for the gauge fields from eq. (\ref{jdef}),
\beq
A_t'(z) = - \frac{\sqrt{g_{zz} |g_{tt}|}}{g_{xx}} \frac{\<J^t\> \xi - B a}{\sqrt{\xi \chi - a^2}}
\label{atsol}
\eeq
where we have introduced the coefficients
\begin{subequations}
\label{coeffdefs}
\bea
\xi & = & |g_{tt}| g_{xx}^2 - (2\p\a')^2 \tilde{F}^{z\m} \tilde{F}^{z}_{\m} \nonumber \\[6pt] & = & |g_{tt}| g_{xx}^2 + (2\p\a')^2 \left ( |g_{tt}| B^2 - g_{xx} E^2 \right )
\label{xidef}
\eea
\bea
\chi & = & |g_{tt}| g_{xx}^3 \N^2 (2\p\a')^4 \cos^6 \th - (2\p\a')^2 \< J_{\m} \> \<J^{\m}\>
\label{chidef} \nonumber \\[6pt] & = & |g_{tt}| g_{xx}^3 \N^2 (2\p\a')^4 \cos^6 \th  + (2\p\a')^2 \left ( |g_{tt}|\<J^t\>^2 - g_{xx} \left(  \<J^x\>^2 + \<J^y\>^2 \right) \right )
\label{chidef}
\eea
\bea
a & = & -(2\p\a')^2 \tilde{F}^{z\m} \< J_{\m}\> \nonumber \\[6pt] & = & (2\p\a')^2 (|g_{tt}| \<J^t\> B + g_{xx} \<J^y\> E)
\label{adef}
\eea
\end{subequations}
Notice that $\xi$ is simply $- det(g_{ab} +(2\p\a')F_{ab})$ in the $(t,x,y)$ subspace, and that $\cos \th(z)$ appears only in $\chi$. We also have
\beq
A_x'(z) = \sqrt{\frac{g_{zz}}{|g_{tt}|}} \frac{\< J^x \> \xi}{\sqrt{\xi \chi - a^2}}, \qquad A_y'(z) =  \sqrt{\frac{g_{zz}}{|g_{tt}|}} \frac{\< J^y \> \xi + E a}{\sqrt{\xi \chi - a^2}}
\label{axaysol}
\eeq
In the original action we may now replace the gauge fields with the conserved charges. The resulting effective action has only the single dynamical field $\th(z)$,
\beq
S = - \N^2 (2\p\a')^2 \int d^4x \cos^6\th g_{xx}^2 \sqrt{g_{zz} |g_{tt}|} \frac{\xi}{\sqrt{\xi \chi - a^2}}
\label{effaction}
\eeq
We may obtain the equation of motion for $\th(z)$ in two ways. We may derive it from the original action eq. (\ref{originaldbi}) and then plug in our gauge field solutions eqs. (\ref{atsol}) and (\ref{axaysol}), or we may Legendre transform to eliminate the gauge fields at the level of the action. The Legendre-transformed action $\hat{S}$ is
\bea
\label{lteffaction}
\hat{S} & = & S - \int d^4x \left ( F_{zt} \frac{\d S}{\d F_{zt}} + F_{zx} \frac{\d S}{\d F_{zx}} + F_{zy} \frac{\d S}{\d F_{zy}} \right ) \\ & = & - \frac{1}{(2\p\a')^2} \int d^4x g_{zz}^{1/2} |g_{tt}|^{-1/2} g_{xx}^{-1} \sqrt{\xi \chi - a^2} \nonumber
\eea
where $\frac{\d \hat{S}}{\d \<J^t \>} = A_t'(z)$, $\frac{\d \hat{S}}{\d \< J^x \>} =  A_x'(z)$ and $\frac{\d \hat{S}}{\d \< J^y \>} = A_y'(z)$ reproduce eqs. (\ref{atsol}) and (\ref{axaysol}).

Specifying the boundary conditions will then determine the D7-brane solution completely. At the horizon, $A_t(z)$ must of course obey $A_t(z_H) = 0$. We are free to choose the leading asymptotic values of the fields near the boundary $z \ra 0$. We first choose the asymptotic value $\th_0$ of $\th(z)$.  The gauge fields asymptotically approach the boundary as
\begin{subequations}
\label{asymptotic}
\beq
A_t(z) = \m - \frac{1}{2} \frac{\< J^t \>}{\N (2\p \a')^2} z^2 + O(z^4)
\eeq
\beq
A_x(z) = -E t + c_x + \frac{1}{2} \frac{\< J^x \>}{\N (2 \p \a')^2} z^2 + O(z^4)
\eeq
\beq
A_y(z) = B x + c_y + \frac{1}{2} \frac{\< J^y \>}{\N (2 \p \a')^2} z^2 + O(z^4)
\eeq
\end{subequations}
where $\m$, $c_x$ and $c_y$ are constants of integration. The leading asymptotic value $\m$ is the $U(1)_B$ chemical potential. For $A_x$ and $A_y$ we impose the boundary condition $c_x = c_y = 0$ because we do not want to source the corresponding operators in the field theory Lagrangian. In what follows, we will not actually make use of this boundary condition, however, since we will only work with the derivatives (or field strengths) $A_x'(z)$ and $A_y'(z)$.

\section{The Conductivity Tensor}

We focus now on the quantity $\sqrt{\xi \chi - a^2}$ appearing in the effective action eq. (\ref{effaction}). We will find that demanding reality of the effective action allows us to solve for $\langle J^x \rangle$ and $\langle J^y \rangle$, and hence the conductivity, in terms of $E$, $B$ and $\< J^t\>$. The action must be real: an imaginary action signals an instability of the solution.

In eq. (\ref{xidef}) we see that, as a function of $z$, $\xi$ has a zero: $\xi < 0$ at the horizon where $|g_{tt}|=0$, whereas $\xi >0$ near the boundary $z\ra0$. We denote the zero of $\xi$ as $z_*$, for which we can solve (here we must use the explicit form of the metric in eq. (\ref{transportmetric})),
\bea
\label{zstardef}
\frac{z_*^4}{z_H^4} & = & e^2 - b^2 + \sqrt{(e^2 - b^2)^2 + 2 (e^2 + b^2) + 1}\\ & & - \sqrt{ \left ( (e^2 - b^2) + \sqrt{(e^2 - b^2)^2 + 2 (e^2 + b^2) + 1} \right )^2 - 1}  \nonumber
\eea
where we have defined the dimensionless quantities
\beq
e = \frac{1}{2} (2\p\a') E z_H^2 = \frac{E}{\frac{\p}{2}\sqrt{\lam}T^2}, \qquad b = \frac{1}{2}(2\p\a')B z_H^2 = \frac{B}{\frac{\p}{2} \sqrt{\lam}T^2}
\eeq
and converted to field theory quantities. Knowing that $\xi$ is the $(t,x,y)$ part of $- det(g_{ab} +(2\p\a')F_{ab})$, we will interpret $z_*$ as an effective horizon on the D7-brane worldvolume, or in other words, as the location in $z$ where the determinant of the D7-brane action vanishes. Notice also that when $E=0$ we find $z_* = z_H$, so in this case the effective horizon coincides with the actual horizon of the background geometry. As we increase $E$, then, we should think of $z_*$ as starting at the horizon and then moving up the D7-brane, toward the boundary.

We will also need $g_{xx}^2(z_*) = \p^4 T^4 {\cal F}(e,b)$ where
\beq
{\cal F}(e,b) = \frac{1}{2} \left ( 1+ e^2 - b^2 + \sqrt{(e^2 - b^2)^2 + 2 (e^2 + b^2) + 1}\right )
\eeq
For later use notice that ${\cal F}(0,b) = 1$.

In fact all three functions, $\xi$, $\chi$ and $a$ must share the same zero $z_*$. From eq. (\ref{chidef}) we see that at the horizon $\chi < 0$ while at the boundary $\chi > 0$, so $\chi$ also has a zero. In particular $\xi \chi > 0$ at the horizon and at the boundary. If $\xi$ and $\chi$ have distinct zeroes, then in the region between those zeroes one would change sign while the other would not, hence in that region $\xi \chi < 0$ and the effective action would be imaginary. The only consistent possibility is for $\xi$ and $\chi$ to share the zero at $z_*$. We must also have $a^2 < \xi \chi \ra 0$ as $z \ra z_*$, so that $a \ra 0$ at $z_*$ as well.

We thus set all of eqs. (\ref{coeffdefs}) to zero at $z_*$ and solve for $\< J^x \>$ and $\< J^y \>$,
\begin{subequations}
\label{cxcy}
\beq
\label{cx}
\< J^x \> = \frac{E g_{xx}}{g_{xx}^2 + (2\p\a')^2 B^2} \sqrt{(g_{xx}^2 + (2\p\a')^2 B^2) \N^2 (2\p\a')^4 g_{xx} \cos^6 \th(z_*) + (2\p\a')^2 \<J^t\>^2}
\eeq
\beq
\label{cy}
\< J^y \> = -\frac{(2\p\a')^2 \<J^t\>B}{g_{xx}^2 + (2\p\a')^2 B^2} E
\eeq
\end{subequations}
with all functions of $z$ evaluated at $z_*$. Converting to field theory quantities, we find
\begin{subequations}
\label{d7sigma}
\beq
\s_{xx} = \sqrt{\frac{N_f^2 N_c^2 T^2}{16 \p^2} \frac{{\cal F}^{3/2}}{b^2 + {\cal F}}\cos^6 \th(z_*) + \frac{\rho^2 {\cal F}}{(b^2 + {\cal F})^2}}
\eeq
\beq
\s_{xy} = \frac{\rho b}{b^2 + {\cal F}}
\eeq
\end{subequations}
where we have defined $\rho$ similarly to $e$ and $b$,
\beq
\rho = \frac{\< J^t \>}{\frac{\p}{2} \sqrt{\lam}T^2}
\eeq
but while $e$ and $b$ are dimensionless, $\rho$ has dimension one.

We can interpret our result as follows. Two types of charge carriers contribute to the conductivity. The first are the charge carriers we have introduced explicitly in $\rho$. Taking $\rho=0$ leaves a nonzero $\s_{xx}$, however, so we must have another source of charge carriers. We will guess that these come from pair production in the plasma: although the net charge $\rho$ is zero, the flavor fields are still part of the plasma, just with equal numbers of particles and anti-particles. When we introduce the electric field $E$, the particles go one way and the anti-particles go the other way, producing a net current. Notice that the hypermultiplet mass $m$ is hiding in $\cos \th(z_*)$, whose behavior is consistent with our interpretation: $\cos \th(z_*) \ra 0$ as $m \ra \infty$, and $\cos \th(z_*) = 1$ for $m=0$.

Notice also that our answer depends on the electric field $E$. We are therefore capturing effects beyond linear response: in linear response theory, the conductivity would be constant (as the current would be \textit{linear} in the electric field: $\< J^x \> = \s_{xx} E$, etc.). We can attribute the nonlinearities of our answer to the fact that we started with a nonlinear action, the Born-Infeld action. As explained in section \ref{dbranes}, however, we can linearize the Born-Infeld action, that is, we can Taylor expand in the gauge field and recover the Maxwell action as the leading term. We can thus recover the linear-response answer by linearizing in $E$, which in practical terms means simply setting $E=0$ in our result.

We will now check our answer in two limits, to reproduce the forms in section \ref{prelims}.

To recover eq. (\ref{zerotempconductivity}), we must linearize in the electric field, setting $e=0$, and hence ${\cal F}(0,b) = 1$, in eq. (\ref{d7sigma}). We also restore Lorentz invariance by taking $T \ra 0$. We find $\s_{xx} = 0$ and $\s_{xy} = \< J^t \>/B$, as expected.

To recover eq. (\ref{drude}), we return to finite $T$ and again linearize in the electric field. We additionally take the $m \ra \infty$ limit $\cos \th(z_*) \approx 0$. The conductivity becomes
\beq
\s_{xx} = \frac{\rho}{b^2 + 1}, \qquad \s_{xy} = \frac{\rho b}{b^2 + 1}.
\label{linearlargemass}
\eeq
As shown in section \ref{drag}, in the $m \ra \infty$ limit we identify $\frac{\p}{2} \sqrt{\lam} T^2 = \m_d \, m$, where $\mu_d$ is the drag coefficient. We thus have $\rho = \frac{\< J^t \>}{\m_d \, m}$ and $b = \frac{B}{\m_d \, m}$, so the conductivity indeed has the form expected for quasi-particles propagating through an isotropic, homogeneous medium, eq. (\ref{drude}).

We end with a question: where in the phase diagram of the field theory is our result valid? The short answer is in supergravity language. Our argument relied on the fact that the D7-brane touches the AdS-Schwarzschild horizon, such that the quantities under the square root in the D7-brane action change sign at some $z_*$. Our solution should thus be valid whenever the ground state of the system is a black hole embedding of the D7-brane. Notice in particular that our answer will be valid at zero density, but only in the high-temperature, melted-meson phase, described by black hole embeddings.

\section{The Drag Force}
\label{drag}

In this section we compute the drag force on the charge carriers, in the large-mass limit. In particular, we will compute the product $\mu_d \, m$.

In the $m \ra \infty$ limit where $\cos \th \approx 0$, we expect the flavor excitations to be well-described as a collection of quasi-particles, which will have an equation of motion
\beq
\frac{d \vec{p}}{dt} =  - \m_d \, \vec{p} + \vec{E} + \vec{v} \times \vec{B},
\eeq
with $v$ is the quasi-particle velocity and $\m_d$ is the drag coefficient. Our first goal is to compute the magnitude of the drag force, $\m_d |\vec{p}|$. In the steady-state, $\frac{dp}{dt} = 0$, which implies
\beq
\m_d |\vec{p}| = \sqrt{E^2 + v^2 B^2 + 2 \vec{E}\cdot (\vec{v} \times \vec{B}) }
\eeq
As $m\ra\infty$, we expect pair creation to be suppressed, so only the charge carriers in $\langle J^t \rangle$ should contribute to $\langle \vec{J} \rangle$, hence $\langle \vec{J} \rangle = \langle J^t \rangle \vec{v}$. We immediately read off $v^2 = |g_{tt}|/g_{xx}$ by setting $\chi$ to zero at $z_*$ and dropping the $\cos \th(z_*)$ term. Setting $\xi = 0$ at $z_*$ gives us
\beq
E^2 = \frac{1}{(2\p\a')^2} |g_{tt}|g_{xx} + \frac{|g_{tt}|}{g_{xx}} B^2 = \frac{1}{(2\p\a')^2} g_{xx}^2 v^2 + v^2 B^2,
\eeq
Setting $a=0$ at $z_*$ gives us the component of $\vec{v}$ in the $\hat{y}$ direction,
\beq
v_y = \frac{\<J^y \>}{\< J^t\>} = - \frac{|g_{tt}|}{g_{xx}} \frac{B}{E} = - v^2 \frac{B}{E}.
\eeq
We then have $2\vec{E}\cdot (\vec{v} \times \vec{B}) = 2 E B v_y = - 2 B^2 v^2$. The drag force is then
\beq
\m_d |\vec{p}| = \frac{1}{2\p\a'} g_{xx}(z_*) v
\label{d7force}
\eeq
We can now compute $\m_d \, m$. We employ the relativistic relation $|\vec{p}| = \gamma m v$ with $\gamma = \frac{1}{\sqrt{1-v^2}}$, and find
\beq
\m_d \, m = \frac{1}{2\p\a'} \sqrt{g_{xx}(z_*)^2-|g_{tt}(z_*)|g_{xx}(z_*)} = \frac{1}{\p\a'}z_H^{-2} = \frac{\p}{2} \sqrt{\lam} T^2
\label{mum}
\eeq
This result for $\mu_d \, m$ is identical to the drag force computed from a single-string solution in ref. \cite{Herzog:2006gh}, as advertised. In other words, the value of $\mu_d \, m$ for a single quark or squark, when $\<J^t\> = 0$ and $B=0$, is identical to eq. (\ref{mum}), as expected.

\section*{Appendix: Holographic Renormalization of Worldvolume Gauge Fields}
\setcounter{equation}{0}
\renewcommand{\theequation}{\arabic{equation}}

In this section we study the holographic renormalization of the D7-brane's worldvolume fields, in particular the gauge fields introduced in section \ref{eandbd7branesol}. Our main goal is to identify the constants of motion in eqs. (\ref{jdef}) as the expectation values $\< J^{\mu} \>$.

We find from its equation of motion that $\th(z)$ has the usual asymptotic expansion
\beq
\th(z) = \th_0 z + \th_2 z^3 + O(z^5).
\label{thetasymptotic}
\eeq
where the leading coefficient $\th_0$ is the source for the dual operator, \textit{i.e.} gives the hypermultiplet mass via $\th_0 = (2\p\a') m$.

Plugging eq. (\ref{thetasymptotic}) into the regulated action we find, with nonzero $E$ and $B$, the divergences
\beq
S_{reg} = - \int_{\e}^{z_H} dz L = - \N \int_{\e}^{z_H} dz \left ( z^{-5} - \th_0^2 z^{-3} + \frac{1}{2} (2 \p \a')^2 (B^2 - E^2) z^{-1} + O(z) \right )
\eeq
We again need the counterterms
\beq
L_1 = \frac{1}{4} \N \sqrt{-\g}, \qquad L_2 = - \frac{1}{2} \N \sqrt{-\g} \th(\e)^2, \qquad L_f = \N \frac{5}{12} \sqrt{-\g} \th(\e)^4
\eeq
with $\g_{ij}$ the induced metric at $z=\e$ and $\g$ its determinant. Notice that $\sqrt{-\g} = \e^{-4} + O(\e^4)$. We suppress $\int \, dt \, dx \, dy$ unless stated otherwise. The last divergence requires a new counterterm
\beq
L_F = - \frac{1}{4} \N (2 \p \a')^2 \sqrt{-\g} F^{ij} F_{ij} \log \e  = - \frac{1}{2} \N (2 \p \a')^2 (B^2 - E^2) \log \e + O(\e^4 \log \e)
\eeq
The generating functional of the field theory is then the $\e \ra 0$ limit of $S = S_{reg} + \sum_i L_i$. We want the expectation values $\langle J^t \rangle$, $\langle J^x \rangle$ and $\langle J^y \rangle$. In holo-rg, $\langle J^{\m} \rangle$ is
\beq
\langle J^{\m} \rangle = \lim_{\e \ra 0} \frac{1}{\e^4} \frac{1}{\sqrt{-\g}} \frac{\d S}{\d A_{\m}(\e)}
\eeq
For $\<J^t\>$, we need
\beq
\d S = - \int_{\e}^{z_H} dz \frac{\d L}{\d \partial_z A_t} \partial_z \d A_t = - \frac{\d L}{\d \partial_z A_t} \int_{\e}^{z_H} dz \partial_z \d A_t = - \frac{\d L}{\d \partial_z A_t} \left ( \d A_t(z_H) - \d A_t(\e) \right ),
\eeq
where we have used the fact that $\frac{\d L}{\d \partial_z A_t}$ is $z$-independent on-shell. Enforcing $\d A_t(z_H)=0$ we find $\frac{\d S}{\d A_t(\e)}= \frac{\d L}{\d \partial_z A_t}$ and hence $\langle J^t \rangle = \frac{\d L}{\d \partial_z A_t}$, as advertised.

For $\langle J^x \rangle$, we reinstate $\int dt$ because $A_x$ is time-dependent,
\beq
\d S = - \int dz dt \left ( \frac{\d L}{\d \partial_z A_x} \partial_z \d A_x + \frac{\d L}{\d \partial_t A_x} \partial_t \d A_x \right )
\eeq
We employ precisely the same argument as before for the first term. For the second term we observe that $\frac{\d L}{\d \partial_t A_x}$ is $t$-independent on-shell and hence
\beq
\int dt \frac{\d L}{\d \partial_t A_x} \partial_t \d A_x = \frac{\d L}{\d \partial_t A_x} \int dt \partial_t \d A_x = 0
\eeq
where we demand that the fluctuation be well-behaved (vanishing) at $t = \pm \infty$. The counterterm $L_F$ gives a vanishing contribution to $\langle J^x \rangle$ for the same reason,
\bea
\d L_F & = & - \frac{1}{4} \N (2 \p \a')^2 \sqrt{\g} \g^{ij}\g^{kl} \int dt \frac{\d }{\d \partial_t A_x} \left( F_{ik} F_{jl} \right) \partial_t \d A_x \log \e \\ & = & + \frac{1}{2} \N (2 \p \a')^2 \int dt \dot{A}_x(\e) \partial_t \d A_x \log \e + O(\e^4 \log \e) \nonumber \\ & = & O(\e^4 \log \e) \nonumber
\eea
We then have $\frac{\d S}{\d A_x(\e)}= \frac{\d L}{\d \partial_z A_x}$ and hence $\langle J^x \rangle = \frac{\d L}{\d \partial_z A_x}$.

$\langle J^y \rangle$ is very similar. $A_y$ depends on $x$ so we reinstate $\int dx$. We have
\beq
\d S = - \int dz dx \left ( \frac{\d L}{\d \partial_z A_y} \partial_z \d A_y + \frac{\d L}{\d \partial_x A_y} \partial_x \d A_y \right )
\eeq
The same argument as above applies for the first term, and for the second term we observe that $\frac{\d L}{\d \partial_x A_y}$ is $x$-independent on-shell. Demanding that the fluctuation be well-behaved at $x = \pm \infty$ gives $\int dx \partial_x \d A_y = 0$ and no contribution from $L_F$. We thus have $\langle J^y \rangle = \frac{\d L}{\d \partial_z A_y}$.

%
%
%
%
%
%
%
%
%
%
%
%
%
%
%
\chapter {Conclusion}\label{conclusion}
\setcounter{equation}{0}
\renewcommand{\theequation}{\thesection.\arabic{equation}}

We have covered much ground, so let us summarize the salient points to take away from all of this. We studied the holographic dual of a strongly-coupled non-Abelian gauge theory, focusing on the thermodynamics and transport properties of the fields in the fundamental representation of the gauge group. Specifically, we studied the theory at zero temperature and finite baryon number density, where, using the holographic description, we discovered a second-order phase transition (in the grand canonical ensemble) when the chemical potential was equal to the mass of the flavor fields. The transition is between a zero-density phase with a gapped, discrete spectrum of stable mesons and a finite-density phase with a gapless, continuous spectrum of unstable mesons. At finite temperature, we introduced external electric and magnetic fields and, again using the holographic description, computed the resulting currents of flavor fields, from which we extracted the conductivity tensor. As an added bonus, we computed the drag force acting on the flavor fields.

Much work remains to be done. The most obvious task is to complete the phase diagram of the theory in the full parameter space of $m$, $T$, $\< J^t \>$, $E$ and $B$, in both the canonical and grand canonical ensembles. The D7-brane solutions discussed above are almost certainly insufficient. As mentioned in section \ref{finitedensityphasediagram}, these D7-brane solutions all describe \textit{homogeneous} phases in the gauge theory, whereas we have good reasons to expect that mixed phases may play a pivotal role in certain regions of the phase diagram. As mentioned in section \ref{zeromsol}, additional homogeneous phases may also be present: at low enough temperature and high enough chemical potential, we expect the hypermultiplet scalars to undergo Bose-Einstein condensation, for instance. The D7-brane solutions above do not describe such a condensate. Some regions of the phase diagram are currently not understood at all, for example the regions described by singular D7-brane emebeddings, as mentioned in section \ref{electric}. Additionally, the complete meson spectrum (and/or spectral functions) should be computed in the full parameter space, to provide a complete picture of the physics in any phase. And another obvious direction to go is to continue studying the transport properties of the flavor fields. For example, what current is generated by a thermal gradient (the Nernst effect)?

We argued in the introduction that gauge-gravity duality may teach us something about strongly-coupled non-Abelian gauge theories. More recently, however, a new and exciting application of gauge-gravity duality has emerged. Many \textit{condensed matter} systems may be described by strongly-coupled, scale-invariant field theories, for example cold atoms at unitarity or high-$T_c$ superconductors near their quantum critical points. Gauge-gravity duality has already provided toy models for such systems, and promises much more \cite{Herzog:2007ij,Hartnoll:2007ai,Hartnoll:2007ih,Hartnoll:2007ip,Hartnoll:2008hs,Son:2008ye,Balasubramanian:2008dm}. Only recently (as of the time of this writing), gauge-gravity duality has even described superconductivity \cite{Gubser:2008px,Hartnoll:2008vx,Gubser:2008zu}!

One key feature noticably lacking in these models, however, is a Fermi surface. One big obstacle to realizing a Fermi surface holographically is supersymmetry: gauge-gravity systems usually describe fermionic \textit{and} bosonic charge carriers. Breaking supersymmetry means we are not guaranteed stability, however. To find a holographic description of a finite density of fermionic charge carriers, then, we will probably have to break supersymmetry and then add some other ingredient to ensure stability.

We hasten to point out, however, that looking for a Fermi surface is a fool's errand: generically, the Fermi surface is unstable. For fermions at low temperature and high density, where we expect a Fermi surface, the appropriate degrees of freedom are excitations about the Fermi surface. Standard renormalization arguments then reveal that, for excitations with equal but opposite momenta, under renormalization group flow toward the Fermi surface, any repulsive interaction becomes irrelevant while any attractive interaction becomes relevant. As a result, fermions on opposite sides of the Fermi surface bind into pairs, called Cooper pairs, and the resulting scalar condensate breaks the associated $U(1)$ symmetry (or gauge invariance). This is of course the BCS theory of superconductivity, applicable to electrons in a low-temperature metal or quarks in massless, three-flavor QCD at asymptotically high density. Good explanations of BCS theory appear in refs. \cite{Polchinski:1992ed,Shankar:1993pf,Rajagopal:2000wf,Kaplan:2005es}. Some canonical papers about BCS theory in high-density QCD are refs. \cite{Bailin:1983bm,Alford:1997zt,Rapp:1997zu,Alford:1998mk,Son:1998uk}.

The question we should ask is thus not, ``What is a holographic description of a Fermi surface?'' but rather the more subtle question, ``What is the holographic description of the ground state of a low-temperature, finite-density system of strongly-interacting fermions?'' Any insights that holography can provide into such systems would be valuable for both condensed matter physics and QCD.

We hasten to point out still another caveat: the large-$N_c$ limit changes the physics significantly. For massless QCD at low temperature and high density, \textit{and} in the 't Hooft limit, the ground state of the system may actually be spatially inhomogeneous, forming a so-called ``chiral density wave,'' analogous to similar plane-wave ground states in low-dimensional electron systems \cite{Deryagin:1992rw,Shuster:1999tn}. In practical terms, this would require, for example, studying D7-brane embeddings that depend on Minkowski directions. These are typically very difficult to construct. The rewards for doing so, however, may provide sufficient motivation for us to ``shut up and calculate.''

Gauge-gravity duality is clearly a versatile tool. We hope only that, in this dissertation, we have convinced the reader that gauge-gravity duality is useful for studying strongly-coupled gauge theories, as a tool on par with and complementary to other methods, such as lattice simulation. We believe that holographic methods are worth learning, as they should prove fruitful for a long time to come.

\bibliography{refs}
\bibliographystyle{JHEP}

\vita{Andrew Hill O'Bannon was born in 1979 in Richmond, Virginia. In 2002 he earned a Bachelor of Arts in Physics and the Writing Seminars from the Johns Hopkins University and was awarded a Jack Kent Cooke Foundation Graduate Scholarship. In 2004 he earned a Master of Science in Physics degree from the University of Washington. In 2008 he earned a Doctor of Philosophy in Physics degree from the University of Washington.

In 1997 he won Virginia's state-wide playwriting contest \textit{New Voices for the Theatre} and his winning play \textit{Sing for the Dead} was later filmed as a short video. His short stories and poems have been published in the \textit{Richmond Times-Dispatch} and \textit{Lite} (Balitmore's free literary newspaper), as well as the Johns Hopkins University student publications \textit{Frame of Reference}, \textit{Zeniada} and \textit{j. magazine}.
}

\end{document}